\documentclass[11pt,a4paper]{article}
\usepackage[margin=0.82in]{geometry}
\usepackage[T1]{fontenc}
\usepackage[utf8]{inputenc}
\usepackage{lmodern}
\usepackage{microtype}
\usepackage{amsmath,amssymb,amsfonts,mathtools,bm}
\usepackage{graphicx}
\usepackage{float}
\usepackage[section]{placeins}
\usepackage{booktabs,array,multirow,tabularx}
\usepackage{caption,subcaption}
\usepackage{enumitem}
\usepackage{xcolor}
\usepackage{hyperref}
\usepackage[numbers,compress]{natbib}

\hypersetup{
  colorlinks=true,
  linkcolor=blue!55!black,
  citecolor=blue!55!black,
  urlcolor=blue!55!black
}

\setlist[itemize]{leftmargin=1.35em}
\setlist[enumerate]{leftmargin=1.35em}

\newcommand{\R}{\mathbb{R}}

\newcommand{\abs}[1]{\left\lvert #1 \right\rvert}
\newcommand{\ip}[2]{\left\langle #1,#2 \right\rangle}
\newcommand{\Arg}{\operatorname{Arg}}

\newcommand{\calF}{\mathcal{F}}
\newcommand{\calS}{\mathcal{S}}
\newcommand{\calN}{\mathcal{N}}

\newcommand{\noise}{\mathrm{noise}}

\title{Multimodal Optical Feature Extraction with a Free-Space Photonic Extreme Learning Machine}

\usepackage{authblk}

\author[1,*]{Anushka Kumari}
\author[1,*]{Anushree Khisti}
\author[1,2,*]{Abhinav Choube}
\author[1]{Devansh Satra}
\author[1]{Srivatsa Murali}
\author[1]{Anshuman Kumar\thanks{\texttt{anshuman.kumar@iitb.ac.in}}}

\affil[1]{Laboratory of Optics of Quantum Materials, Department of Physics,
Indian Institute of Technology Bombay, Mumbai 400076, India}
\affil[2]{Department of Physics and Electronics,
Institute for Excellence in Higher Education, Bhopal 462042, India}
\affil[*]{\small These authors contributed equally.}
\date{}

\begin{document}
\maketitle

\begin{abstract}
Photonic extreme learning machines (PELMs) replace a digitally trained hidden layer by a fixed optical transformation, allowing a high dimensional feature map to be generated by physical propagation while only the final readout is learned.  Existing free-space PELM demonstrations have established this principle for image and tabular benchmarks, but a unified multimodal optical feature extractor spanning structurally different data types has remained largely undeveloped.  Here we demonstrate a single free-space PELM platform for image, audio derived, binary tabular, and regression tasks using phase only SLM encoding, Fourier like free space propagation, and camera intensity detection.  The same optical apparatus achieves \(96.56\%\) accuracy on MNIST, \(95.67\%\) on spoken digit audio from log-Mel spectrograms, \(100.00\%\) on Mushroom classification, and \(0.0699\) NRMSE on Abalone regression. To our knowledge, this is the first free space PELM spanning image, audio derived, and tabular tasks in one physical pipeline, and the first PELM implementation of spectrogram based spoken digit classification.  Empirical distance preservation and kernel alignment diagnostics reveal two operating regimes: geometry preserving for image and regression tasks, and distributed class mean accumulation for audio derived spectrograms.  These results establish multimodal PELMs as a practical route toward general purpose optical machine learning.
\end{abstract}

\section{Introduction}

Optical computing has emerged as a serious hardware direction for machine learning because the elementary operations required by modern learning systems, including high-dimensional linear mixing, convolution, projection, and feature expansion, are naturally implemented by interference, diffraction, and propagation. Several distinct photonic machine learning platforms have therefore been developed. Programmable Mach-Zehnder interferometer meshes and coherent nanophotonic circuits implement trainable or programmable linear transformations on chip \cite{reck1994experimental,clements2016optimal,shen2017deep,harris2017quantum,bogaerts2020programmable}. Diffractive optical neural networks use spatially engineered passive layers to implement trained free-space transformations \cite{lin2018diffractive,mengu2020diffractive}, while on chip diffractive photonic processors extend the diffractive computing idea to integrated photonic tensor and graph processing architectures \cite{yan2022graph,huang2022onchip}. Integrated photonic tensor cores and wavelength multiplexed processors exploit microcombs, phase change materials, non volatile photonic memories, and high speed modulation to accelerate convolution and matrix vector operations \cite{rios2015integrated,feldmann2019alloptical,feldmann2021parallel,lin2024tfln}. Photonic reservoir computers use multimode propagation, delay systems, speckle dynamics, or nonlinear optical nodes as fixed dynamical reservoirs with a trained electronic readout \cite{brunner2013parallel,paudel2020classification,rafayelyan2020large}. Recent work has also begun to explore multimodal and on-chip training paradigms in photonic neural networks \cite{hughes2018insitu,gu2021l2ight,Cheng2024}. Broader reviews have emphasized that optical inference can be performed through deeply different physical mechanisms, ranging from coherent integrated photonics to deep optics, computational imaging, and neuromorphic photonics \cite{wetzstein2020inference,shastri2021photonics}.

Despite this progress, many photonic neural networks remain difficult to deploy as general purpose laboratory scale learning systems.  Interferometric meshes require precise calibration of many phase shifters.  Diffractive networks require task specific design of physical or virtual diffractive layers.  Phase change and resonator based processors are powerful but depend on integrated fabrication and device level control.  Photonic reservoir computers avoid full end to end training, but recurrent or dynamical implementations often require careful temporal encoding and may be naturally suited to time-series tasks rather than arbitrary static modalities.  These considerations motivate a complementary question: how far can one go with an extremely simple optical feature extractor consisting only of a phase encoder, free-space propagation, spatial filtering, and a camera?

Extreme learning machines (ELMs) provide the algorithmic framework for this question \cite{huang2006elm}.  An ELM maps data to a high dimensional hidden representation using fixed random or structured weights, and only the final linear readout is trained.  The optical analogue is a photonic extreme learning machine (PELM), in which the hidden feature map is produced physically.  Pierangeli et al.\ \cite{pierangeli2021pelm} demonstrated a free-space PELM in which an SLM phase encodes the input and an embedding mask, lens propagation generates the optical mixing, camera intensity supplies the nonlinear features, and ridge regression learns the readout.  That work established that free-space propagation can compete with digital kernel machines on image and tabular benchmarks without nanofabricated optical networks or optical backpropagation.

The present work advances the PELM framework in multiple directions. First, we demonstrate a single free-space PELM pipeline across genuinely distinct data modalities: MNIST image classification, spoken digit audio classification from log-Mel spectrograms, Mushroom binary tabular classification, and Abalone tabular regression.  ELMs have been applied digitally to many individual data types, and photonic reservoirs have been used for temporal waveform and audio related tasks \cite{paudel2020classification}, but to the best of our knowledge no previous free-space PELM has demonstrated image, audio derived, and tabular learning within one fixed physical optical pipeline.  We are also not aware of a prior PELM implementation using Mel spectrogram phase encoding for spoken digit classification.

Second, we provide empirical feature diagnostics across all modalities.  Pairwise distance preservation, centered kernel alignment (CKA), and class mean separation are measured directly from the optical reservoir outputs and used to characterize two distinct operating regimes: one in which the optical feature map preserves input geometry (MNIST, Abalone) and one in which high dimensional readout accumulates distributed class signals despite weak global distance preservation (FSDD).  A detailed theoretical analysis connecting these observations to random projection theory, intensity induced quadratic features, and matched filter accumulation is deferred to a companion publication.

The resulting message is that a PELM should not be viewed as implementing a single universal kernel independent of modality.  Instead, the optical hardware provides a high dimensional physical feature map whose utility depends on the interaction among input encoding, embedding mask, intensity detection, and readout.  For image like and regression tasks, distance preservation and local geometry can be directly useful.  For audio derived spectrograms, class structure may survive as distributed mean shifts rather than as globally preserved distances.  Characterizing these two regimes empirically, and demonstrating that the same fixed hardware supports both, is the central contribution of the present work.

\FloatBarrier
\section{Photonic extreme learning machine architecture}
\label{sec:architecture}

\subsection{ELM readout model}

Let \(\{(x_i,t_i)\}_{i=1}^N\) be a supervised dataset, with inputs \(x_i\in\R^d\) and targets \(t_i\).  A conventional ELM defines a fixed nonlinear feature map \(g:\R^d\to\R^M\), constructs the hidden layer matrix
\begin{equation}
H = \begin{bmatrix} g(x_1)^\top \\ g(x_2)^\top \\ \vdots \\ g(x_N)^\top \end{bmatrix}\in\R^{N\times M},
\end{equation}
and learns only the readout matrix \(\beta\).  For one hot classification targets or regression targets collected in \(T\), the primal ridge solution is
\begin{equation}
\beta = (H^\top H+\lambda I_M)^{ -1}H^\top T,
\label{eq:ridge primal}
\end{equation}
where \(\lambda>0\) is the ridge parameter.  When it is advantageous to work in the sample dimension, the equivalent dual solution is
\begin{equation}
\beta = H^\top (H H^\top+\lambda I_N)^{ -1}T.
\label{eq:ridge dual}
\end{equation}
Inference is then \(\hat{T}=H_{\mathrm{test}}\beta\), with the class determined by the largest output component for classification.  In a PELM, \(g(x)\) is not explicitly computed by a digital hidden layer; it is measured from an optical experiment.

\subsection{Optical feature map}

Each input is rescaled to a phase pattern and displayed on a phase only SLM.  After reshaping an input into a two dimensional array \(x_{pq}\), the displayed phase is
\begin{equation}
\varphi_{pq}(x)=x_{pq}+W_{pq},\qquad x_{pq}\in[0,\pi],
\label{eq:phase -mask}
\end{equation}
where \(W\) is a fixed embedding mask kept unchanged during both training and testing.  The incident field immediately after the SLM is approximately
\begin{equation}
E^{\mathrm{in}}_{pq}(x)=\exp[i\varphi_{pq}(x)] = \exp[i(x_{pq}+W_{pq})].
\label{eq:input field}
\end{equation}
The free-space optical path and lenses implement a structured linear mixing of this complex field. We write the sampled field at the selected camera modes as
\begin{equation}
u(x)=\calS\calF\left[\exp(i(x+W))\right],
\label{eq:field propagation}
\end{equation}
where \(\calF\) denotes the fourier like propagation operator and \(\calS\) denotes spatial selection of the first diffracted order and the finite camera readout region.  The camera measures intensity rather than complex field amplitude. After binning and normalization, the feature vector is
\begin{equation}
g(x)=\calN\left\{G\left(\abs{u_1(x)}^2\right),\ldots,G\left(\abs{u_M(x)}^2\right)\right\}\in\R^M,
\label{eq:optical feature}
\end{equation}
where \(M=4096\) in the reported experiments, \(\calN\) denotes the applied feature normalization, and \(G\) is the camera response. A useful phenomenological model for the saturating response is
\begin{equation}
G(I)\simeq \frac{I}{I+I_s},
\label{eq:camera -sat}
\end{equation}
with saturation intensity \(I_s\). Even in the weakly saturated regime, the modulus square operation introduces a nonlinear transformation of the phase encoded input.

Raw camera features are preprocessed before training. First, per sample row mean subtraction removes the global background offset:
\begin{equation}
    \tilde{G}_i = G_i  - \frac{1}{M}\sum_{m=1}^{M} G_{im},
\end{equation}
where $M$ is the total number of readout features. For kernel analysis, 
an additional column mean centering step is applied to remove the dataset level mean from each feature channel:
\begin{equation}
    \hat{G}_{im} = \tilde{G}_{im}  - \frac{1}{N}\sum_{j=1}^{N} \tilde{G}_{jm},
\end{equation}
where $m = 1, \ldots, M$ indexes the feature dimension and the sum runs 
over all $N$ training samples. Finally, hyperspherical ($\ell_2$) 
normalization is applied to the centered features:
\begin{equation}
    g(x_i) = \frac{\hat{G}_i}{\|\hat{G}_i\|_2 + \varepsilon},
\end{equation}
where $\epsilon = 10^{ -12}$ is a small numerical constant added for stability. This operation maps the features onto a unit hypersphere, hereby making the inner product $\ip{g(x_i)}{g(x_j)}$ correspond to the cosine similarity between two encoded inputs. This normalization also improves the numerical conditioning of the ridge regression readout. For the 
ridge regression readout (lambda cross validation and testing), only the per sample centering and $\ell_2$ normalization steps are applied, without the additional column mean subtraction.

\begin{figure}[h]
\centering
\includegraphics[width=\linewidth]{\detokenize{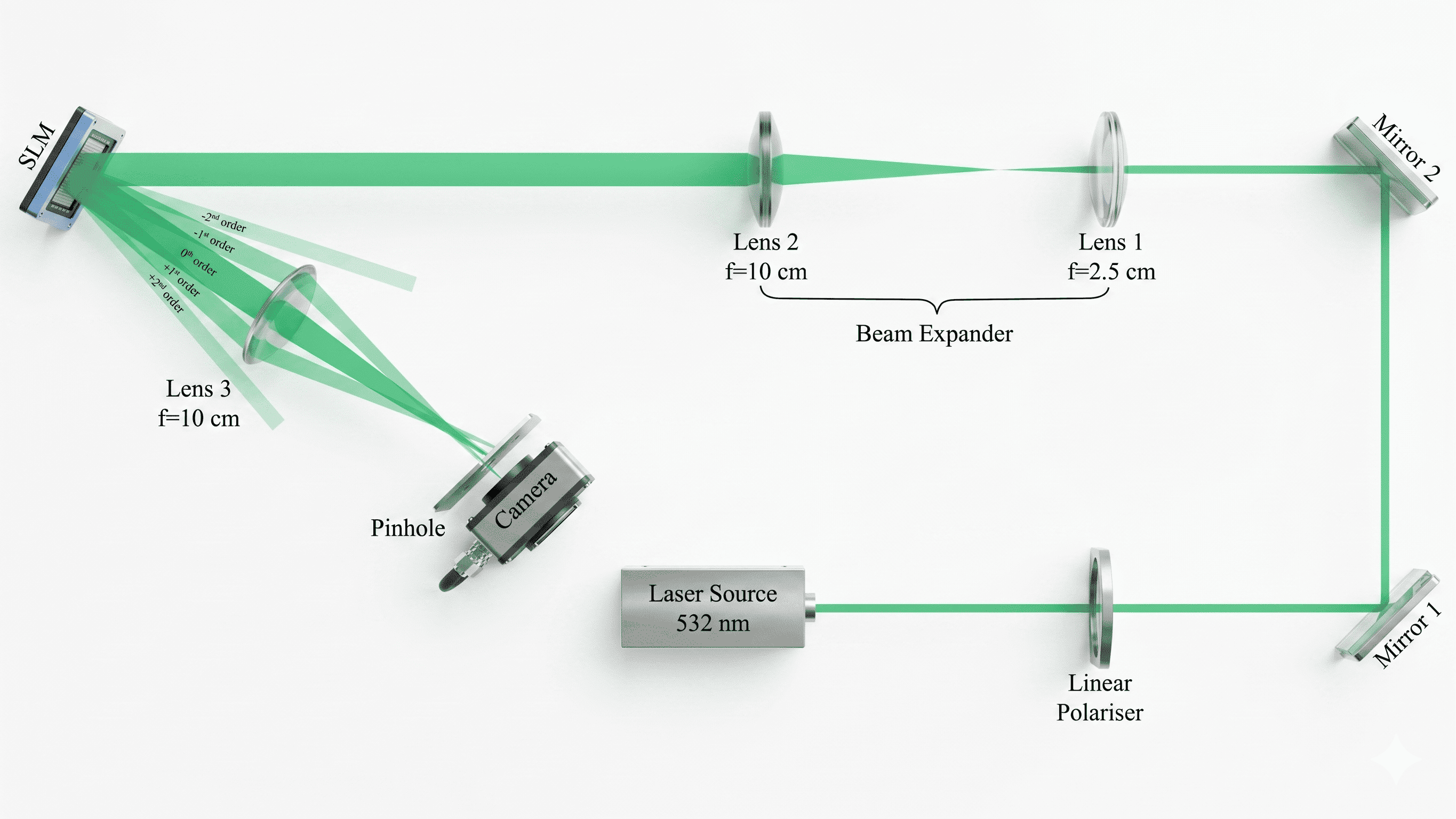}}
\caption{Experimental free-space PELM.  A 532 nm laser is polarisation -controlled and phase modulated by a spatial light modulator. The optical path implements Fourier like free-space mixing, an iris selects the informative first diffracted order and suppresses the zero order background, and a camera records intensity features for the ridge regression readout.}
\label{fig:setup}
\end{figure}

\subsection{Experimental implementation}

The experimental setup consists of a green continuous wave laser, a phase only spatial light modulator (SLM), a \(4f\) optical propagation system, an iris for spatial filtering, and a CMOS camera. The SLM serves both as the phase encoder and as a diffraction grating. During initial alignment, the unmodulated zero order beam produced a strong background contribution that could easily dominate the recorded camera signal if left unsuppressed. In contrast, the first diffracted order carried the phase modulated information and generated a substantially richer speckle like intensity distribution. As a result, both the iris aperture and the polarizer orientation became important alignment parameters throughout the experiments.

Camera frames were converted into \(M = 4096\) readout features by selecting a central region of interest and binning local \(10 \times 10\) pixel blocks into a \(64 \times 64\) feature grid. Although the camera sensor has a native resolution of \(1440 \times 1080\) pixels, only a cropped region of \(1280 \times 1024\) pixels was used during processing. This cropping was necessary because the optical field followed an approximately Gaussian intensity profile, leaving pixels near the sensor edges weakly illuminated or completely dark. Including these regions artificially lowered the average frame intensity and distorted the measured saturation statistics. Before ridge regression, feature normalization was applied to reduce sensitivity to laser, camera response fluctuations, and overall intensity variation. Over long acquisition runs, stable operation additionally required continuous camera streaming, careful bit depth scaling, and sufficient liquid crystal settling time after each SLM update. Reducing the settling interval from 0.10\,s to 0.05\,s consistently degraded classification performance, likely because the liquid crystal phase state was not fully relaxed.

\section{Encoding strategies and multimodal protocol}
\label{sec:protocol}

\subsection{Noise embedding}

The baseline embedding is a random phase mask
\begin{equation}
W^{\noise}_{pq}\sim \mathcal{U}(0,\rho),
\label{eq:noise embed}
\end{equation}
optionally with a finite spatial correlation length set by block wise or smoothed disorder.  This mask plays the role of a fixed random bias or fan in matrix.  It breaks symmetries of the pure Fourier transform, lowers coherent artifacts, and increases the effective rank of the optical feature map.

\subsection{Fourier embedding}

As an alternative to using purely random phase noise, the free space PELM can also employ a Fourier based embedding strategy. Instead of introducing only a single spatial frequency component, this method adds a broadband spatial carrier composed of multiple frequency components to the input phase pattern. The embedding mask is generated by summing \(N_f\) discrete spatial frequencies along a fixed direction, where each component is assigned a random phase offset \(\phi_n \sim \mathcal{U}(0, 2\pi)\):
\begin{equation}
W^{\mathrm{Fourier}}_{pq} = \Arg \left[ \sum_{n=1}^{N_f} \exp\left\{ i \left[ 2\pi n (X_{pq} + Y_{pq}) + \phi_n \right] \right\} \right],
\label{eq:fourier -embed}
\end{equation}
with \((X_{pq}, Y_{pq})\) representing the normalized spatial coordinates of the SLM.

Introducing this directional carrier shifts the modulated optical field away from the optical axis, allowing the information carrying first diffracted order to be more cleanly separated from the unmodulated zero order background. Compared with the fully stochastic noise mask, the Fourier embedding imposes a more structured and multi-periodic phase modulation during optical mixing. This increases the spectral richness of the resulting phase mask and, in turn, broadens the effective dimensionality of the optical feature space 
\begin{figure}[H]
\centering
\includegraphics[width=1\linewidth]{\detokenize{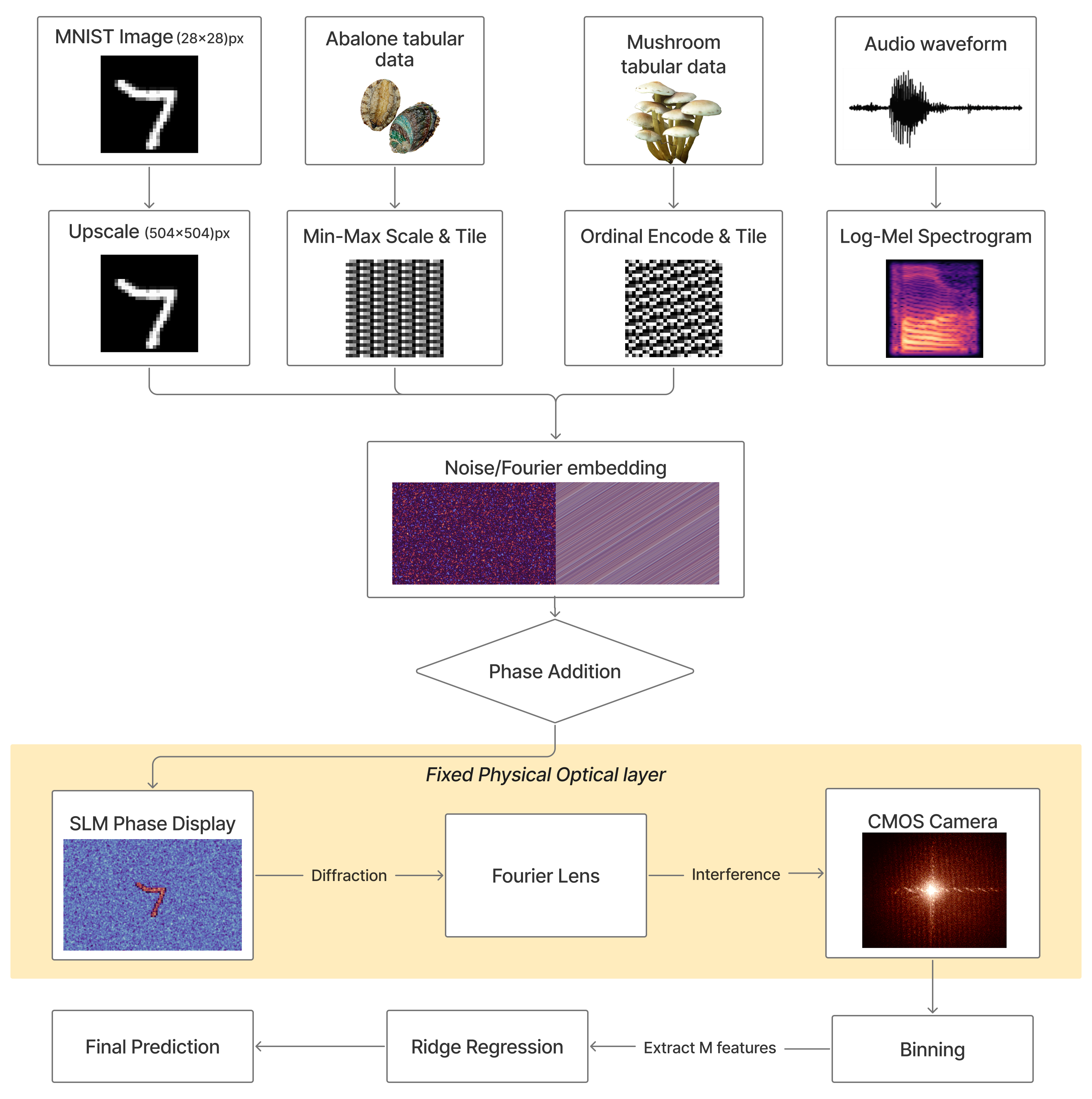}}
\caption{Overview of the multimodal PELM pipeline used in this work. Image data, audio derived spectrograms, and tabular inputs are first converted into phase patterns for SLM display and then combined with fixed embedding mask. The encoded optical field undergoes free space propagation, after which the resulting intensity patterns are captured by the camera and converted into feature vectors. Only the final readout layer is trained digitally, while the optical transformation itself remains fixed}
\label{fig:workflow}
\end{figure}

\begin{figure}[H]
\centering
\begin{subfigure}{0.49\linewidth}
\centering
\includegraphics[width=\linewidth]{\detokenize{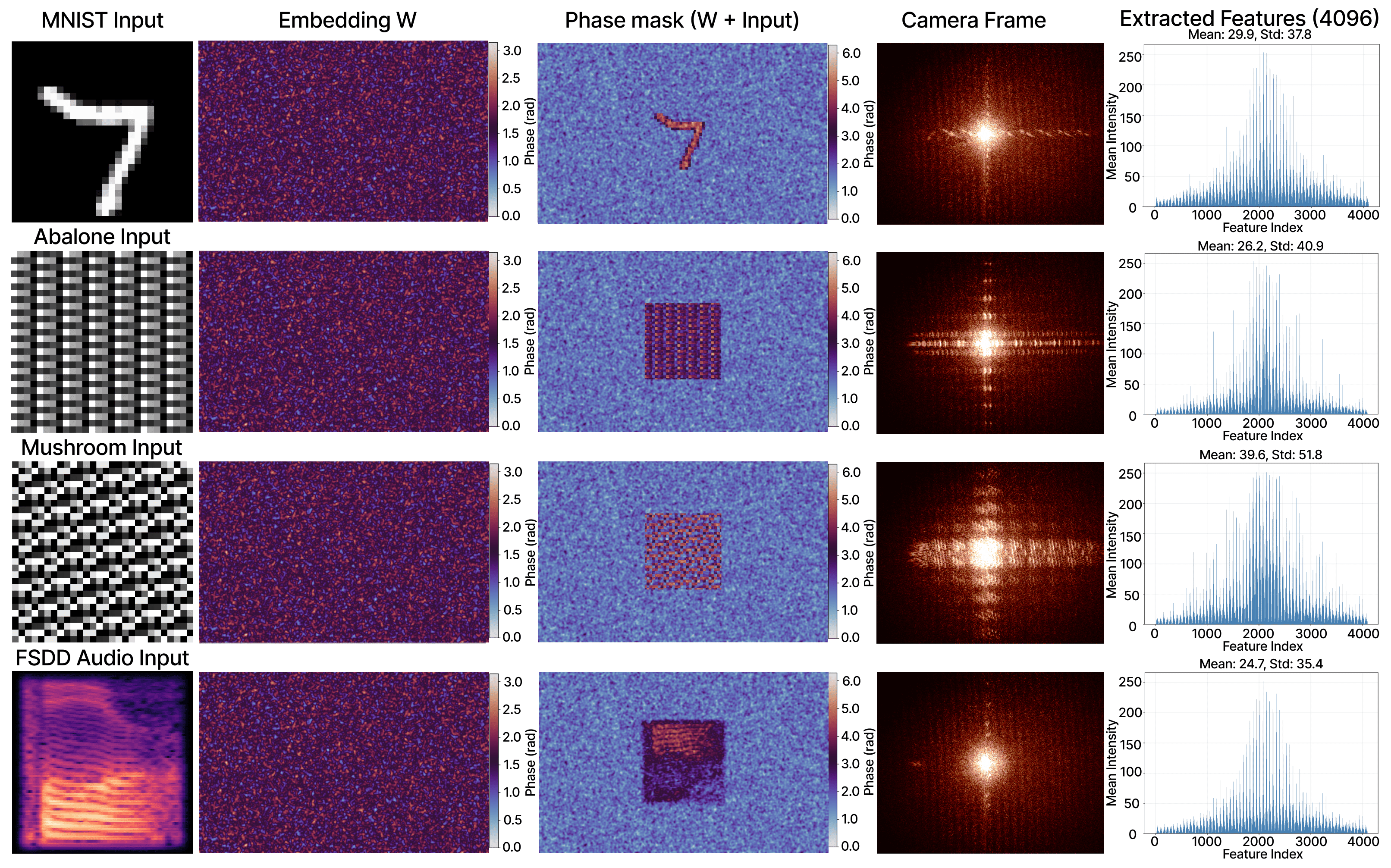}}
\caption{Noise embedding}
\end{subfigure}
\begin{subfigure}{0.49\linewidth}
\centering
\includegraphics[width=\linewidth]{\detokenize{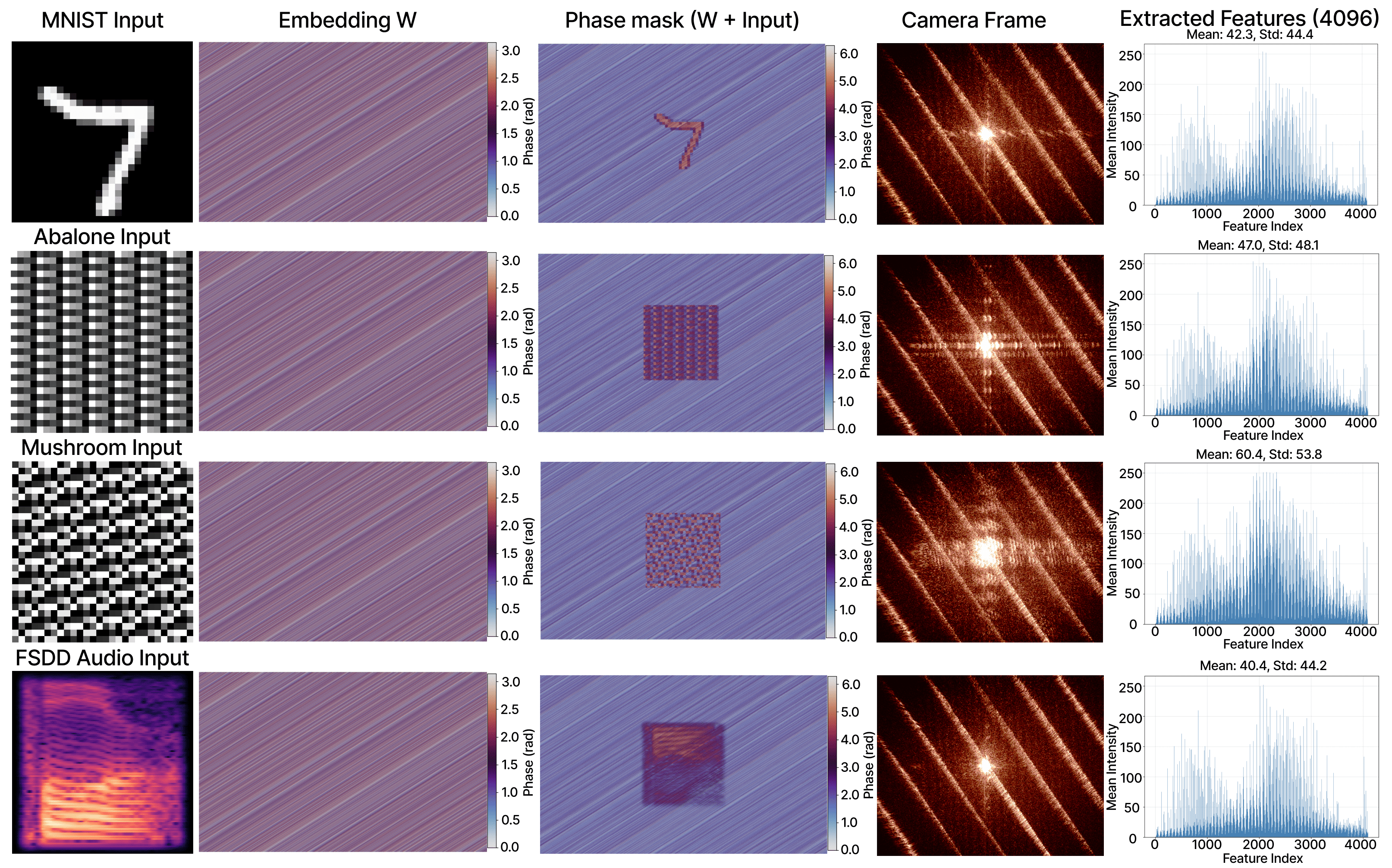}}
\caption{Fourier embedding}
\end{subfigure}

\caption{ Comparison of the two optical encoding approaches used in the experiments: (a) noise embedding and (b) Fourier embedding. Each row illustrates a representative stage of the pipeline, including the original input pattern, the corresponding phase mask displayed on the SLM, and the resulting camera recorded intensity distribution after optical propagation.}
\label{fig:embedding -visual}
\end{figure}

\subsection{Datasets}

Four datasets representing different data modalities were evaluated in this work. MNIST was used as the primary image classification benchmark and consists of 60,000 training images together with 10,000 test images distributed across ten handwritten digit classes. Each grayscale image has a spatial resolution of \(28 \times 28\) pixels, making it well suited for direct optical phase encoding and free-space propagation experiments.

The Free Spoken Digit Dataset (FSDD)~\cite{fsdd2017} was employed as an audio classification task. The dataset contains \(3{,}000\) spoken recordings of digits \(0\text{ - }9\), sampled at \(8\,\text{kHz}\), with individual recordings typically ranging from \(0.5\) to \(1\) second in duration. Audio recordings were converted into log-Mel spectrogram representations before phase encoding, following the use of Mel scale spectral representations in speech processing~\cite{davis1980comparison}, thereby forming a ten class audio derived classification task. For the experimental runs, $2,700$ recordings were used for training, and remaining $300$ recordings for final performance evaluation, with balanced class distributions across all spoken digits.

The Mushroom dataset~\cite{mushroom_dataset} was used as a binary classification task to distinguish edible from poisonous mushrooms based on 22 categorical tabular attributes describing mushroom characteristics such as cap shape, cap surface, cap color, bruising, odor, gill attachment, gill spacing, stalk shape, habitat, and population type. Of the original 8,124 samples, 2,480 samples containing missing stalk root information were excluded, yielding 5,644 valid records. Since the dataset consisted entirely of categorical attributes, each feature was encoded independently using deterministic per feature categorical index encoding to preserve category identity without introducing artificial ordinal relationships. The encoded features were subsequently min-max scaled to the range \([0,255]\) and tiled into \(28 \times 28\) two dimensional spatial representations for optical phase modulation. The valid records were then divided into a training set of 4,124 samples and a testing set of 1,520 samples.

The Abalone dataset~\cite{abalone_dataset} was used for the regression experiments and contains 4,177 samples described by eight input features related to the physical properties of abalone specimens. These features include a categorical sex label (Male, Female, or Infant), shell length, shell diameter, shell height, whole weight, shucked weight, viscera weight, and shell weight. Here, whole weight refers to the total mass of the abalone before processing, shucked weight represents the mass of the edible meat after removal from the shell, viscera weight corresponds to the internal organ mass, and shell weight denotes the dry shell mass after processing. The regression target is the number of shell rings, which is commonly used as an approximate estimate of abalone age. The dataset was divided into 3,480 training samples and 696 testing samples, corresponding to an approximate 80/20 split.

To maintain compatibility with the optical processing pipeline, nonimage inputs were reshaped or tiled into two dimensional phase patterns before being displayed on the spatial light modulator (SLM). The same free space propagation setup and readout training procedure were then used across all datasets to ensure a consistent framework. The regularization parameter \(\lambda\) was optimized independently for each dataset using validation based hyperparameter sweeps.

\section{Experimental Results}
\label{sec:results}

The experimental performance was evaluated across two embedding strategies: the random noise embedding and the Fourier embedding.  For both strategies, the ridge regularization parameter \(\lambda\) (Eq.~\ref{eq:ridge primal}) was optimized independently for each dataset by a logarithmic grid search sweeping \(\lambda\) from \(10^{ -5}\) to \(10^{1}\).  The optimal \(\lambda\) was selected using cross validation performance on the training split, highest accuracy for classification tasks (MNIST, FSDD, Mushroom) and the lowest NRMSE for regression (Abalone).
The optimal $\lambda$ values are dataset dependent and reflect the interaction between the optical feature geometry and the ridge readout; the test set was held out throughout and evaluated once after final model selection. The close agreement between cross validation and test performance across all datasets (Table~\ref{tab:performance}) confirms that the selected $\lambda$ values do not lead to overfitting.

Table~\ref{tab:performance} summarizes the optimized classification results.  For each dataset, the reported accuracy corresponds to the best value obtained over the \(\lambda\) sweep, with the corresponding value of \(\lambda\) shown explicitly.  The comparison to earlier free-space PELM work is not included as a table column, since the goal here is to establish a unified multimodal optical feature extraction framework rather than to frame the work as a direct benchmark against a single previous implementation. Nevertheless, to the best of our knowledge, the MNIST accuracy reported here is among the highest, reported for a free-space PELM architecture using an SLM camera optical reservoir and a trained linear readout.

\begin{table}[H]
\centering
\caption{Optimized experimental performance of the free-space PELM. The ridge parameter \(\lambda\) was swept independently for each dataset and embedding. Both the noise and Fourier embeddings achieve comparable performance across all datasets, with the Fourier embedding providing a small but consistent improvement in most cases. The best classification accuracies obtained with the Fourier embedding are \(96.56\%\) on MNIST, \(95.67\%\) on FSDD spoken digits, and \(100.00\%\) on Mushroom classification, while the regression performance on Abalone remains within a small difference in NRMSE between embeddings.}
\label{tab:performance}

\small
\renewcommand{\arraystretch}{1.18}

\begin{tabularx}{\linewidth}{@{}lXlccc@{}}
\toprule

Dataset & Task & Embedding & Best \(\lambda\) & Test performance & CV performance \\

\midrule

MNIST 
& 10 class handwritten digit classification 
& Noise 
& \(7.278\times10^{ -4}\)
& \(96.35\%\) accuracy 
& \(95.86\%\) accuracy \\

MNIST 
& 10 class handwritten digit classification 
& Fourier 
& \(7.278\times10^{ -4}\) 
& \(96.56\%\) accuracy
& \(96.46\%\) accuracy \\

FSDD 
& 10 class spoken digit classification from log-Mel spectrograms 
& Noise 
& \(1.887\times10^{ -3}\)
& \(93.00\%\) accuracy
& \(91.56\%\) accuracy \\

FSDD 
& 10 class spoken digit classification from log-Mel spectrograms 
& Fourier 
& \(1.887\times10^{ -3}\) 
& \(95.67\%\) accuracy
& \(95.22\%\) accuracy \\

Mushroom 
& Binary tabular classification 
& Noise 
& \(1.00\times10^{ -5}\) 
& \(100.00\%\) accuracy
& \(100.00\%\) accuracy \\

Mushroom 
& Binary tabular classification 
& Fourier 
& \(1.00\times10^{ -5}\) 
& \(100.00\%\) accuracy
& \(100.00\%\) accuracy \\

Abalone 
& Tabular regression 
& Noise 
& \(4.89\times10^{ -3}\) 
& \(0.0699\) NRMSE
& \(0.0770\) NRMSE \\

Abalone 
& Tabular regression 
& Fourier 
& \(4.89\times10^{ -3}\)
& \(0.0704\) NRMSE
& \(0.0768\) NRMSE \\

\bottomrule
\end{tabularx}

\end{table}

The MNIST experiments show that the optical feature map remains highly effective on a large scale image classification benchmark. After optimizing the ridge regularization parameter \(\lambda\), the PELM achieved a test accuracy of \(96.35\%\) using the noise embedding and \(96.56\%\) using the Fourier embedding.

A similar pattern was observed across the other datasets, where the difference in performance between the two embeddings remained relatively small. In most cases, the variation was within \(1\text{  - }2\%\) for classification accuracy and only a few thousandths in NRMSE for regression tasks. These results suggest that both embeddings generate strong optical feature representations, although the Fourier embedding consistently provides a slight performance advantage across most benchmarks.

For MNIST, the corresponding cross validation (CV) accuracies were \(95.86\%\) for the noise embedding and \(96.46\%\) for the Fourier embedding. The close agreement between CV and test performance indicates that the model generalizes consistently across validation folds rather than benefiting from a favorable train test split. During hyperparameter optimization, the CV scores were computed using 5 fold cross validation. The training set was divided into five folds, with four folds used for training and the remaining fold used for validation in each run . The final CV accuracy was obtained by averaging the validation performance across all folds. StratifiedKFold was used for classification datasets to preserve class balance within each split, while standard KFold was used for regression experiments.

The confusion matrices shown in Fig.~4 further shows that both embeddings maintain a strongly diagonal classification structure. However, the Fourier embedding produces slightly fewer off diagonal digit confusions, indicating a slight improvement in class separability within the optical feature space.

\begin{figure}[H]
\centering
\begin{subfigure}{0.49\linewidth}
\centering
\includegraphics[trim={0cm 0cm 0cm 1.5cm}, clip, width=\linewidth]{\detokenize{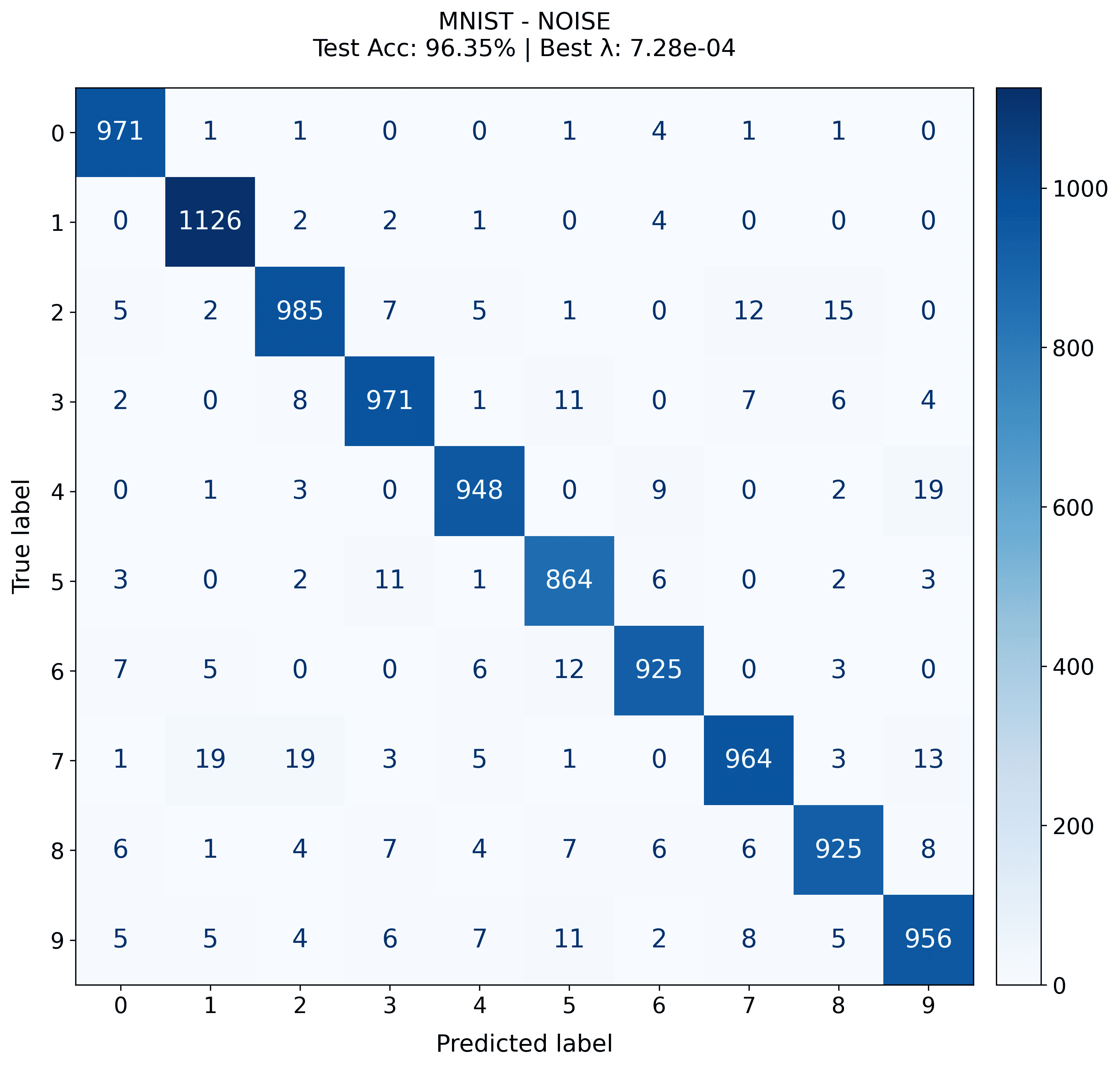}}
\caption{Noise embedding, \(96.35\%\) test accuracy, CV accuracy \(95.86\%\), best \(\lambda\) is \(\ 7.278\times10^{ -4}\).}
\end{subfigure}
\hfill
\begin{subfigure}{0.49\linewidth}
\centering
\includegraphics[trim={0cm 0cm 0cm 1.5cm}, clip, width=\linewidth]{\detokenize{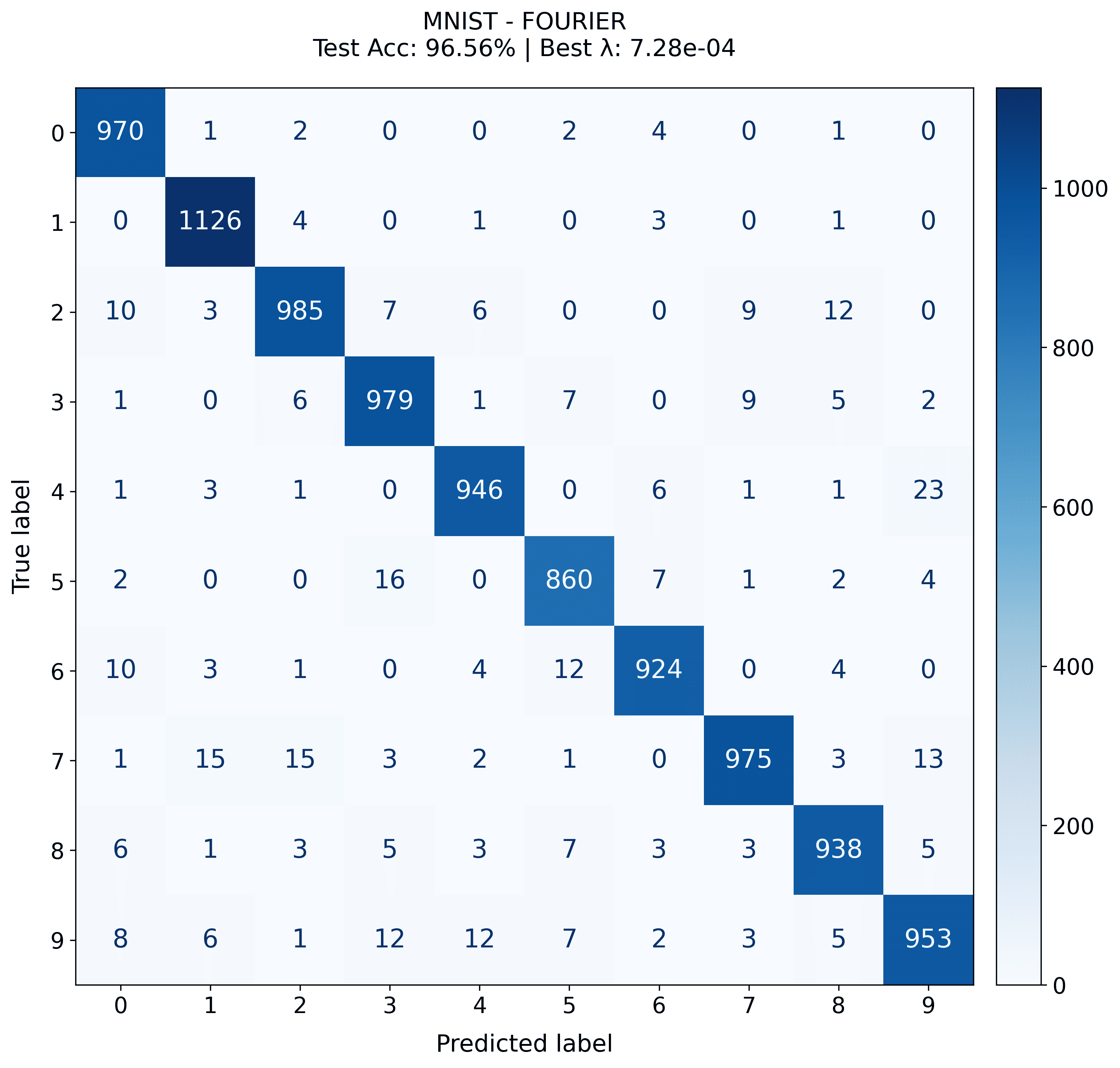}}
\caption{Fourier embedding, \(96.56\%\) test accuracy, CV accuracy \(96.46\%\), best \(\lambda\) is \(\ 7.278\times10^{ -4}\).}
\end{subfigure}

\caption{MNIST confusion matrices obtained after optimizing the ridge regularization parameter \(\lambda\). Both embeddings produce a strongly diagonal classification structure, showing  reliable digit recognition across classes. The Fourier embedding achieves slightly higher overall accuracy and shows fewer off diagonal misclassifications compared to the noise embedding, suggesting improved class separation in the optical feature space.}
\label{fig:mnist confusions}
\end{figure}

The FSDD experiment tests whether the same optical hardware can process a non-image modality.  Each spoken digit is first represented as a log-Mel spectrogram and then encoded as a phase pattern on the SLM.  This is a significantly different task from MNIST ,the two axes of the spectrogram represent time and frequency rather than spatial coordinates, and therefore the relation between input geometry and SLM plane geometry is less direct.  Even under this encoding mismatch, the PELM reaches \(93.00\%\) accuracy with noise embedding and \(95.67\%\) accuracy with Fourier embedding.  This result is one of the central multimodal findings of the work.  It shows that free-space optical propagation can extract discriminative features not only from spatial image data but also from audio derived time frequency representations.

\begin{figure}[H]
\centering
\begin{subfigure}{0.49\linewidth}
\centering
\includegraphics[trim={0cm 0cm 0cm 1.5cm}, clip, width=\linewidth]{\detokenize{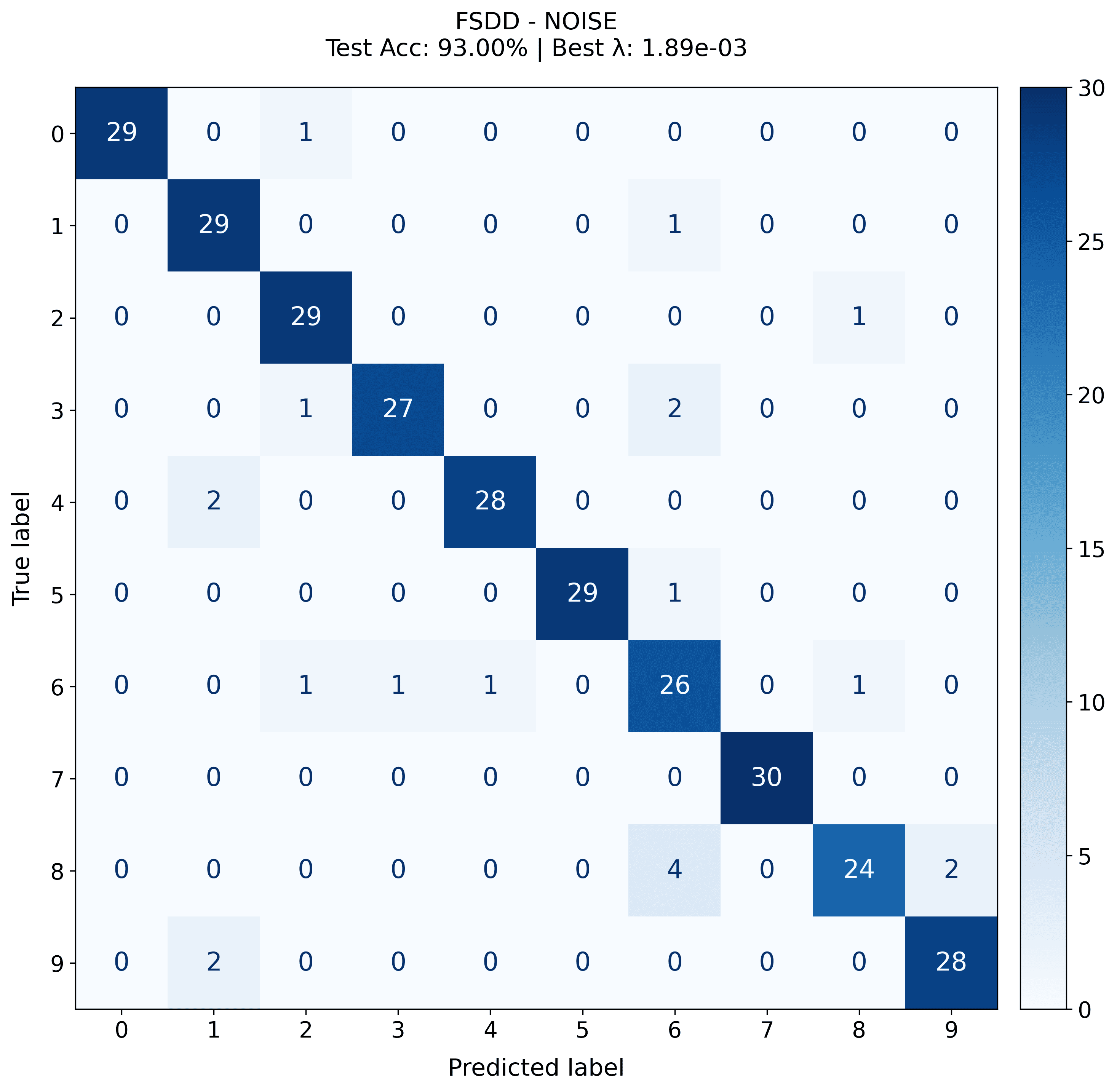}}
\caption{Noise embedding, \(93.00\%\) test accuracy, CV accuracy \(91.56\%\), best \(\lambda=1.887392\times10^{ -3}\).}
\end{subfigure}
\hfill
\begin{subfigure}{0.49\linewidth}
\centering
\includegraphics[trim={0cm 0cm 0cm 1.5cm}, clip, width=\linewidth]{\detokenize{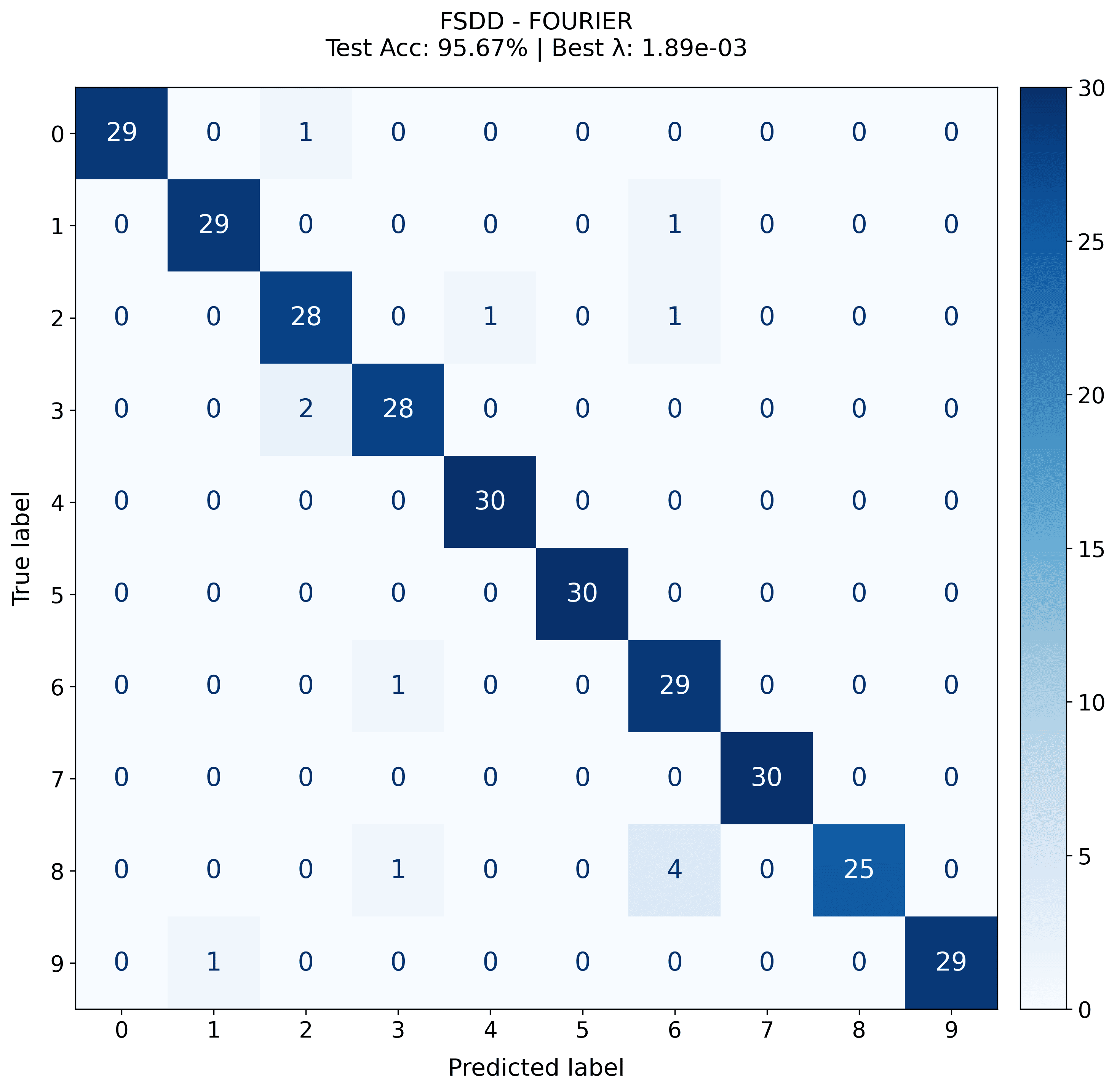}}
\caption{Fourier embedding, \(95.67\%\) test accuracy, CV accuracy \(95.22\%\), best \(\lambda=1.887\times10^{ -3}\).}
\end{subfigure}
\caption{FSDD spoken digit confusion matrices after optimizing the ridge parameter.  The PELM performs ten class audio derived classification using log-Mel spectrograms encoded as optical phase patterns.  The Fourier embedding improves the accuracy to \(95.67\%\).}
\label{fig:fsdd confusions}
\end{figure}

The Mushroom dataset provides a binary tabular classification benchmark.  Since the task is nearly linearly separable after suitable encoding, both embeddings perform close to saturation.  Both the noise and Fourier embeddings reach \(100.00\%\) classification.  The excellent performance indicates that the optical feature map does not degrade tabular class information when the categorical features are encoded into a stable phase representation.  At the same time, the strong dependence on the ridge parameter shows that for smaller tabular datasets, readout regularization can be as important as the optical embedding itself.

\begin{figure}[H]
\centering
\begin{subfigure}{0.49\linewidth}
\centering
\includegraphics[trim={0cm 0cm 0cm 1.5cm}, clip, width=\linewidth]{\detokenize{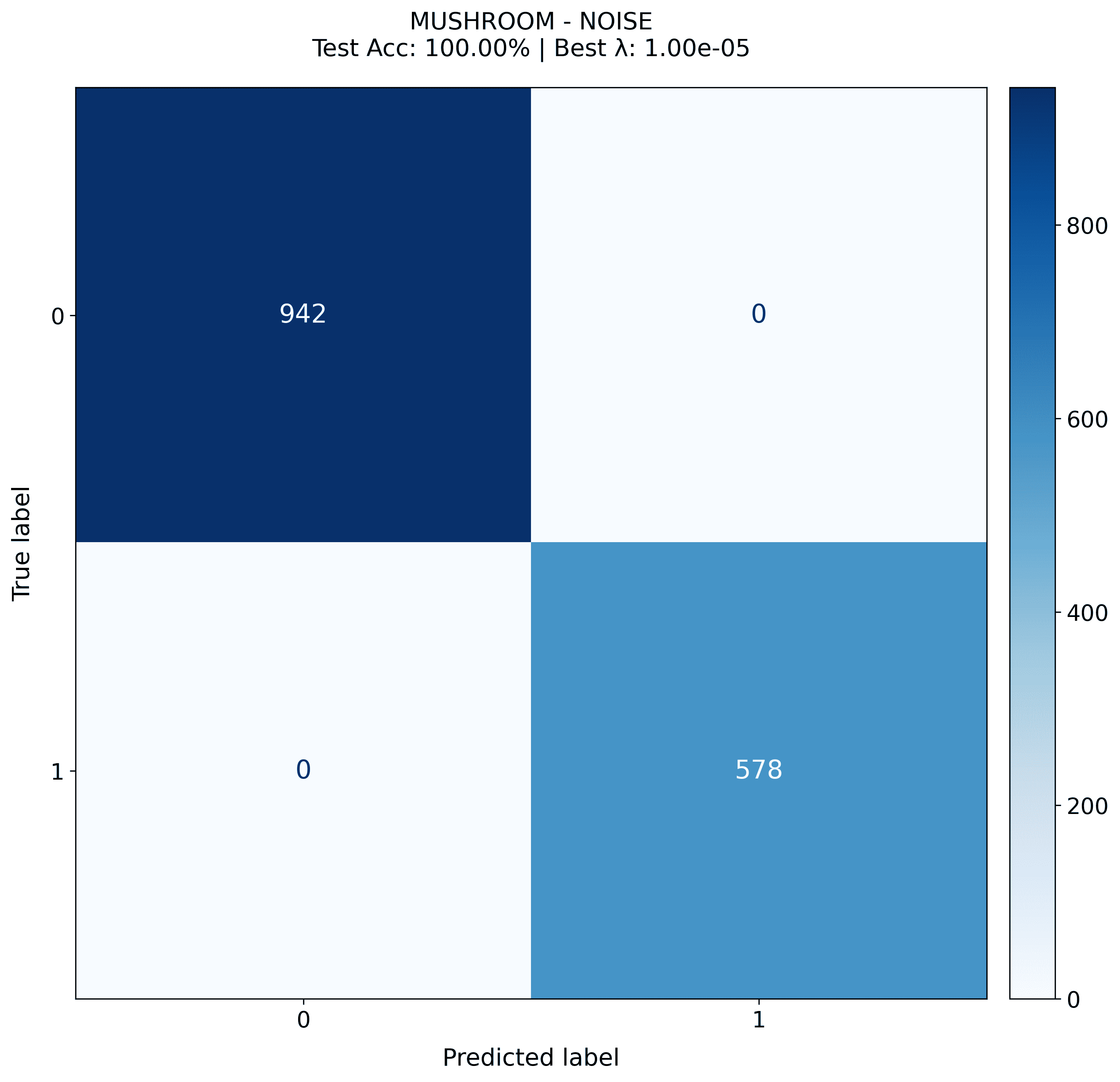}}
\caption{Noise embedding, \(100\%\), best \(\lambda=1.00\times10^{ -5}\).}
\end{subfigure}
\hfill
\begin{subfigure}{0.49\linewidth}
\centering
\includegraphics[trim={0cm 0cm 0cm 1.5cm}, clip, width=\linewidth]{\detokenize{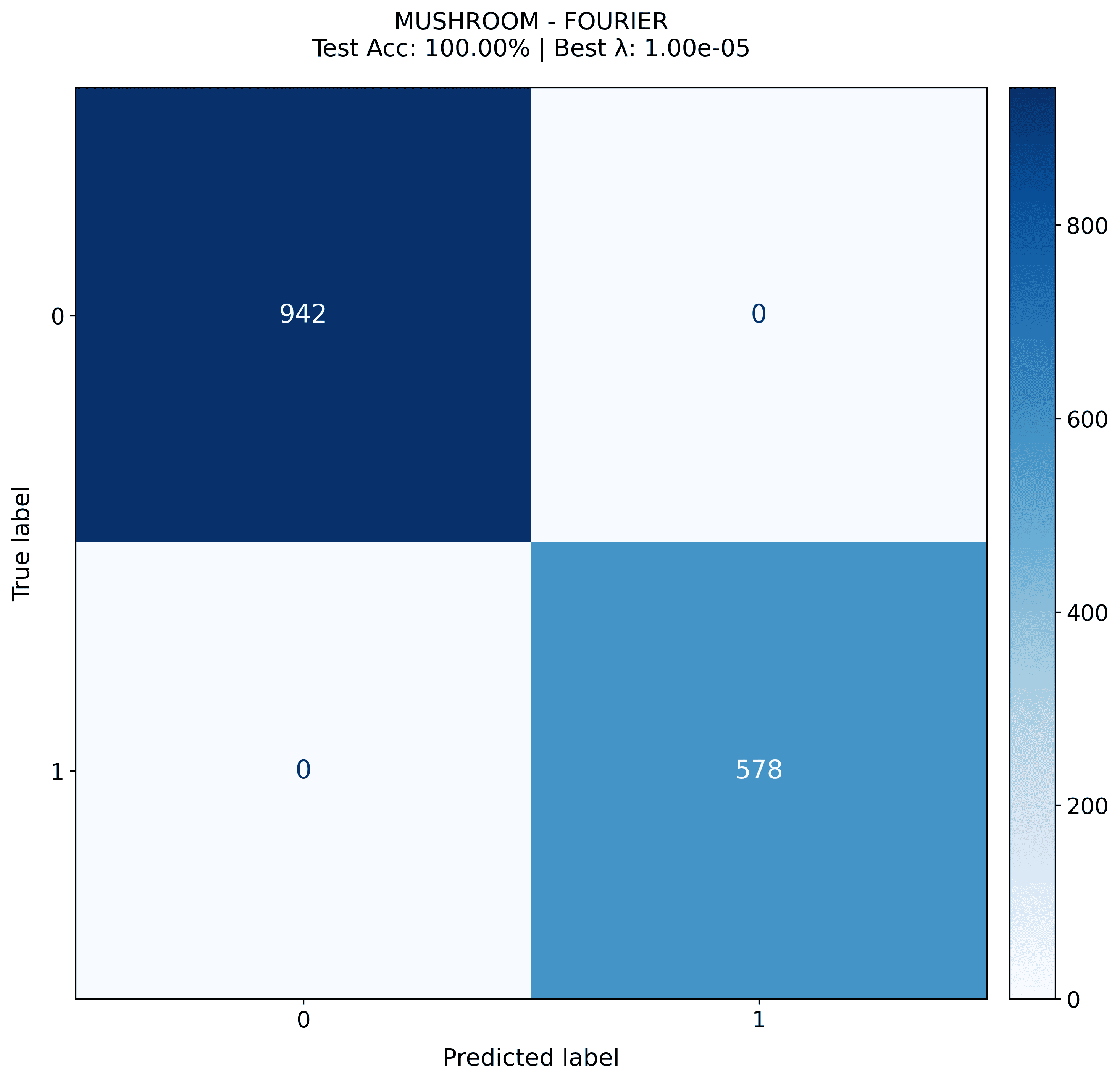}}
\caption{Fourier embedding, \(100\%\), best \(\lambda=1.00\times10^{ -5}\).}
\end{subfigure}
\caption{Mushroom binary classification confusion matrices after optimizing the ridge parameter \(\lambda\). Both embeddings achieve \(100.00\%\) classification accuracy, indicating that the optical feature map preserves the separability of the tabular features extremely well.}
\label{fig:mushroom confusions}
\end{figure}

Across all three classifications, the Fourier embedding provides performance comparable to or slightly better than the noise embedding after optimizing \(\lambda\).  This trend suggests that structured spectral embeddings can provide a more useful optical feature basis than purely random phase disorder for the present SLM camera PELM configuration. Importantly, the improvement is observed across image, audio derived, and tabular data, supporting the view that the embedding mask should be treated as a central design rather than as a minor experimental detail. The physical origin of this behavior, and its relationship to prior reports on random optical embeddings, remains under investigation.

The Abalone benchmark is a regression task; its optimized performance is reported using the normalized root mean square error in Table~\ref{tab:performance}.

The full \(\lambda\) optimization curves for all reported datasets and embeddings are provided in the Supplementary Information, Sec.~\ref{app:lambda sweeps}.  These sweeps show that the optimal regularization strength is dataset dependent: MNIST is relatively stable over a broad range of \(\lambda\), whereas smaller tabular and audio derived datasets exhibit sharper optima.

\section{Empirical feature space diagnostics}
\label{sec:diagnostics}

In this paper, we focus on the experimental evidence for multimodal optical feature extraction rather than presenting a complete mathematical theory of the mechanism.  A detailed theory of distance preservation, kernel formation, and high dimensional readout accumulation is being developed and will be reported in a subsequent publication. Here, we restrict ourselves to directly measured diagnostics obtained from the optical feature matrices: pairwise distance preservation, kernel alignment, class separation metrics, and low dimensional feature visualizations. These diagnostics provide an empirical view of how the same free-space PELM hardware supports image, audio derived, tabular classification, and regression tasks.

For each input \(x_i\), the optical system produces a normalized feature vector \(g(x_i)\in\mathbb{R}^{M}\), where \(M\) denotes the number of binned camera readout channels. The diagnostics in this section compare the structure of the original input representation with that of the experimentally measured optical feature representation. To ensure computational tractability and reduce class imbalance bias, these diagnostics were evaluated on controlled subsets of the training data. Pairwise distance preservation analysis was computed using 50 random samples across all datasets to optimize scatter plot legibility. Centered kernel alignment (CKA) and class separation metrics were evaluated on class balanced subsets (20 samples per class) for discrete classification datasets. Conversely, the continuous target manifold of the Abalone regression task was evaluated using an expanded random subset of 2000 samples; this high density pool is mathematically required to maintain proper target density across its integer milestones, preventing metric sparsity and ensuring a stable median distance heuristic during continuous auto RBF kernel construction. Unless stated otherwise, the main text diagnostics are shown for the Fourier embedding, which achieved the strongest optimized performance across the classification benchmarks in Sec.~\ref{sec:results}. The corresponding diagnostics for the noise embedding are provided in the Supplementary Information.

\subsection{Experimental optical kernel characterization}
\label{subsec:kernel characterization}

Before evaluating the task specific utility of the optical feature space, we first verify the fundamental mathematical operation executed by the physical hardware. The intensity detection process of the PELM is theoretically predicted to implement a nonlinear kernel governed by the angular geometry of the encoded inputs. To test this, we experimentally measured the empirical optical kernel from raw pairwise feature similarities and compared it against several exact double centered theoretical kernels. These include the analytical phase kernel (\(K_{phase}\)), the complex Gaussian approximation (\(K_{Gaussian}\)), arc cosine kernels corresponding to infinite width ReLU networks (\(K_1\), \(K_2\)), and an angular Radial Basis Function (RBF) kernel.

\begin{figure}[H]
\centering
\includegraphics[width=\linewidth]{\detokenize{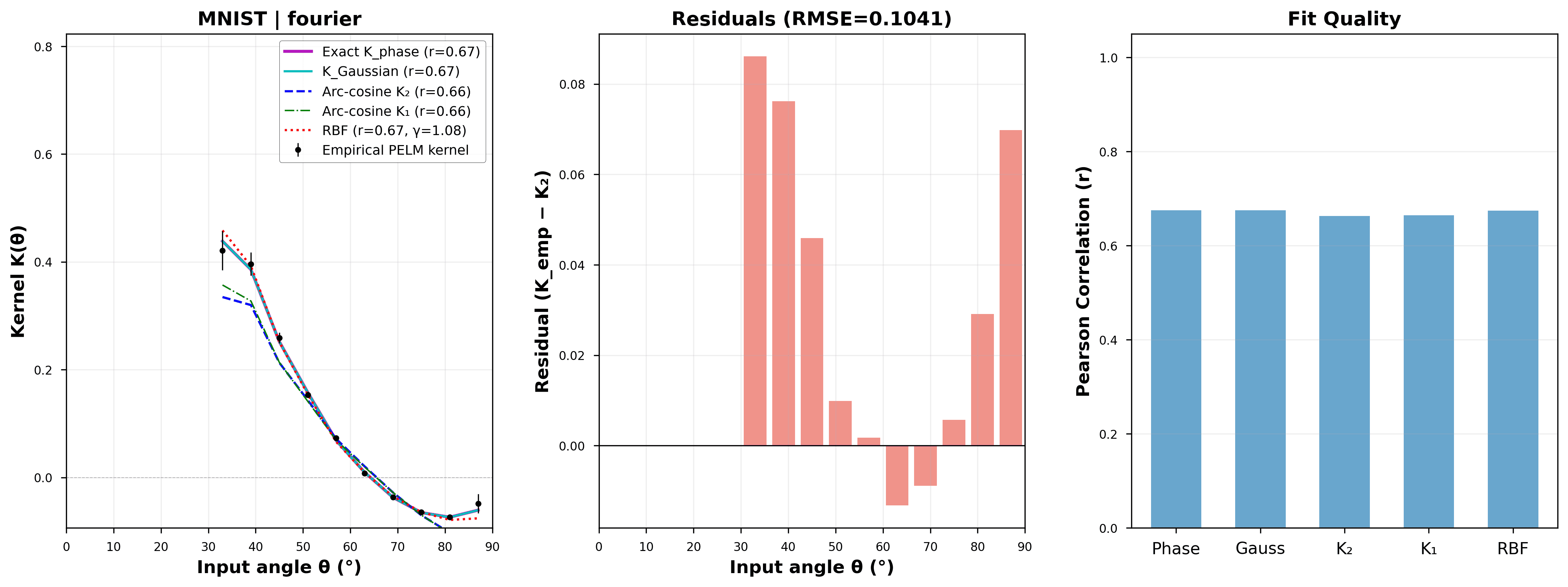}}
\caption{Experimental kernel characterization for the MNIST dataset using the Fourier embedding. The empirical PELM kernel is binned by input angle \(\theta\) and compared to exact double centered theoretical predictions. The high Pearson correlation confirms that the optical hardware, aided by the structured spatial carrier of the Fourier mask, physically computes an angular RBF and \(K_{phase}\) kernel.}
\label{fig:kernel characterization}
\end{figure}

As shown in Fig.~\ref{fig:kernel characterization} for the MNIST dataset using the Fourier embedding, the measured empirical kernel exhibits a strong Pearson correlation with the angular RBF, validating that the physical optics faithfully compute the expected nonlinear geometric projection. Notably, the structured spatial carrier wave of the Fourier embedding efficiently diffracts the information bearing signal away from the unmodulated background, yielding a highly robust and well defined empirical kernel fit. The arc cosine \(K_2\) kernel is also plotted to serve as a baseline comparison to standard digital neural networks. Because this hardware characterization proves the system naturally executes an angular RBF like projection, we adopt the angular RBF for the downstream machine learning diagnostics (such as Centered Kernel Alignment) in the subsequent sections. The kernel fit $r$ varies across datasets and embeddings, with natural 
image and tabular inputs introducing confounds that suppress the measured 
correlation relative to the underlying empirical kernel fit, reflecting 
in-situ hardware kernel quality under realistic measurement conditions. 
The Pearson correlation of $r \approx 0.67$ obtained from natural MNIST 
image pairs reflects this in situ kernel fit quality. Natural images are 
not designed to probe the angular kernel: the binning by $\theta$ averages 
over many input pairs drawn from the full MNIST distribution, whose complex 
spatial structure and varying content introduce confounds that suppress the 
measured correlation relative to the true fit quality. A dedicated angular 
sweep experiment using purpose constructed input pairs at controlled, 
known angles yields $r = 0.990$ for the angular RBF kernel (see 
companion publication), confirming that the lower in situ value is a 
measurement artifact rather than a failure of the kernel model. Together, 
the two measurements establish that the optical hardware faithfully 
implements the angular RBF kernel, and Fig.~7 should be interpreted as a 
conservative lower bound on the true fit quality rather than a precision 
characterization. Full experimental kernel fits for the remaining datasets and the noise embedding are provided in the Supplementary Information (Sec.~\ref{app:kernel fits}).

\subsection{Relative distance structure preservation}
\label{subsec:distance preservation empirical}

A first diagnostic asks whether distances in the original input representation are preserved after optical propagation and camera detection.  For a set of sample pairs \((x_i,x_j)\), we compute the original space distance
\begin{equation}
d_{\mathrm{in}}(i,j)=\|x_i -x_j\|_2,
\end{equation}
and the optical feature distance
\begin{equation}
d_{\mathrm{opt}}(i,j)=\|g(x_i) -g(x_j)\|_2.
\end{equation}
Both distance lists are standardized before plotting, and Pearson and Spearman correlations are used as empirical measures of relative distance preservation. For computational consistency across datasets, the distance-preservation diagnostics were computed using randomly selected subsets of \(50\) samples per dataset. This is not meant as a proof of a Johnson-Lindenstrauss type guarantee for the physical apparatus.  Rather, it is a direct experimental measurement of whether the optical reservoir approximately preserves pairwise structure for each modality.

\begin{figure}[H]
\centering
\includegraphics[width=\linewidth]{\detokenize{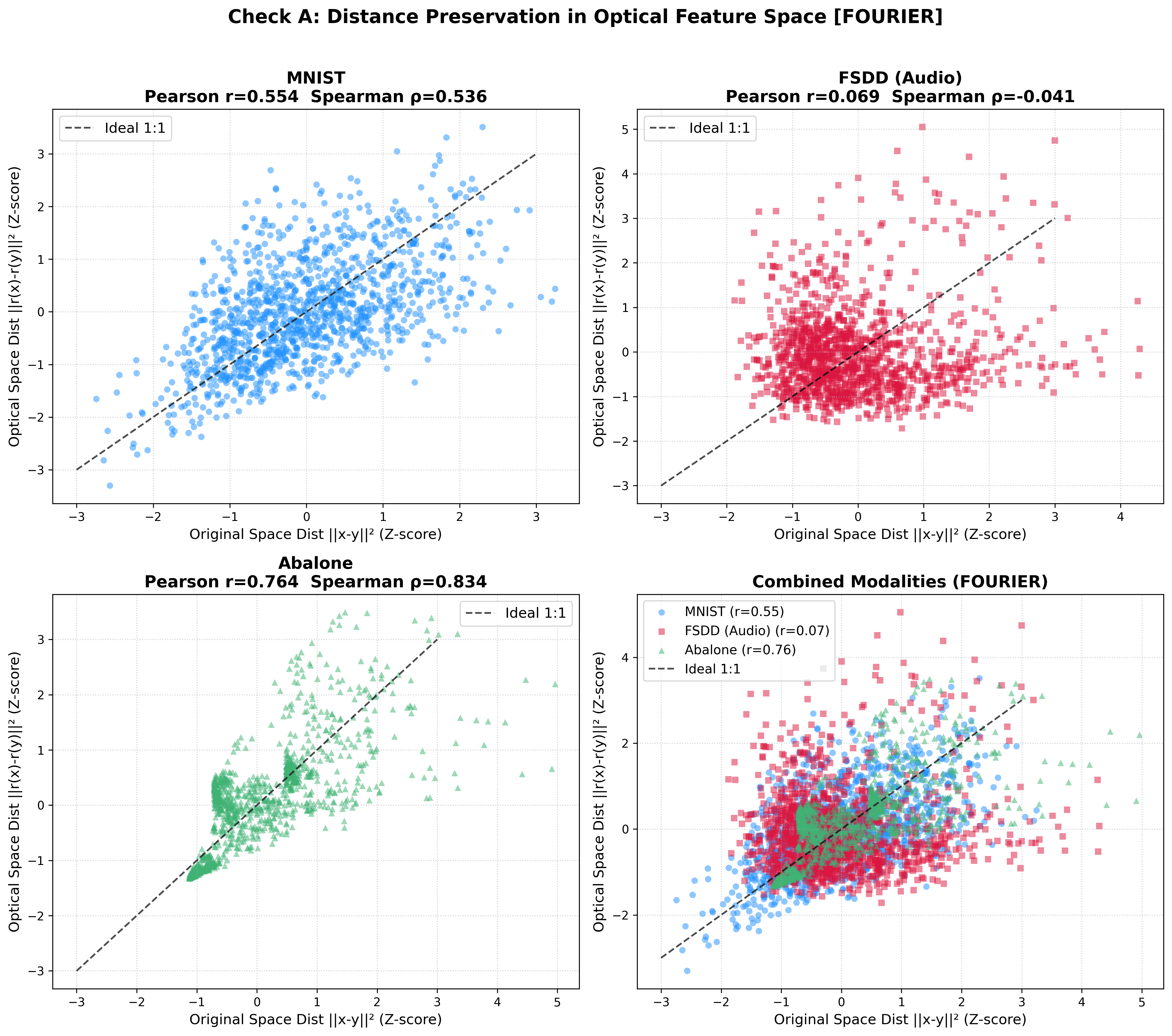}}
\caption{Empirical relative distance structure preservation for the Fourier embedding. Pairwise distances in the original representation are compared with distances in the experimentally measured optical feature space.  MNIST shows moderate positive distance preservation, Abalone shows strong monotonic preservation, while FSDD shows weak global distance preservation despite high classification accuracy.  The integrated panel emphasizes that a single optical feature extractor can function across various empirical regimes according to the modality.}
\label{fig:distance preservation}
\end{figure}

Figure~\ref{fig:distance preservation} reveals three separate empirical regimes.  The MNIST dataset demonstrates a moderate level of positive distance preservation, indicated by a Pearson correlation coefficient of \(r=0.554\) and a Spearman rank correlation of \(\rho=0.536\).  This implies that the optical feature space maintains some aspects of the geometry inherent to the original digit manifold.  In contrast, the Abalone regression task exhibits significantly stronger distance preservation, with a Pearson correlation of \(r=0.764\) and a Spearman rank correlation of \(\rho=0.834\).  This finding aligns our observation that the optical representation retains a continuous structure that is advantageous for regression.

FSDD shows a different behavior. The spoken digit signals are represented as log Mel spectrograms, whose axes correspond to time and frequency rather than physical spatial coordinates. When these spectrograms are encoded onto the SLM plane, global Euclidean distances in the input space are not well preserved in the resulting optical feature space. This is reflected in the FSDD panel, where Pearson correlation is \(r = 0.069\) and Spearman correlation is \(\rho =  -0.041\). Despite this weak distance preservation, the optical system still achieves a classification accuracy of \(95.67\%\) on FSDD using the Fourier embedding. This indicates that while preserving global pairwise distances can be beneficial, it is not a strict requirement for achieving high classification performance in the PELM readout.

\subsection{Kernel alignment and class separation diagnostics}
\label{subsec:empirical kernel separation}
To probe class structure more directly, we compute empirical kernel and separation diagnostics from the measured optical features. Kernel methods provide a natural language for comparing nonlinear feature maps \cite{scholkopf2002learning}, while kernel target alignment and centered kernel alignment quantify how well a feature space Gram matrix agrees with a target or label Gram matrix \cite{cristianini2001alignment,kornblith2019similarity}. We use these diagnostics empirically: they are not treated as a complete theory of PELM operation, but as quantitative summaries of how the measured optical representation organizes samples.

The linear optical Gram matrix is
\begin{equation}
K^{\mathrm{lin}}_{ij}=g(x_i)^\top g(x_j),
\end{equation}
and an RBF optical Gram matrix is constructed as
\begin{equation}
K^{\mathrm{RBF}}_{ij}
=
\exp\left[ -\gamma \|g(x_i) -g(x_j)\|_2^2\right],
\end{equation}
with \(\gamma\) fixed by the scale of the optical features.  We compare these empirical kernels with the label kernel using centered kernel alignment (CKA),
\begin{equation}
\mathrm{CKA}(K,L)=
\frac{\langle H_c K H_c,H_c L H_c\rangle_F}
{\|H_c K H_c\|_F\|H_c L H_c\|_F},
\qquad
H_c=I -\frac{1}{N}\mathbf{1}\mathbf{1}^\top .
\label{eq:cka -diagnostic}
\end{equation}
In practice, feature vectors are L2 normalized before computing \(K^{\mathrm{lin}}\), so the linear kernel entry reduces to the cosine similarity \(K^{\mathrm{lin}}_{ij}=g(x_i)^\top g(x_j)/(\|g(x_i)\|\|g(x_j)\|)\); this follows the recommendation of Kornblith et al.\ \cite{kornblith2019similarity} and makes the CKA values invariant to global intensity fluctuations in the optical measurement.

In addition to global CKA, we compute class wise separation metrics.  These quantify whether a given class is better separated from the remaining samples in the measured optical feature space.  The separation plots therefore test a more task specific question than pairwise distance preservation: even if the full geometry is not preserved, are the measured optical features organized in a way that the final linear readout can exploit?

For each class, separation was quantified using a Cohen's-\(d\) style standardized separation metric,
\begin{equation}
S_c =
\frac{\mu_{\mathrm{within}} -\mu_{\mathrm{between}}}
{\sqrt{\frac{1}{2}\left(\sigma_{\mathrm{within}}^2+\sigma_{\mathrm{between}}^2\right)}},
\end{equation}
where \(\mu_{\mathrm{within}}\) and \(\mu_{\mathrm{between}}\) denote the mean within class and between class similarities, respectively, and \(\sigma_{\mathrm{within}}\), \(\sigma_{\mathrm{between}}\) are the corresponding standard deviations. Larger values indicate stronger class separation in the measured optical feature space.

MNIST exhibits stronger global alignment, with linear CKA $=0.3624$ and RBF CKA $=0.3781$, consistent with the corresponding distance preservation diagnostics. FSDD shows weaker global alignment (linear CKA $=0.2962$, RBF CKA $=0.3085$) despite maintaining high classification accuracy, indicating that the ridge readout can exploit class discriminative structure even when the global feature geometry appears only weakly organized.

The corresponding Mushroom diagnostics, shown in the Supplementary Information, exhibit strong binary separation. Finally, for the Abalone regression task, we employ a continuous target similarity kernel to evaluate manifold preservation, complementing the pairwise distance preservation analysis. The corresponding figures for all datasets are provided in Supplementary Section C.3.

\subsection{Low dimensional feature visualizations}
\label{subsec:tsne empirical}

We further visualize the measured optical feature matrices using t-SNE \cite{maaten2008tsne}. These plots are not employed as standalone quantitative proof, because t-SNE has the potential to skew overall distances and is susceptible to the arrangement of local neighborhoods. Nevertheless, they offer a helpful qualitative perspective on the arrangement of high dimensional camera attributes across various modalities.

\begin{figure}[H]
\centering
\begin{subfigure}{0.49\linewidth}
\centering
\includegraphics[width=\linewidth]{\detokenize{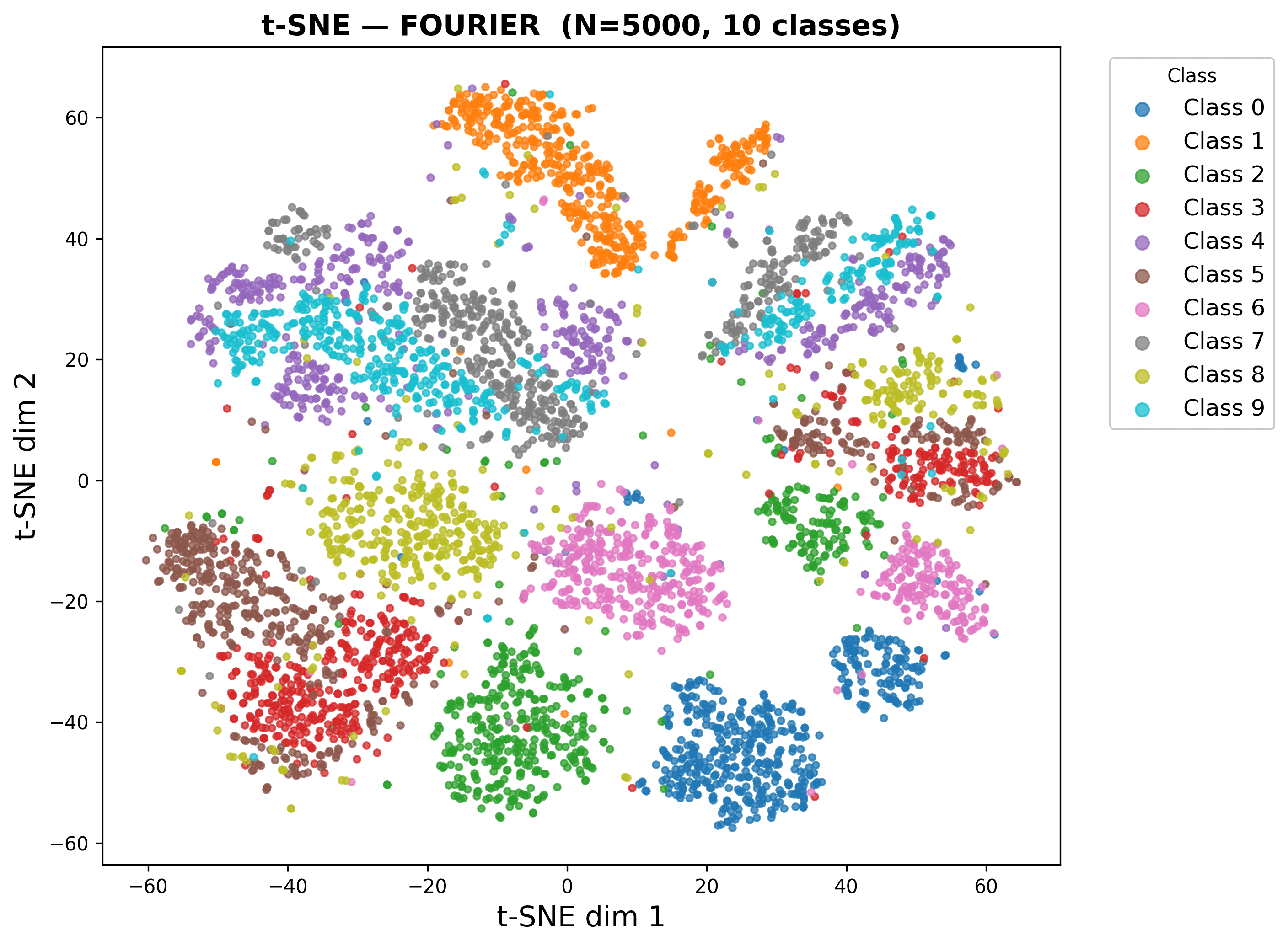}}
\caption{MNIST}
\end{subfigure}
\hfill
\begin{subfigure}{0.49\linewidth}
\centering
\includegraphics[width=\linewidth]{\detokenize{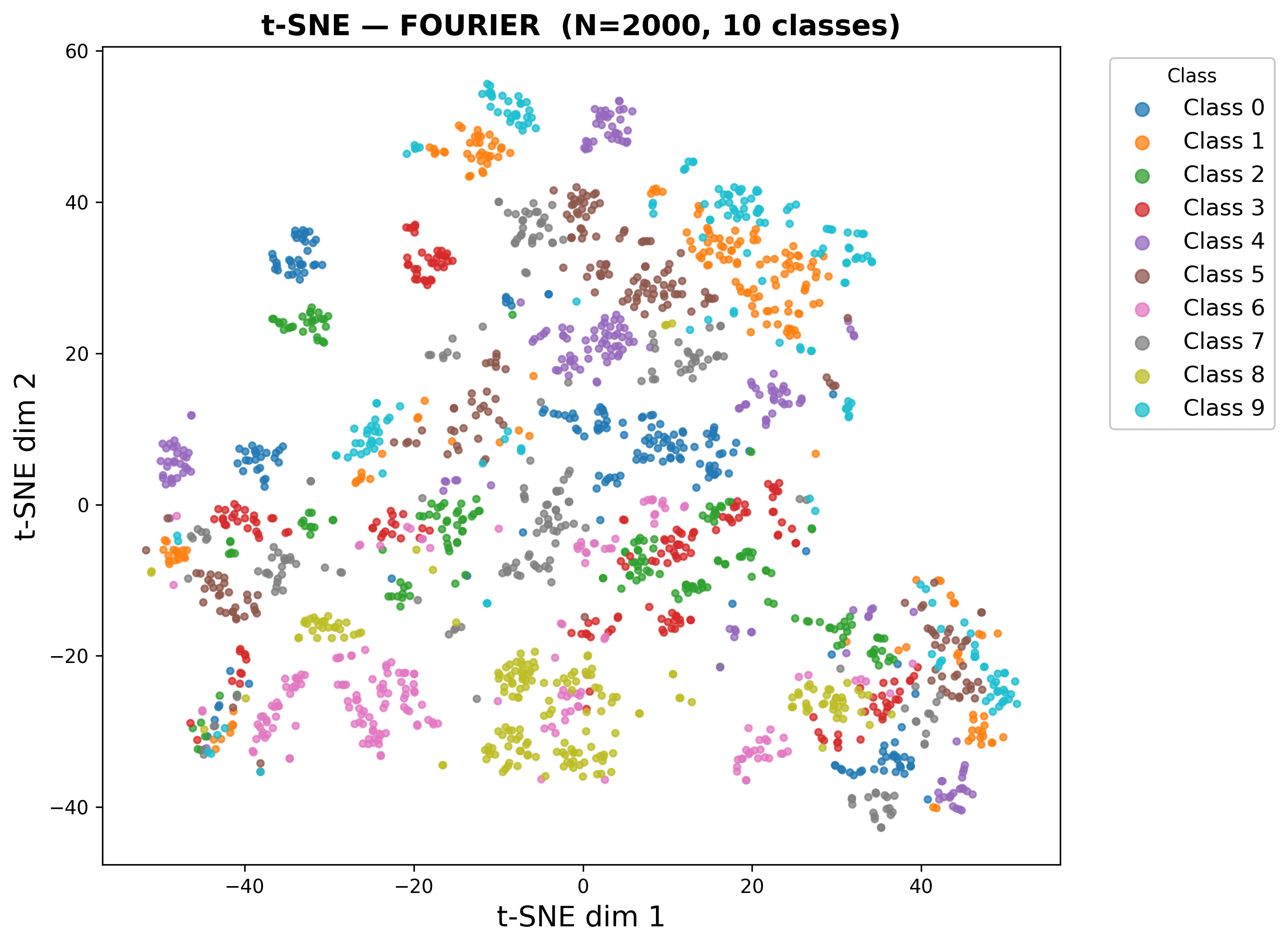}}
\caption{FSDD}
\end{subfigure}

\vspace{0.8em}

\begin{subfigure}{0.49\linewidth}
\centering
\includegraphics[width=\linewidth]{\detokenize{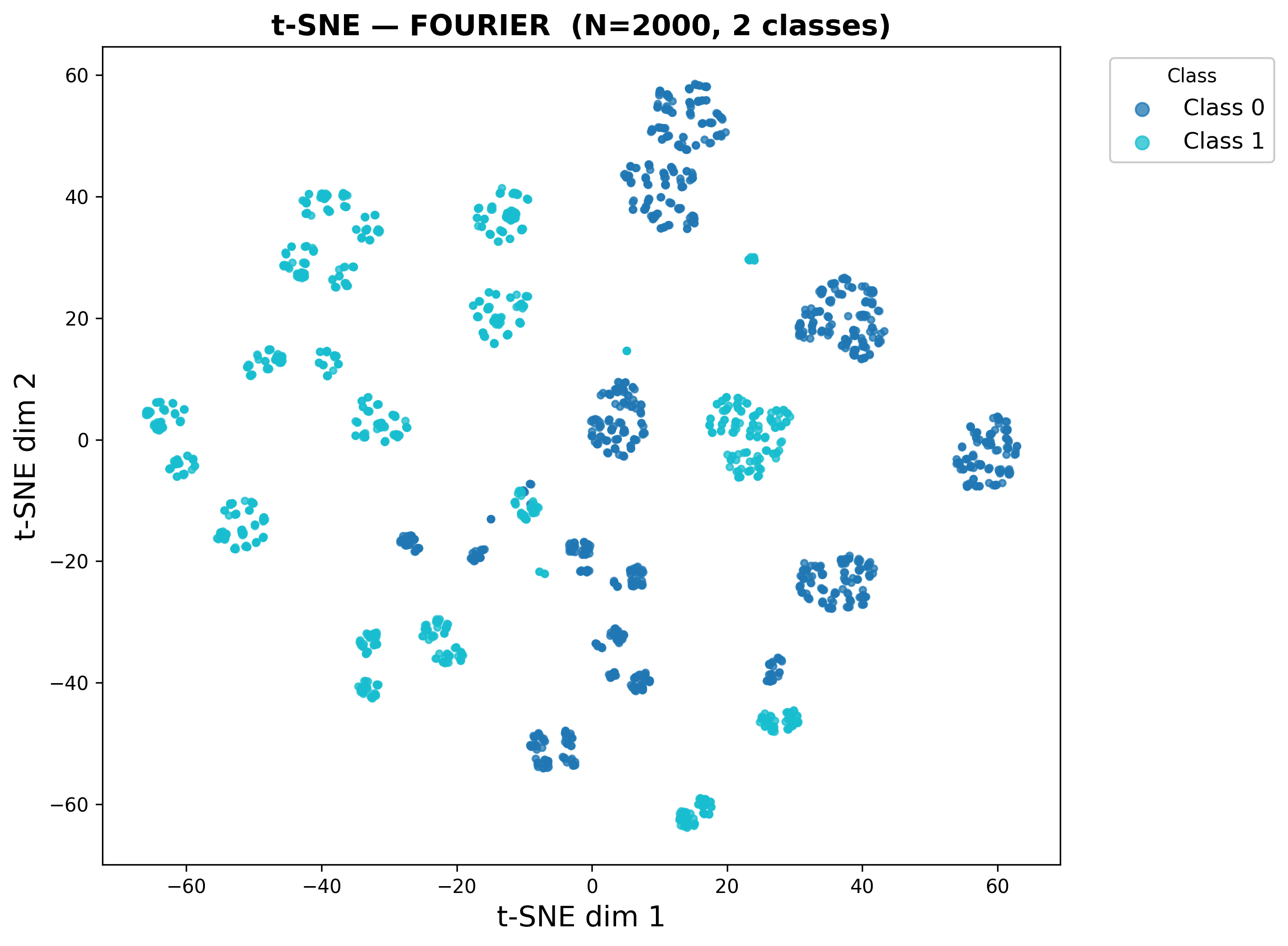}}
\caption{Mushroom}
\end{subfigure}
\hfill
\begin{subfigure}{0.49\linewidth}
\centering
\includegraphics[width=\linewidth]{\detokenize{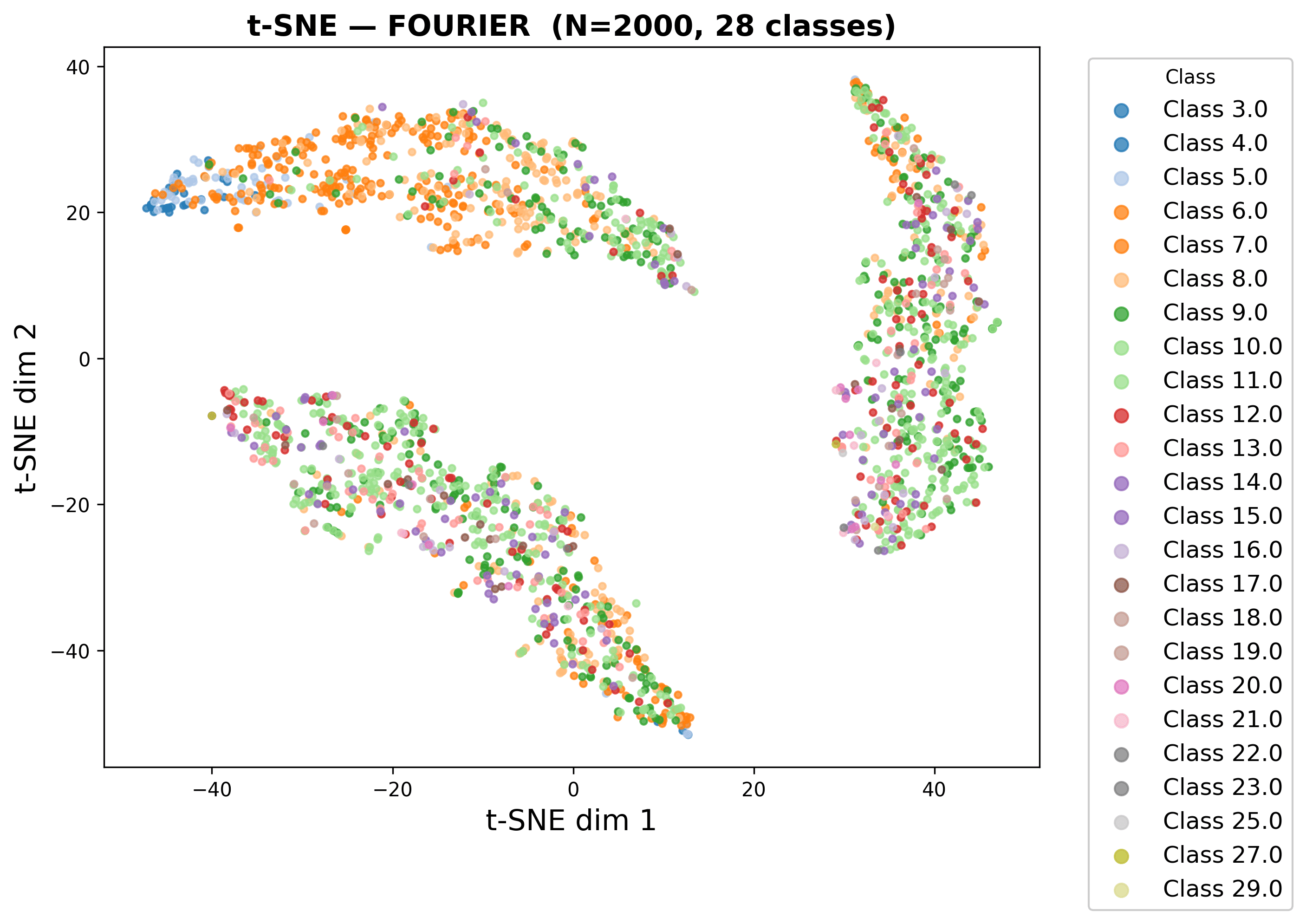}}
\caption{Abalone}
\end{subfigure}

\caption{t-SNE graphic representations of optical characteristics that were empirically recorded for the Fourier embedding. MNIST and Mushroom demonstrate clearer class organization, whereas FSDD appears more globally blended in two dimensions, even with a strong ten class accuracy. Abalone exhibits continuous structures that depend on the target, rather than separate class clusters, aligning with its regression oriented nature.}
\label{fig:tsne main}
\end{figure}

The t-SNE visualizations illustrated in Fig. 9 strengthen the structural understandings derived from both distance preservation and CKA assessments. The MNIST dataset displays clearly defined topological groupings based on numeral digits, while the Mushroom dataset reveals apparent and linearly separated binary classifications. In contrast, the FSDD dataset seems to be more globally scattered across two dimensions, showing that its distinguishing characteristics are spread throughout the entire 4096 dimensional optical space instead of being limited to a narrow low dimensional area. Regarding the Abalone regression task, the samples are categorized by color according to specific integer target milestones related to shell rings, which range between 3.0 and 29.0. Instead of creating isolated categories, the visualization indicates a smoothly transitioning continuous structure where nearby integer targets retain close structural relationships within the latent space, visually verifying the regression maintaining attributes of the optical reservoir.

\FloatBarrier

\section{Discussion}
\label{sec:discussion}

The main result of this work is the experimental demonstration that a single free-space PELM can be used as a multimodal optical feature extractor.  The same physical pipeline, consisting of phase only SLM encoding, free-space optical propagation, first order spatial filtering, camera intensity detection, and ridge regression readout, supports image classification, audio derived spectrogram classification, tabular binary classification, and tabular regression.  This distinguishes the present work from earlier free-space PELM demonstrations that focused on more conventional image and tabular benchmarks.

The audio result is particularly important.Log-Mel spectrograms are not native optical images; they are time-frequency representations of speech signals.  The robust performance of FSDD indicates that the optical system is not simply utilizing spatial image structure.  Rather, when the audio signal is incorporated as a phase pattern, free-space propagation along with intensity detection produces a high dimensional representation that encompasses sufficient class discriminative information for a linear readout.

Different modalities interact differently with the optical feature space, according to the empirical diagnostics. Measurable relative distance preservation is shown by MNIST and Abalone, suggesting that some of the input geometry survives the optical transformation.  On the other hand, as optical characteristics expand, Mushroom gets closer to near separability.  Despite having limited low dimensional clustering and poor global distance preservation, FSDD nevertheless retains a high degree of accuracy. Pairwise geometry preservation is helpful, but it is not the sole path to good PELM operation. This is a key finding of the manuscript.

The Fourier embedding consistently improves the optimized classification performance over the noise embedding for the completed experimental runs. This indicates that the embedding mask should be treated as a central optical design parameter rather than as a minor perturbation of the input phase. In the current work, we restrict ourselves to the completed noise and Fourier embedding experiments.  More advanced embedding constructions and their detailed theoretical analysis will be addressed in future work.

In addition to the baseline preprocessing pipeline consisting of per sample DC removal, feature centering, and hyperspherical ($L_2$) normalization, we also explored alternative digital conditioning strategies, including power law intensity scaling and channel wise standardization. Preliminary empirical diagnostics suggest that such transformations can further improve feature space conditioning metrics, including Centered Kernel Alignment (CKA) and kernel signal to noise ratios (SNR), particularly for heterogeneous multimodal datasets. These observations indicate that the measured optical feature manifold remains amenable to additional digital conditioning prior to readout optimization. Although these advanced preprocessing techniques were not employed in the main benchmarks reported, they indicate a promising avenue for future enhancements.

Practical limitations also exist. Rather than the actual speed of light propagation, the existing system is limited by the SLM update frequency, liquid crystal settling time, camera exposure time, data transfer speeds, and environmental stability. Performance is greatly impacted by variables such iris alignment, polarizer modifications, camera bit-depth conversion, feature normalization, and SLM settling time. These difficulties are essential for guaranteeing reproducibility and scaling towards faster optical inference, but they do not constitute conceptual constraints of PELMs.

Overall, the results suggest that free-space PELMs provide a low cost, promising framework for optical machine learning. Their benefit is in enabling high dimensional feature expansion by a passive optical transformation while retaining a simple and reliable electronic readout, rather than in replacing all digital computations. In situations where data is intrinsically optical, image like, or can be efficiently represented as two dimensional phase patterns, this hybrid optical electronic architecture is very attractive. When data are inherently visual, image like, or can be effectively represented as two dimensional phase patterns, this hybrid optical electronic structure is particularly appealing.

\FloatBarrier

\section{Conclusion}
\label{sec:conclusion}

We have devloped a multimodal free-space photonic extreme learning machine using phase only SLM encoding, Fourier like optical propagation, camera intensity detection, and ridge regression readout. With optimized ridge regularization, the system achieves \(96.56\%\) accuracy on MNIST using the Fourier embedding, \(95.67\%\) accuracy on FSDD spoken -digit classification from log-Mel spectrograms, and \(100.00\%\) accuracy on Mushroom binary classification, together with the optimized Abalone regression performance reported in Table~\ref{tab:performance}. The Supplementary Information contains the complete \(\lambda\) optimization curves.

The empirical diagnostics show that the same optical hardware can support different learning mechanisms across modalities.Some tasks, such as MNIST and Abalone, show measurable relative distance preservation in optical feature space.Others, such as FSDD, achieve high accuracy despite weak global distance preservation, indicating that the trained readout can exploit discriminative high dimensional structure not visible in simple geometric diagnostics.These results establish free space PELMs as experimentally viable multimodal optical feature extractors and motivate future work on embedding design, theoretical modeling, and faster hardware implementations.

\section*{Author Contributions}

Anushka Kumari contributed to the MNIST and FSDD experiments, optical kernel analysis, feature space diagnostics, and manuscript preparation. Anushree Khisti contributed to the Abalone and Mushroom experiments and participated in the MNIST studies. Abhinav Choube contributed to the experimental setup, hardware alignment, and execution of the experimental analyses. Devansh Satra assisted with manuscript writing and preparation. Srivatsa Murali contributed to the initial experimental setup and system development. Anshuman Kumar supervised the project, guided the research direction, experimental design, theoretical interpretation, and manuscript development.

\FloatBarrier


\appendix

\section{Supplementary experimental details}
\label{app:exp details}

\begin{table}[H]
\centering
\caption{Representative hardware and acquisition parameters used in the free-space PELM implementation.}
\label{tab:hardware}
\renewcommand{\arraystretch}{1.15}
\begin{tabularx}{\linewidth}{@{}lXX@{}}
\toprule
Component & Representative specification & Role \\
\midrule
Laser & 532 nm green diode, 5 mW class & Coherent illumination \\
SLM & Phase only Holoeye ERIS class, 1920\(\times\)1200, 8 \(\mu\)m pitch & Phase encoding \\
Optics & 4\(f\) style propagation path with iris & Fourier like mixing and first order selection \\
Camera & CMOS scientific camera, finite bit depth intensity readout & Nonlinear feature detection \\
Readout & 4096 binned spatial channels & ELM feature vector \\
\bottomrule
\end{tabularx}
\end{table}

The iris strongly affects the measured feature quality by suppressing the zero order background and selecting the modulated first order field.Feature normalization minimizes the impact of variations in global laser power and camera response characteristics. The settling time of the SLM needs to be sufficient for the liquid crystal phase state to stabilize following each update;reducing this settling time was shown to negatively affect classification performance. For stable acquisition, it was also necessary to have continuous streaming from the camera, accurate bit depth conversion, timely updates to the SLM buffer, and a methodical sequence for shutting down the hardware.

\section{Supplementary results: ridge parameter optimization}
\label{app:lambda sweeps}

The readout layer is trained using ridge regression,
\begin{equation}
\beta = \left(H^\top H+\lambda I\right)^{ -1}H^\top T,
\end{equation}
where \(H\in\mathbb{R}^{N\times M}\) is the measured optical feature matrix, \(T\) is the target matrix, and \(\lambda\) controls the strength of \(\ell_2\) regularization.The parameter \(\lambda\) stabilizes the inversion of \(H^\top H\) and controls the bias variance tradeoff of the readout.  For each dataset and embedding strategy, \(\lambda\) was swept and the optimized performance was reported in the main text.  Classification tasks are evaluated using accuracy, while Abalone regression is evaluated using NRMSE.

\subsection{MNIST}

\begin{figure}[H]
\centering
\begin{subfigure}{0.49\linewidth}
\centering
\includegraphics[trim={0cm 0cm 0cm 0.75cm}, clip, width=\linewidth]{\detokenize{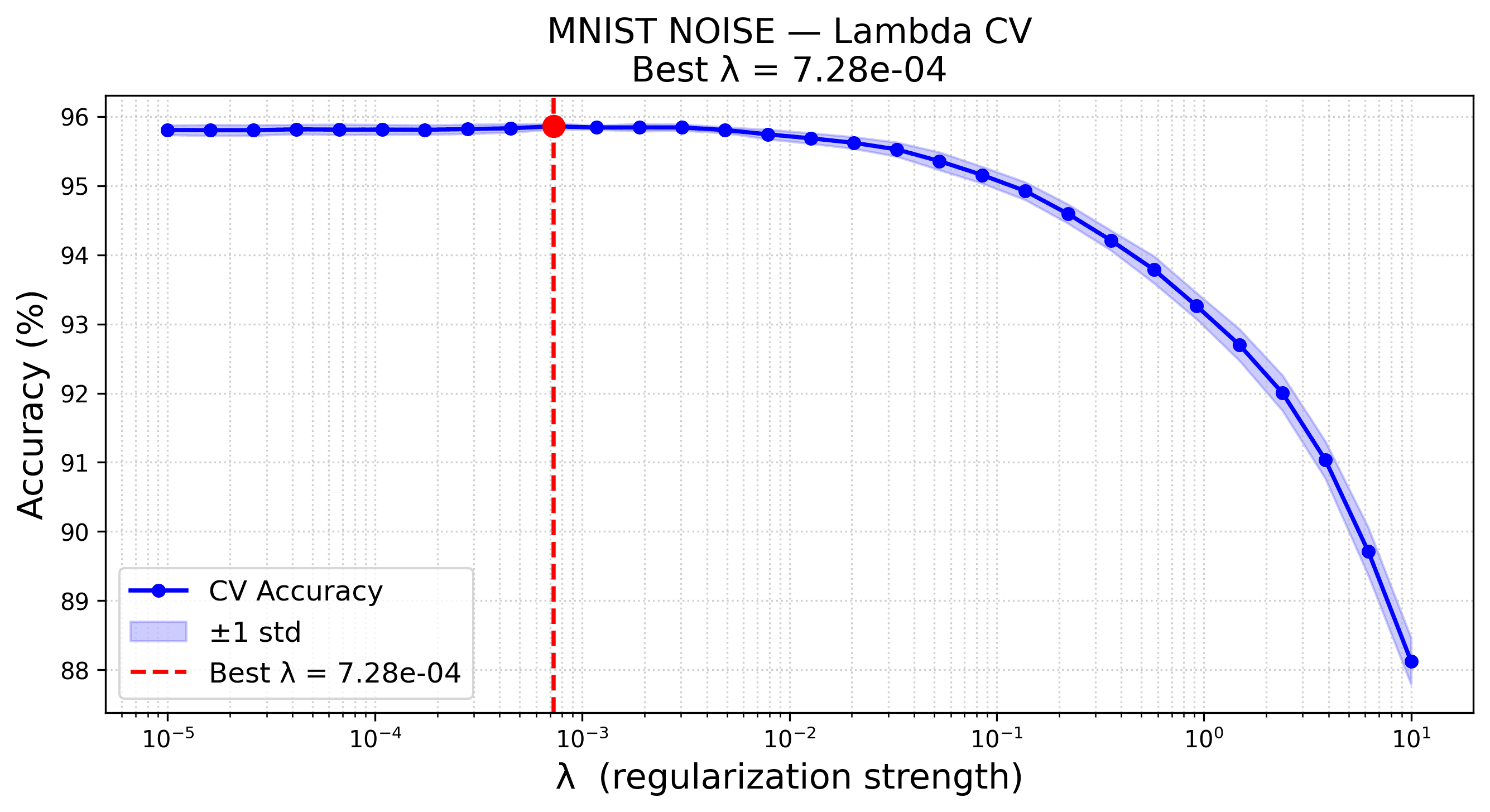}}
\caption{Noise embedding.}
\end{subfigure}
\hfill
\begin{subfigure}{0.49\linewidth}
\centering
\includegraphics[trim={0cm 0cm 0cm 0.75cm}, clip, width=\linewidth]{\detokenize{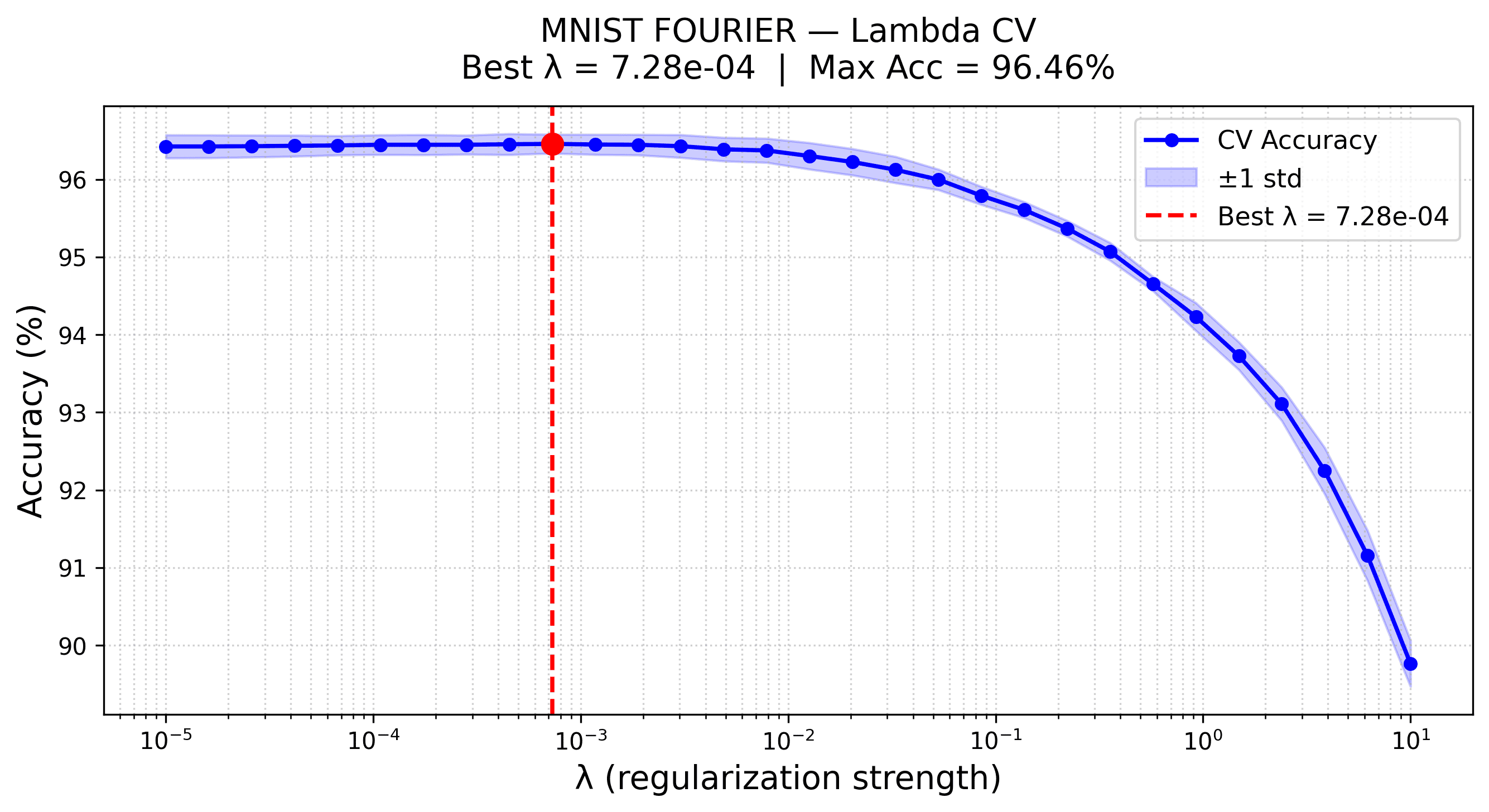}}
\caption{Fourier embedding.}
\end{subfigure}
\caption{Ridge parameter optimization for MNIST classification. The Fourier embedding gives slightly improved performance, reaching \(96.56\%\) accuracy at \(\lambda=7.278\times10^{ -4}\).}
\label{fig:lambda mnist}
\end{figure}

\subsection{FSDD spoken digit classification}

\begin{figure}[H]
\centering
\begin{subfigure}{0.49\linewidth}
\centering
\includegraphics[trim={0cm 0cm 0cm 0.75cm}, clip, width=\linewidth]{\detokenize{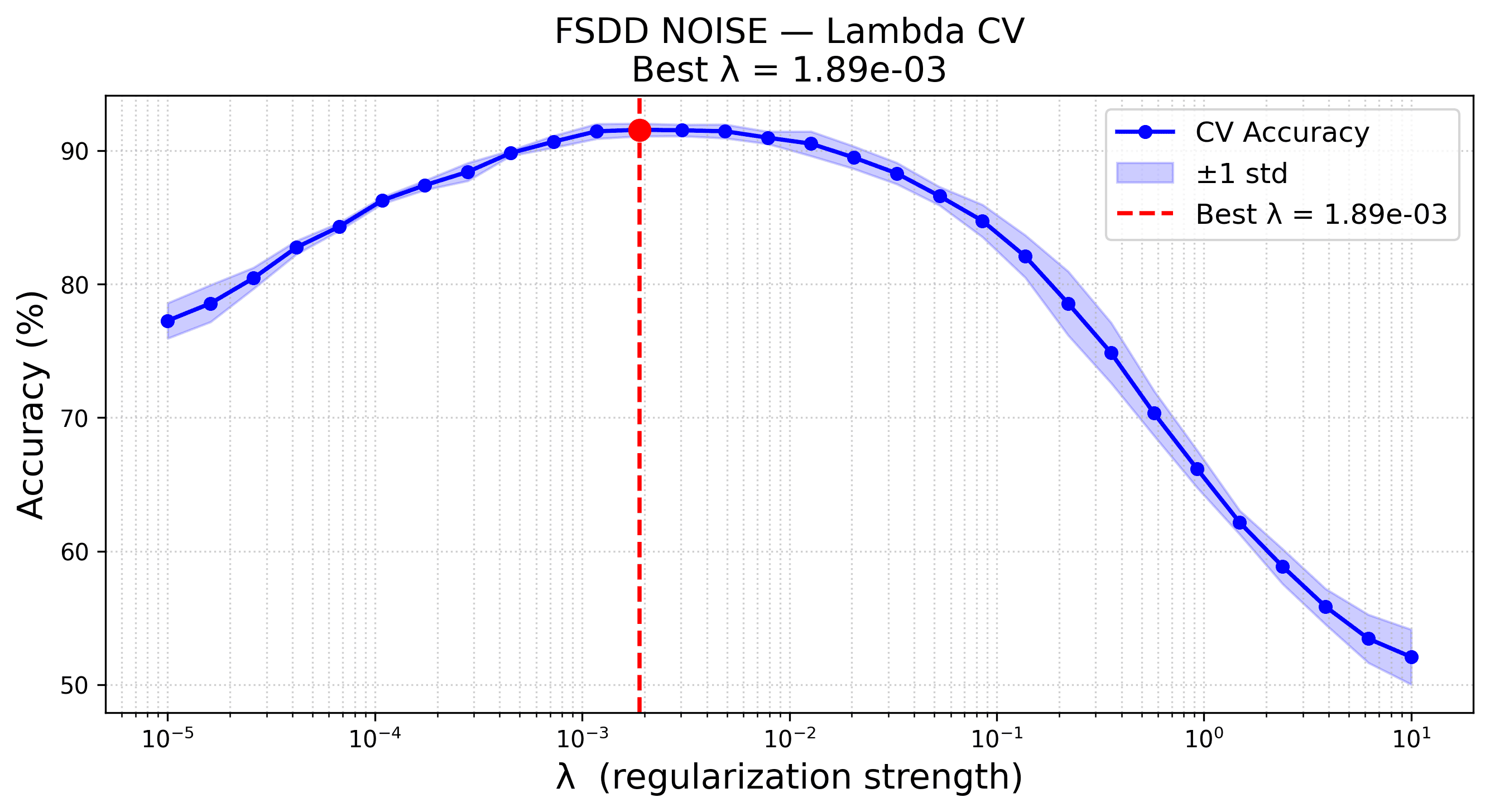}}
\caption{Noise embedding.}
\end{subfigure}
\hfill
\begin{subfigure}{0.49\linewidth}
\centering
\includegraphics[trim={0cm 0cm 0cm 0.75cm}, clip, width=\linewidth]{\detokenize{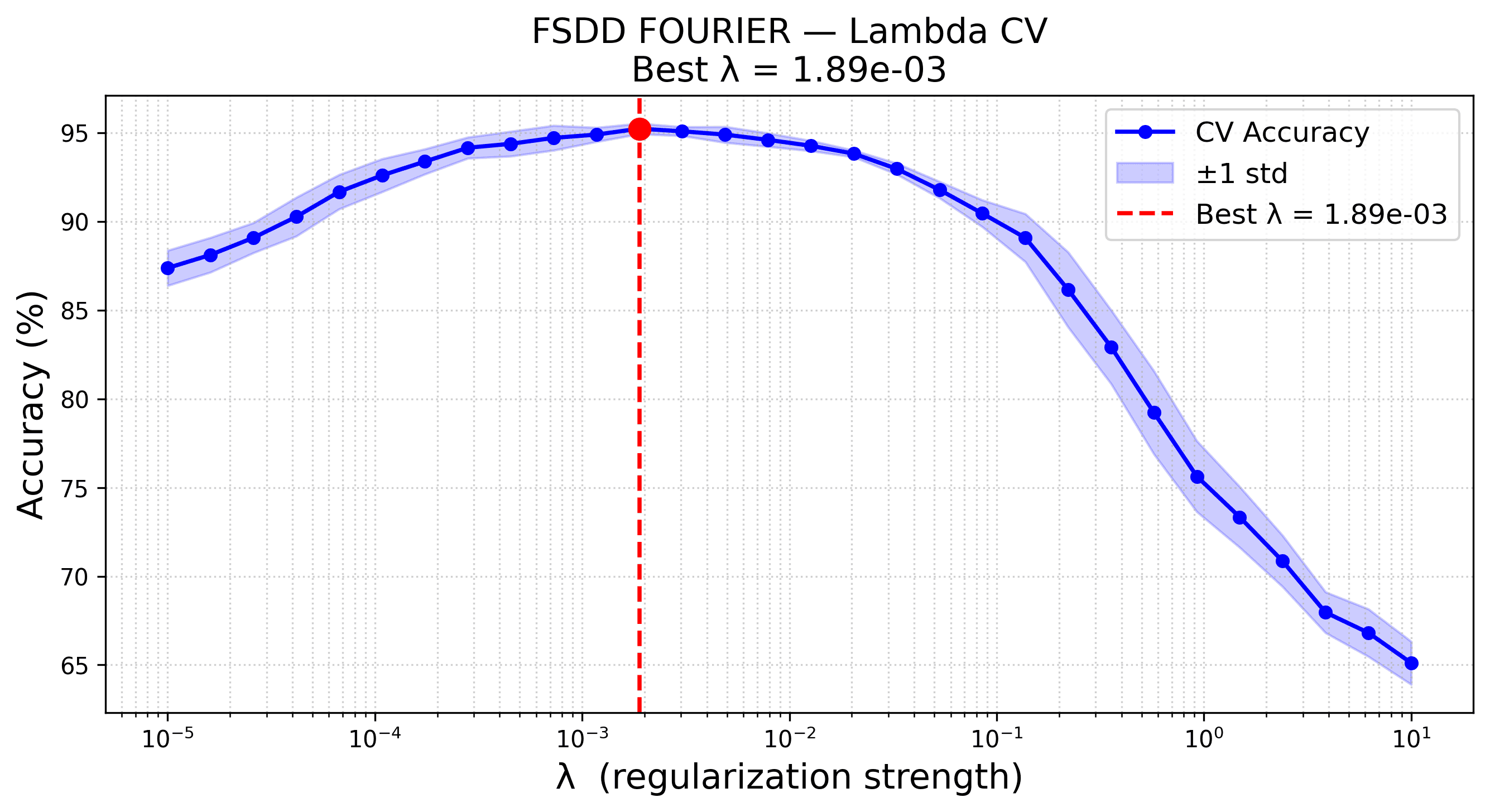}}
\caption{Fourier embedding.}
\end{subfigure}
\caption{Ridge parameter optimization for FSDD spoken digit classification. The Fourier embedding reaches \(95.67\%\) accuracy at \(\lambda=1.887\times10^{ -3}\).}
\label{fig:lambda -fsdd}
\end{figure}

\subsection{Mushroom binary classification}

\begin{figure}[H]
\centering
\begin{subfigure}{0.49\linewidth}
\centering
\includegraphics[trim={0cm 0cm 0cm 0.75cm}, clip, width=\linewidth]{\detokenize{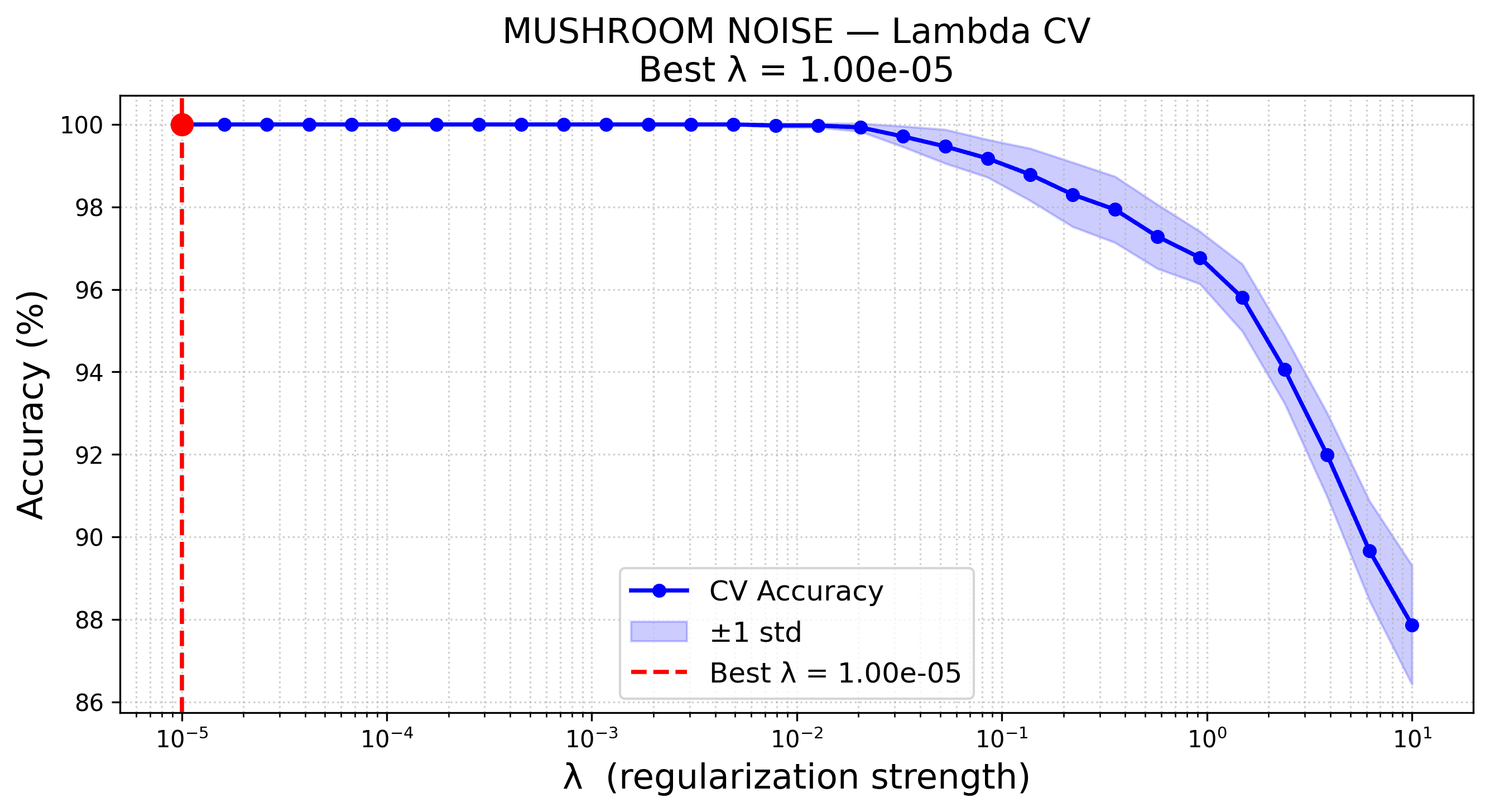}}
\caption{Noise embedding.}
\end{subfigure}
\hfill
\begin{subfigure}{0.49\linewidth}
\centering
\includegraphics[trim={0cm 0cm 0cm 0.75cm}, clip, width=\linewidth]{\detokenize{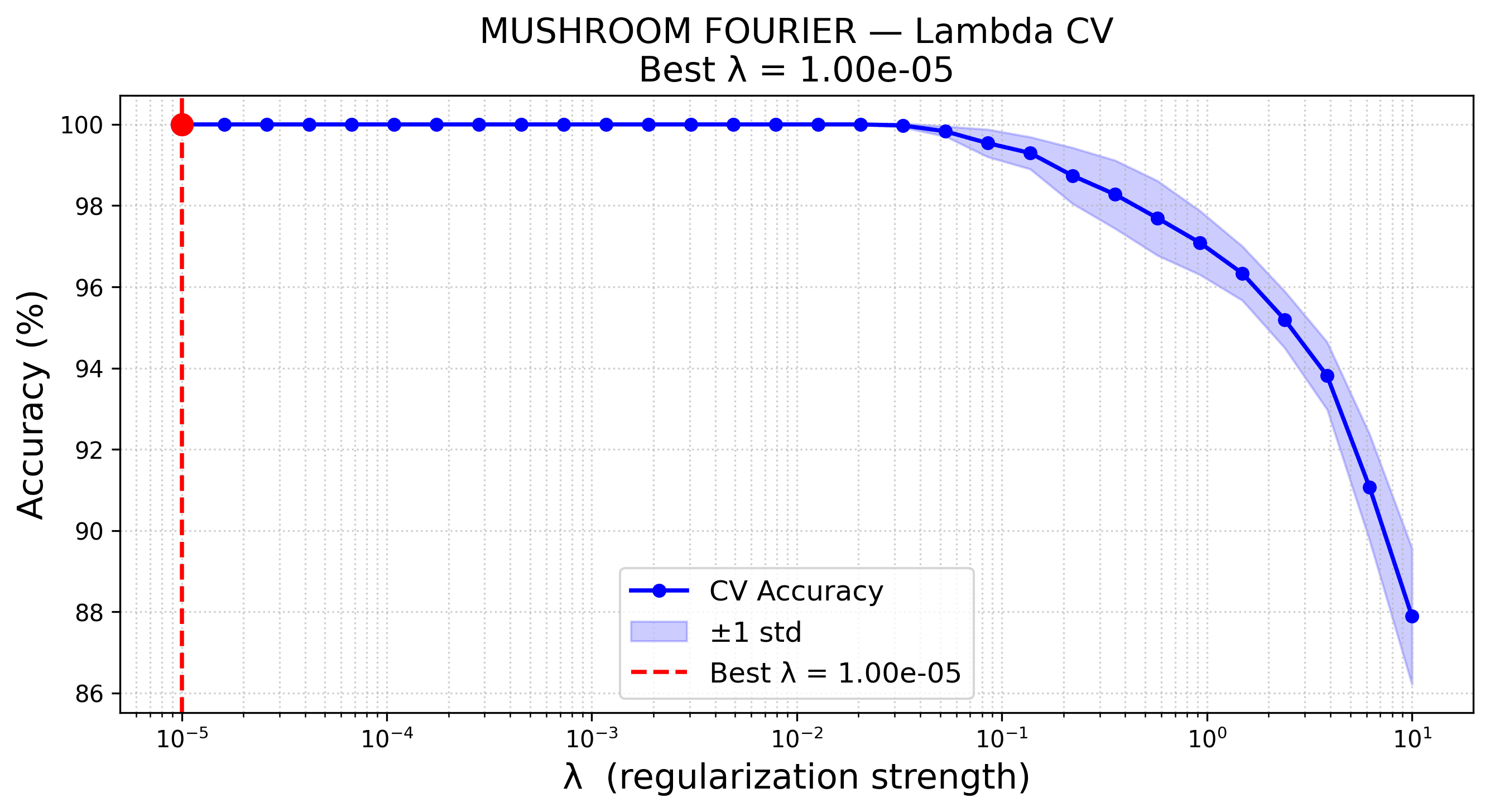}}
\caption{Fourier embedding.}
\end{subfigure}
\caption{Ridge parameter optimization for Mushroom binary classification. Both embeddings operate at saturation and achieve \(100.00\%\) classification accuracy.}
\label{fig:lambda mushroom}
\end{figure}

\subsection{Abalone regression}

\begin{figure}[H]
\centering
\begin{subfigure}{0.49\linewidth}
\centering
\includegraphics[trim={0cm 0cm 0cm 0.75cm}, clip, width=\linewidth]{\detokenize{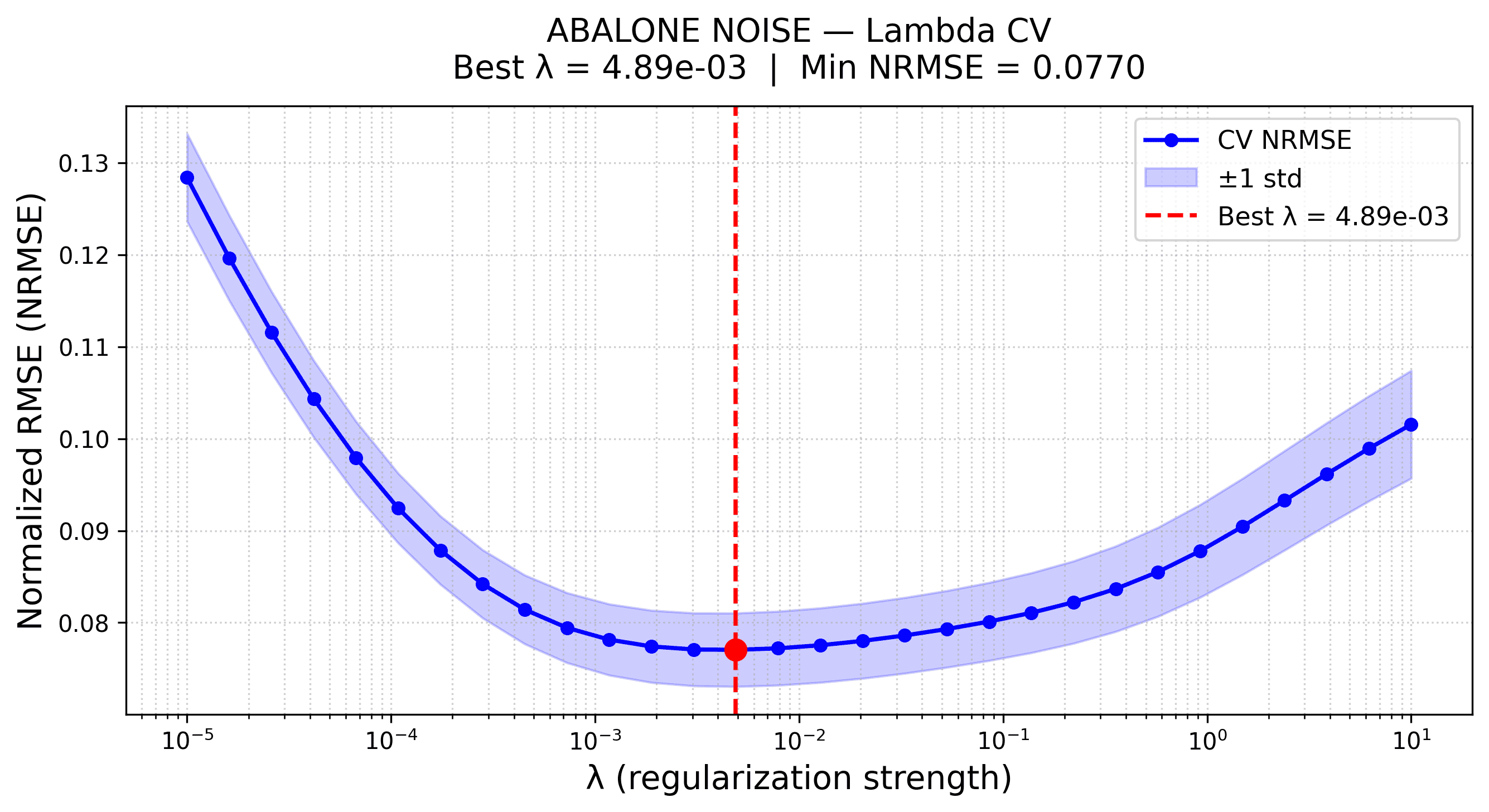}}
\caption{Noise embedding.}
\end{subfigure}
\hfill
\begin{subfigure}{0.49\linewidth}
\centering
\includegraphics[trim={0cm 0cm 0cm 0.75cm}, clip, width=\linewidth]{\detokenize{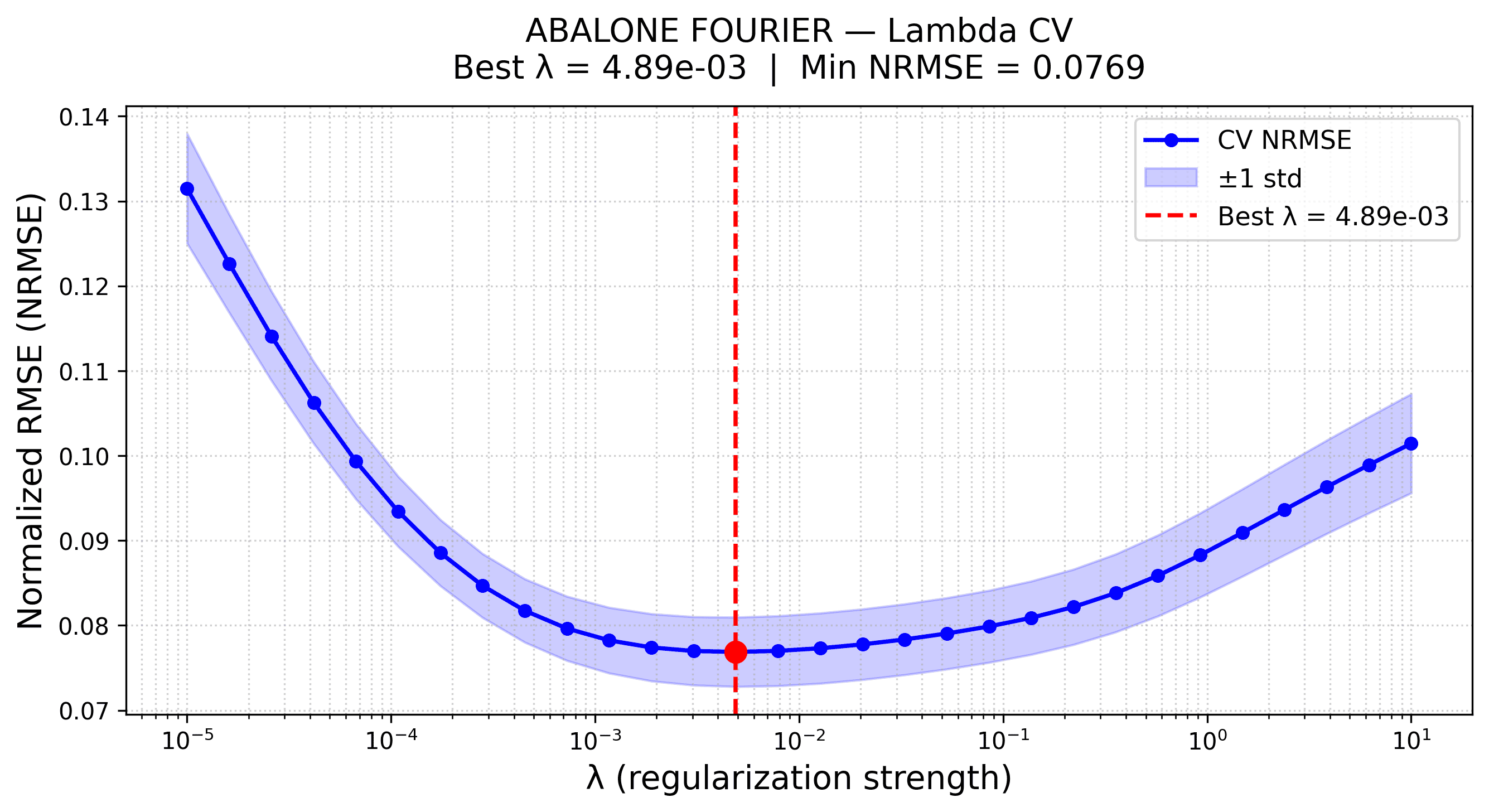}}
\caption{Fourier embedding.}
\end{subfigure}
\caption{Ridge parameter optimization for Abalone regression.  The metric is NRMSE, so lower values correspond to better performance.  The optimized values are reported in Table~\ref{tab:performance}.}
\label{fig:lambda abalone}
\end{figure}

These sweeps show that optical feature extraction and readout regularization must be considered jointly.  The embedding controls the measured feature geometry, while \(\lambda\) controls how strongly the linear readout can exploit weak directions in that feature space.

\section{Supplementary feature space diagnostics}
\label{app:feature diagnostics}

This section contains the empirical diagnostics for the noise embedding together with additional comparative diagnostics for both embedding strategies beyond those presented in the main text.

\subsection{Distance preservation }

\begin{figure}[H]
\centering
\begin{subfigure}{1\linewidth}
\centering
\includegraphics[trim={0cm 0cm 0cm 1.5cm}, clip, width=0.82\linewidth]{\detokenize{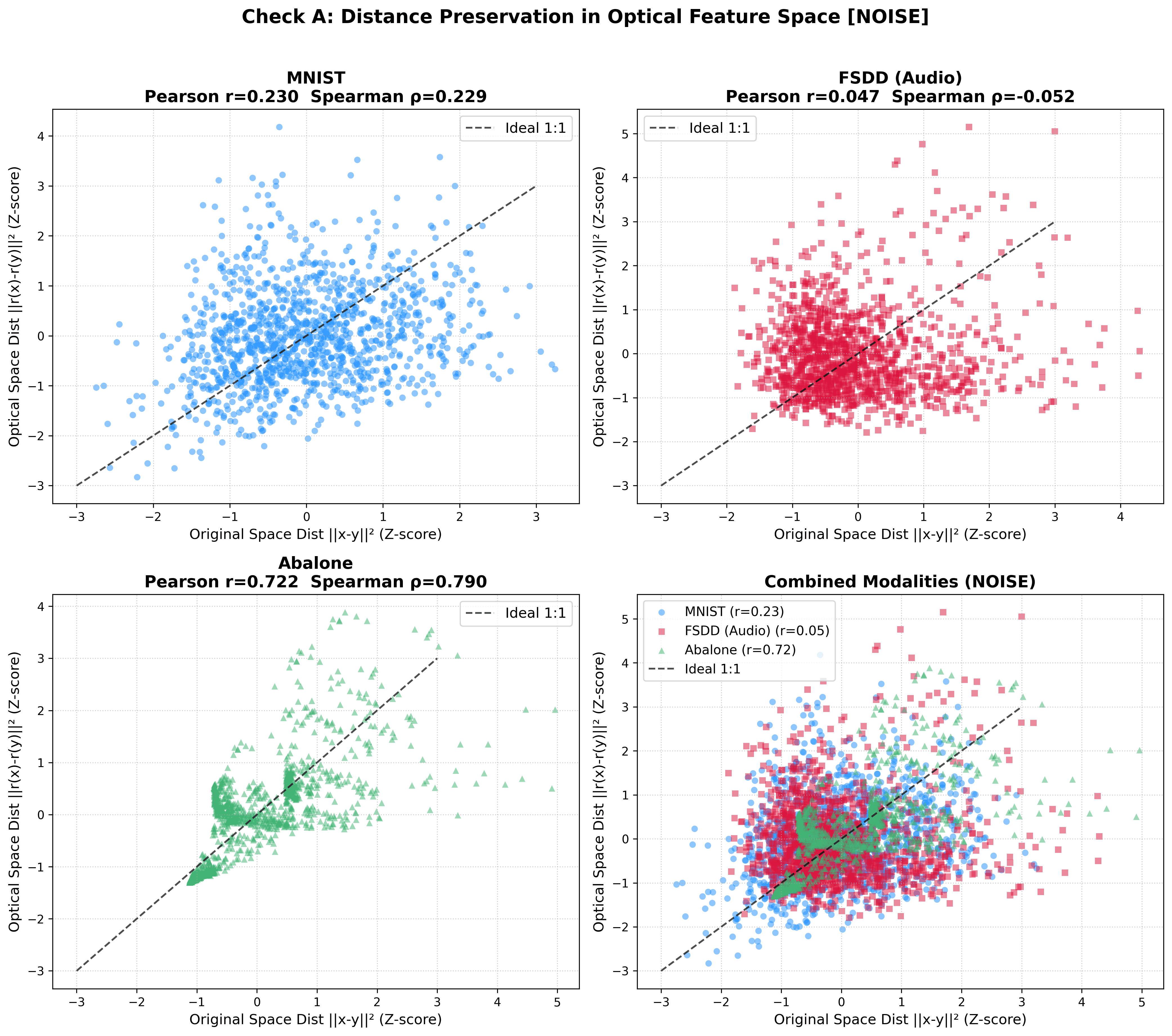}}
\end{subfigure}
\caption{Relative distance preservation diagnostics for Noise Embedding.}  
\label{fig:si distance preservation}
\end{figure}

\subsection{Supplementary Experimental Kernel Fits}
\label{app:kernel fits}

To characterize the optical manifold mapping performed by the PELM, we compare the empirical optical kernel against five analytical models. We define the angular separation between $L_2$ normalized input vectors $\mathbf{x}_i, \mathbf{x}_j$ as $\theta_{ij} = \cos^{ -1}(\text{clip}(\mathbf{x}_i^\top \mathbf{x}_j,  -1, 1))$. The empirical kernel is estimated as $K_{\mathrm{emp}} = \mathbf{g}_i^\top \mathbf{g}_j$, where $\mathbf{g}_i$ denotes the centered and normalized optical feature vector. In equation of $K_{\mathrm{phase}}$ $d$ denotes the dimensionality of the input vectors $x$.

We evaluate the following theoretical kernel geometries:

\begin{align}
    K_{\mathrm{phase}}(\mathbf{x}_i, \mathbf{x}_j) &= 1 + (\mathbf{x}_i^\top \mathbf{x}_j)^2  - \frac{1}{d}\sum_{k=1}^{d} x_{ik}^2 x_{jk}^2, \label{eq:kphase} \\
    K_{\mathrm{Gauss}}(\theta) &= 1 + \cos^2\theta, \label{eq:kgauss} \\
    K_{2}(\theta) &= \frac{\sin\theta + (\pi -\theta)\cos\theta}{\pi}, \label{eq:k2} \\
    K_{1}(\theta) &= \frac{\pi -\theta}{\pi}, \label{eq:k1} \\
    K_{\mathrm{RBF}}(\theta) &= \exp( -\gamma \theta^2). \label{eq:krbf}
\end{align}

All theoretical kernels are subsequently double centered prior to comparison with the empirical optical kernel.Kernel agreement is quantified using Pearson correlation coefficient ($r$) and root mean square error (RMSE) after least squares scaling.

The empirical PELM kernel is binned by input angle $\theta$ and compared against the corresponding centered theoretical predictions. Residual and fit quality panels are omitted in the supplementary plots for visual clarity, as the Pearson correlation values are directly reported within the figure legends.
\begin{figure}[H]
\centering
\begin{subfigure}{0.49\linewidth}
\centering
\includegraphics[trim={0cm 0cm 25cm 0cm}, clip, width=\linewidth]{\detokenize{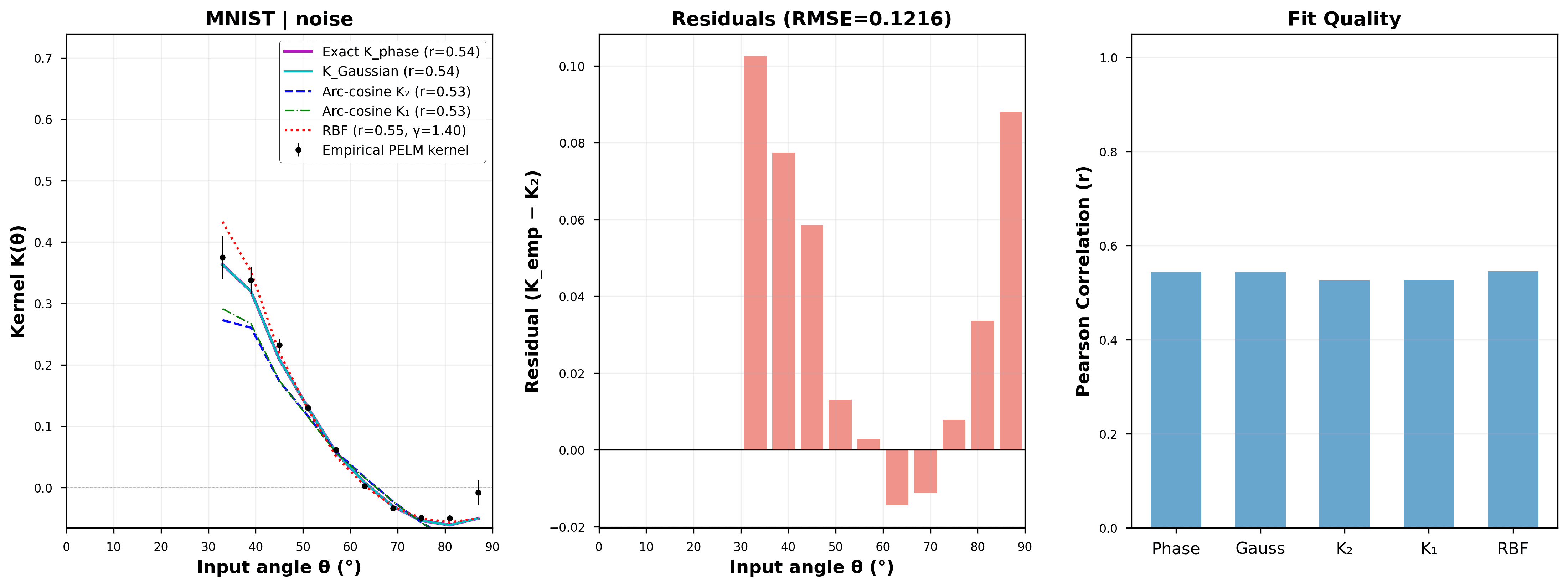}}
\caption{MNIST (Noise)}
\end{subfigure}
\hfill
\begin{subfigure}{0.49\linewidth}
\centering
\includegraphics[trim={0cm 0cm 25cm 0cm}, clip, width=\linewidth]{\detokenize{f_kernel_comparison_mnist.png}}
\caption{MNIST (Fourier)}
\end{subfigure}
\caption{Experimental kernel characterizations for the MNIST dataset.}
\label{fig:si kernel fits mnist}
\end{figure}

\begin{figure}[H]
\centering
\begin{subfigure}{0.49\linewidth}
\centering
\includegraphics[trim={0cm 0cm 25cm 0cm}, clip, width=\linewidth]{\detokenize{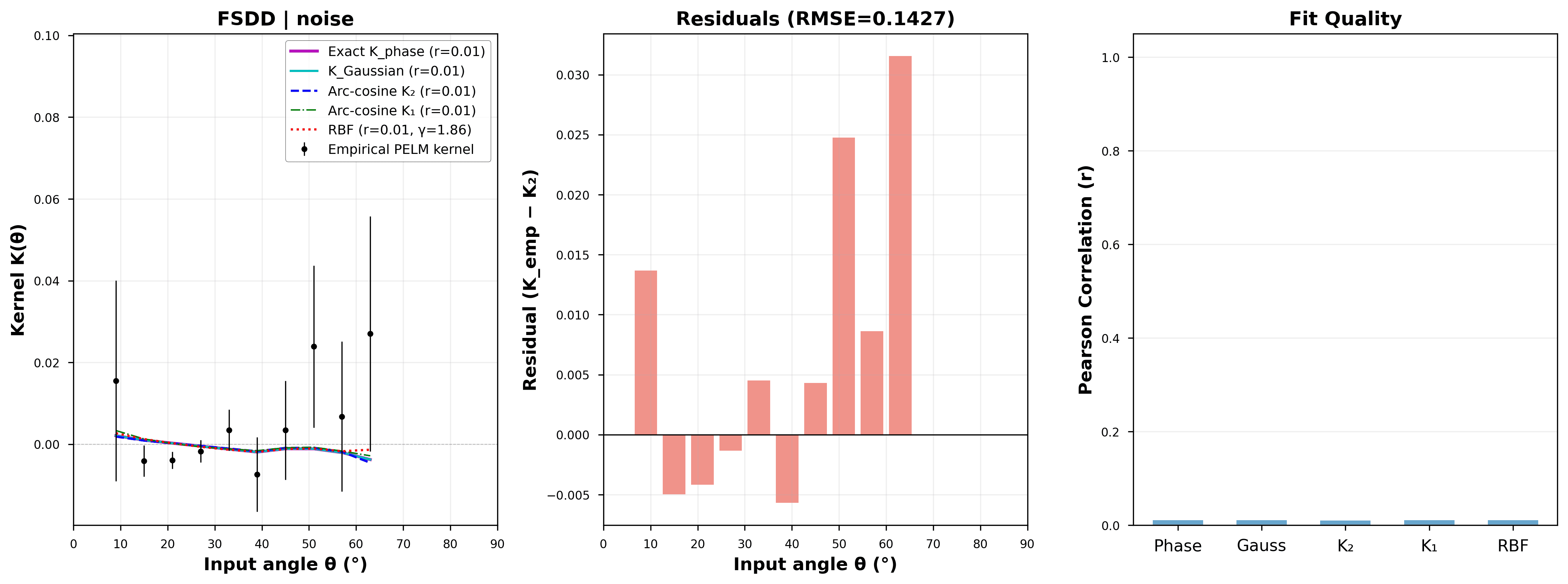}}
\caption{FSDD (Noise)}
\end{subfigure}
\hfill
\begin{subfigure}{0.49\linewidth}
\centering
\includegraphics[trim={0cm 0cm 25cm 0cm}, clip, width=\linewidth]{\detokenize{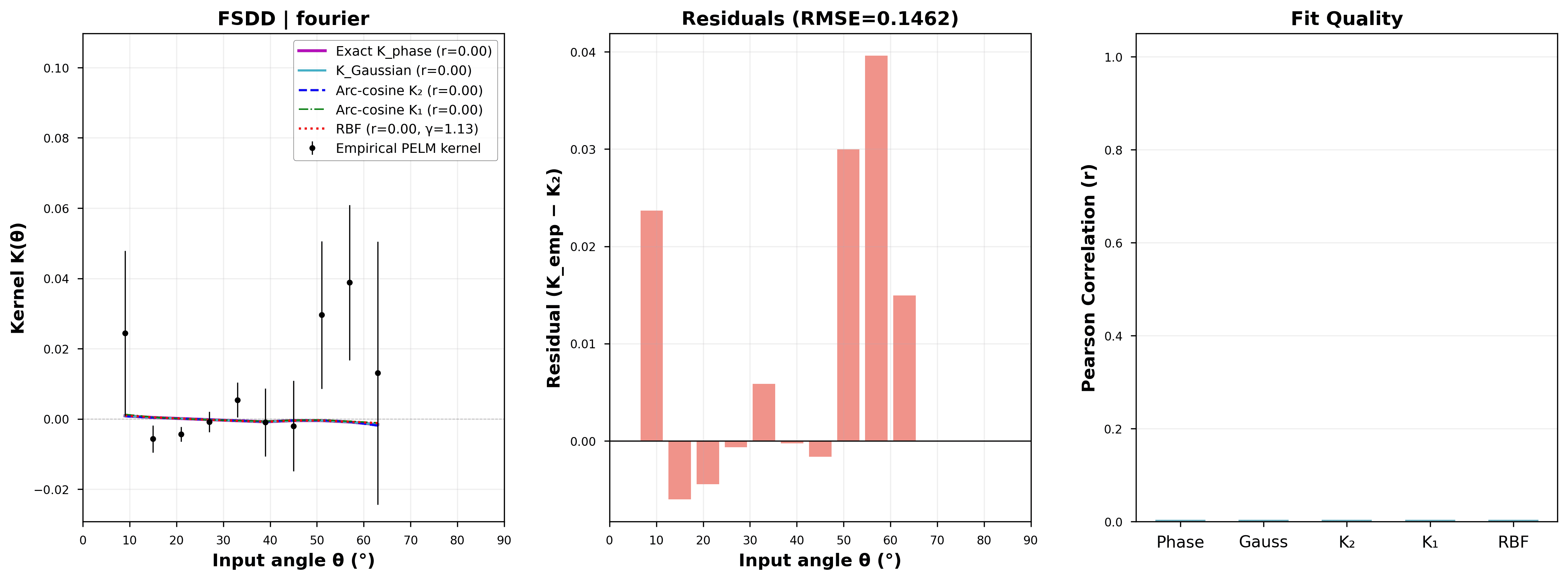}}
\caption{FSDD (Fourier)}
\end{subfigure}
\caption{Experimental kernel characterizations for the FSDD dataset. Note the flat angular distribution due to the high dimensionality of the audio spectrograms causing distance concentration.}
\label{fig:si kernel fits fsdd}
\end{figure}

\begin{figure}[H]
\centering
\begin{subfigure}{0.49\linewidth}
\centering
\includegraphics[trim={0cm 0cm 25cm 0cm}, clip, width=\linewidth]{\detokenize{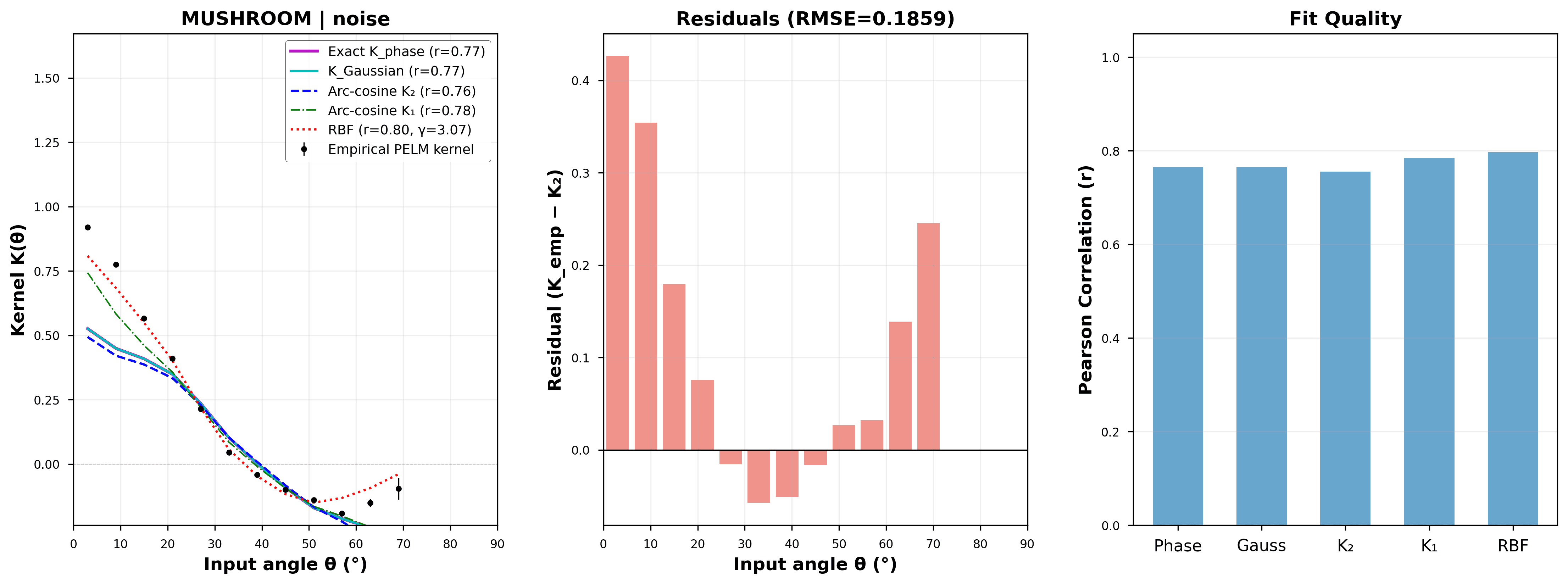}}
\caption{Mushroom (Noise)}
\end{subfigure}
\hfill
\begin{subfigure}{0.49\linewidth}
\centering
\includegraphics[trim={0cm 0cm 25cm 0cm}, clip, width=\linewidth]{\detokenize{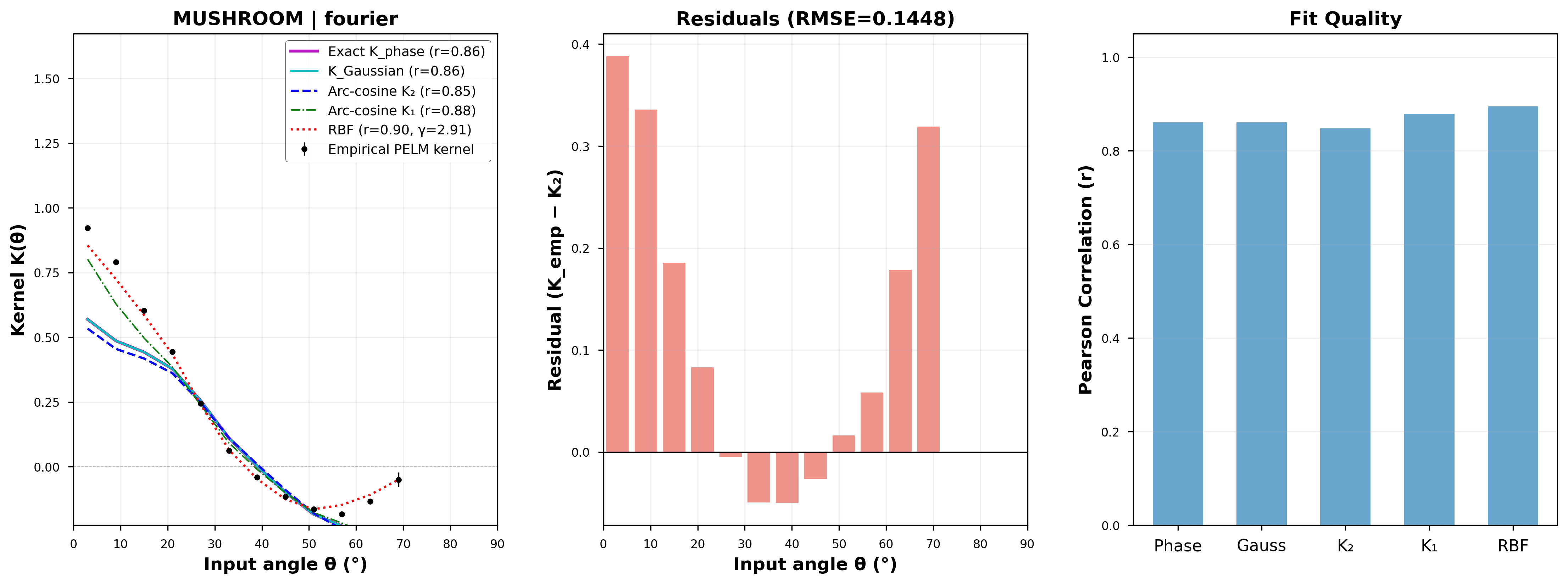}}
\caption{Mushroom (Fourier)}
\end{subfigure}
\caption{Experimental kernel characterizations for the Mushroom dataset.}
\label{fig:si kernel fits mushroom}
\end{figure}

\begin{figure}[H]
\centering
\begin{subfigure}{0.49\linewidth}
\centering
\includegraphics[trim={0cm 0cm 25cm 0cm}, clip, width=\linewidth]{\detokenize{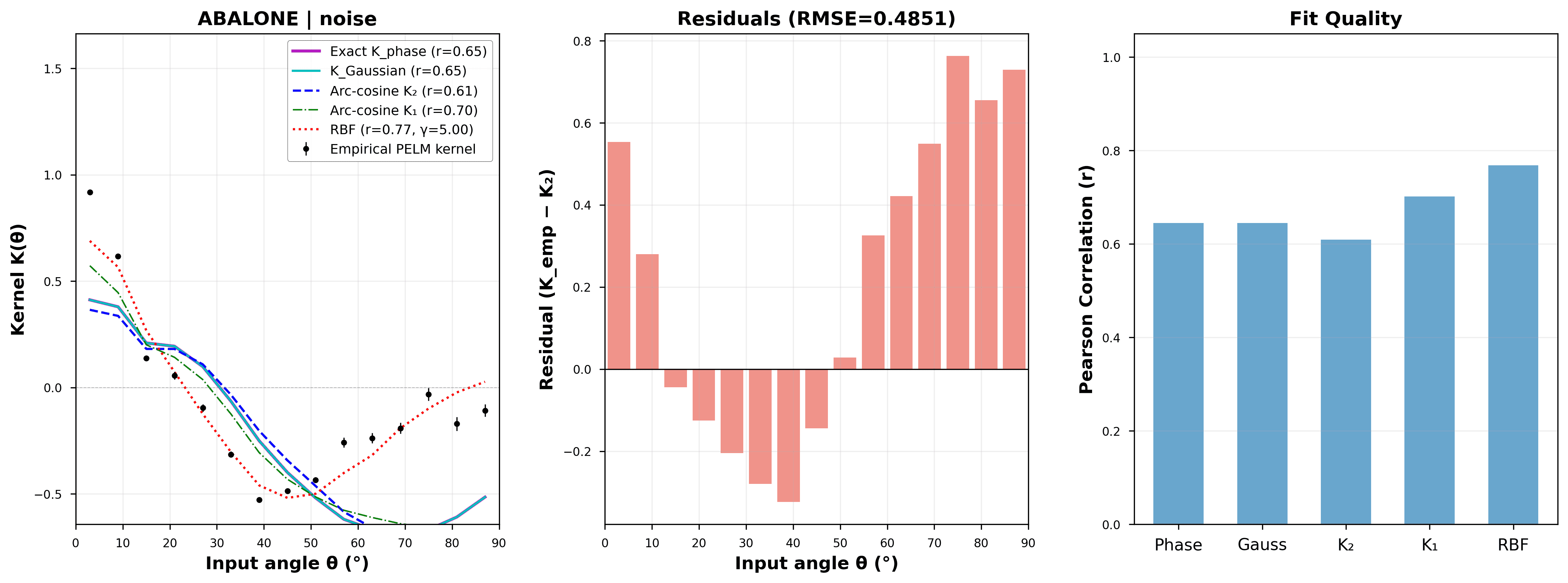}}
\caption{Abalone (Noise)}
\end{subfigure}
\hfill
\begin{subfigure}{0.49\linewidth}
\centering
\includegraphics[trim={0cm 0cm 25cm 0cm}, clip, width=\linewidth]{\detokenize{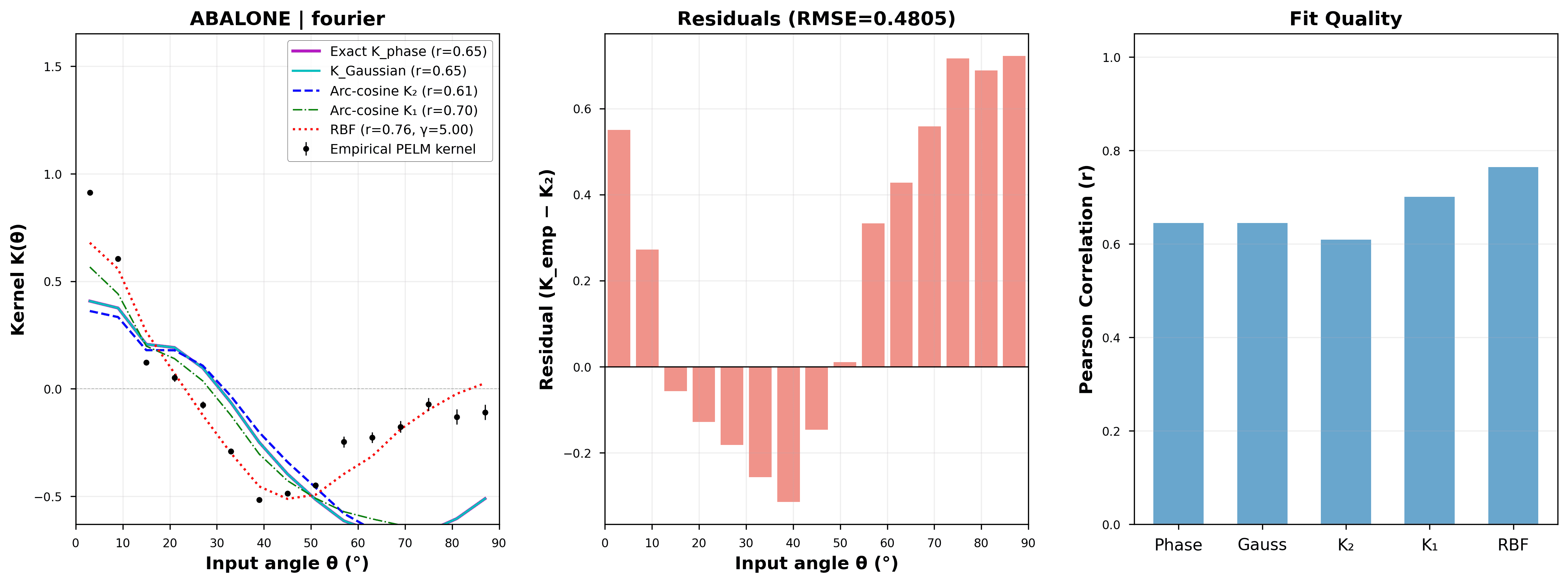}}
\caption{Abalone (Fourier)}
\end{subfigure}
\caption{Experimental kernel characterizations for the Abalone dataset.}
\label{fig:si kernel fits abalone}
\end{figure}
\subsection{Per class separation diagnostics}

\begin{figure}[H]
\centering
\begin{subfigure}{0.45\linewidth}
\centering
\includegraphics[width=\linewidth]{\detokenize{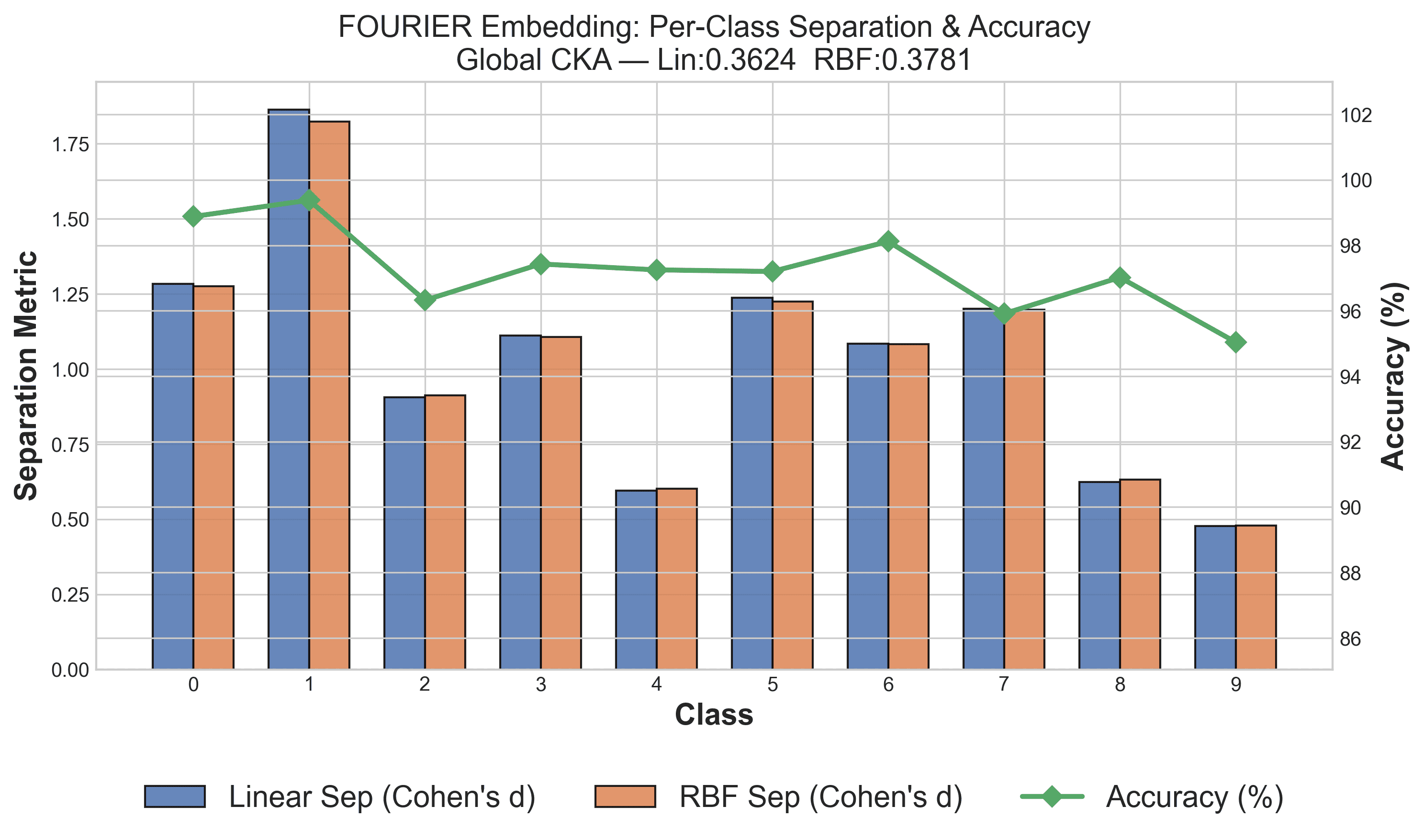}}
\caption{MNIST}
\end{subfigure}
\hfill
\begin{subfigure}{0.45\linewidth}
\centering
\includegraphics[width=\linewidth]{\detokenize{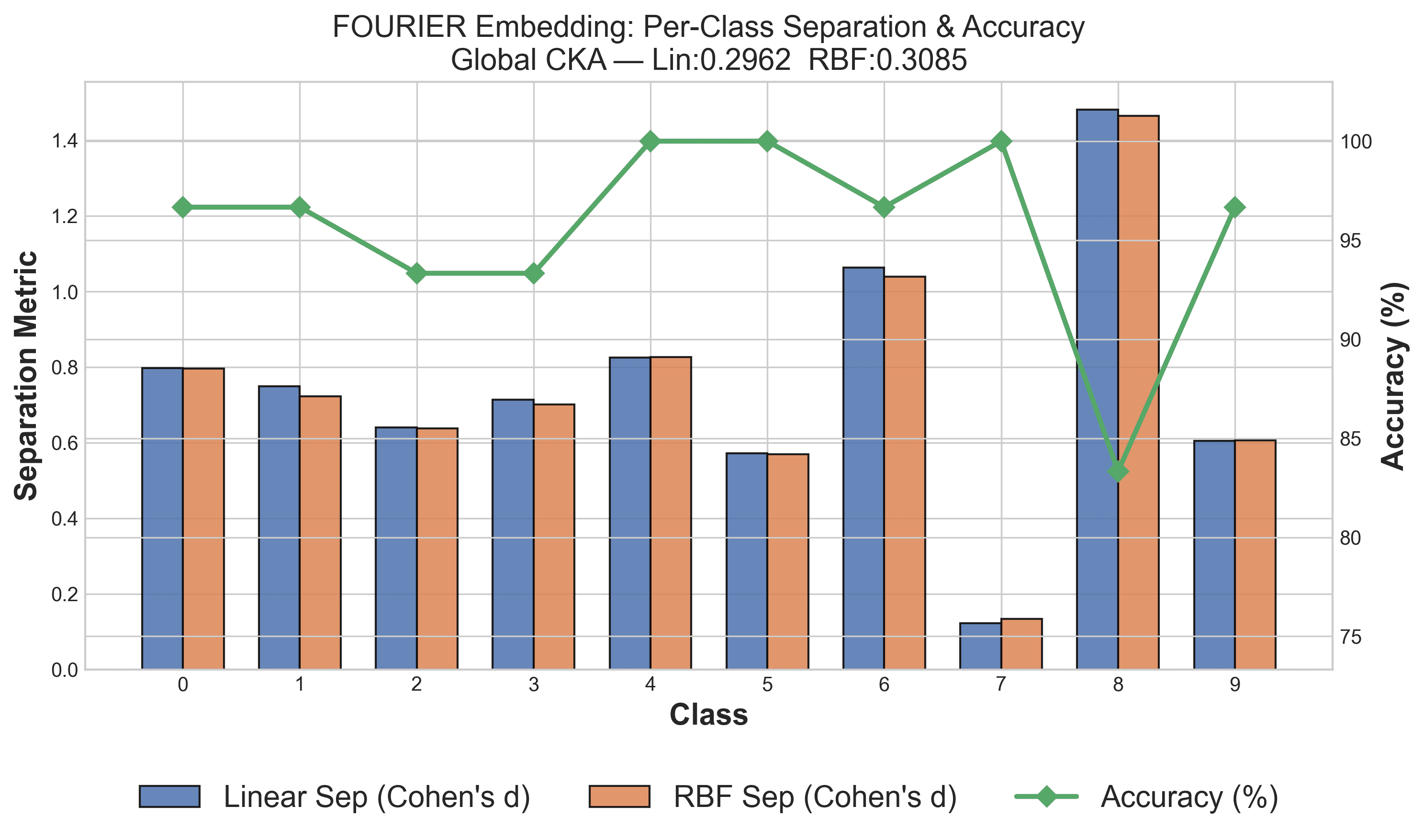}}
\caption{FSDD}
\end{subfigure}

\vspace{0.8em}

\begin{subfigure}{0.45\linewidth}
\centering
\includegraphics[width=\linewidth]{\detokenize{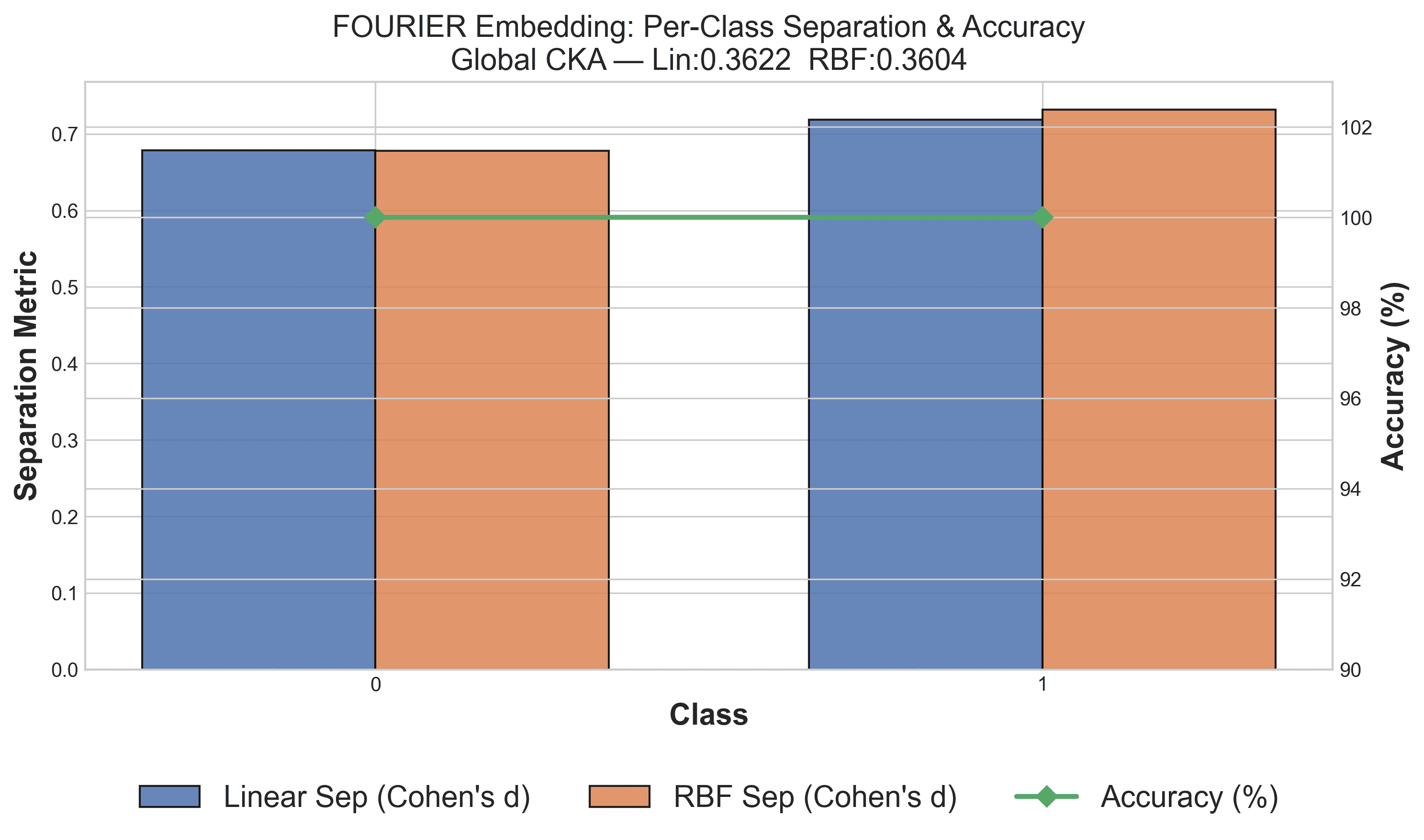}}
\caption{Mushroom}
\end{subfigure}
\hfill

\caption{Per class or per target separation diagnostics for the Fourier embedding.  MNIST and Mushroom show stronger visible class organization, while FSDD has weaker global separation despite high classification accuracy. Abalone is excluded from these bar plots as categorical class separation metrics (Cohen's $d$) are inapplicable to continuous regression variables.}
\label{fig:si separation individual fourier}
\end{figure}

\begin{figure}[H]
\centering
\begin{subfigure}{0.45\linewidth}
\centering
\includegraphics[width=\linewidth]{\detokenize{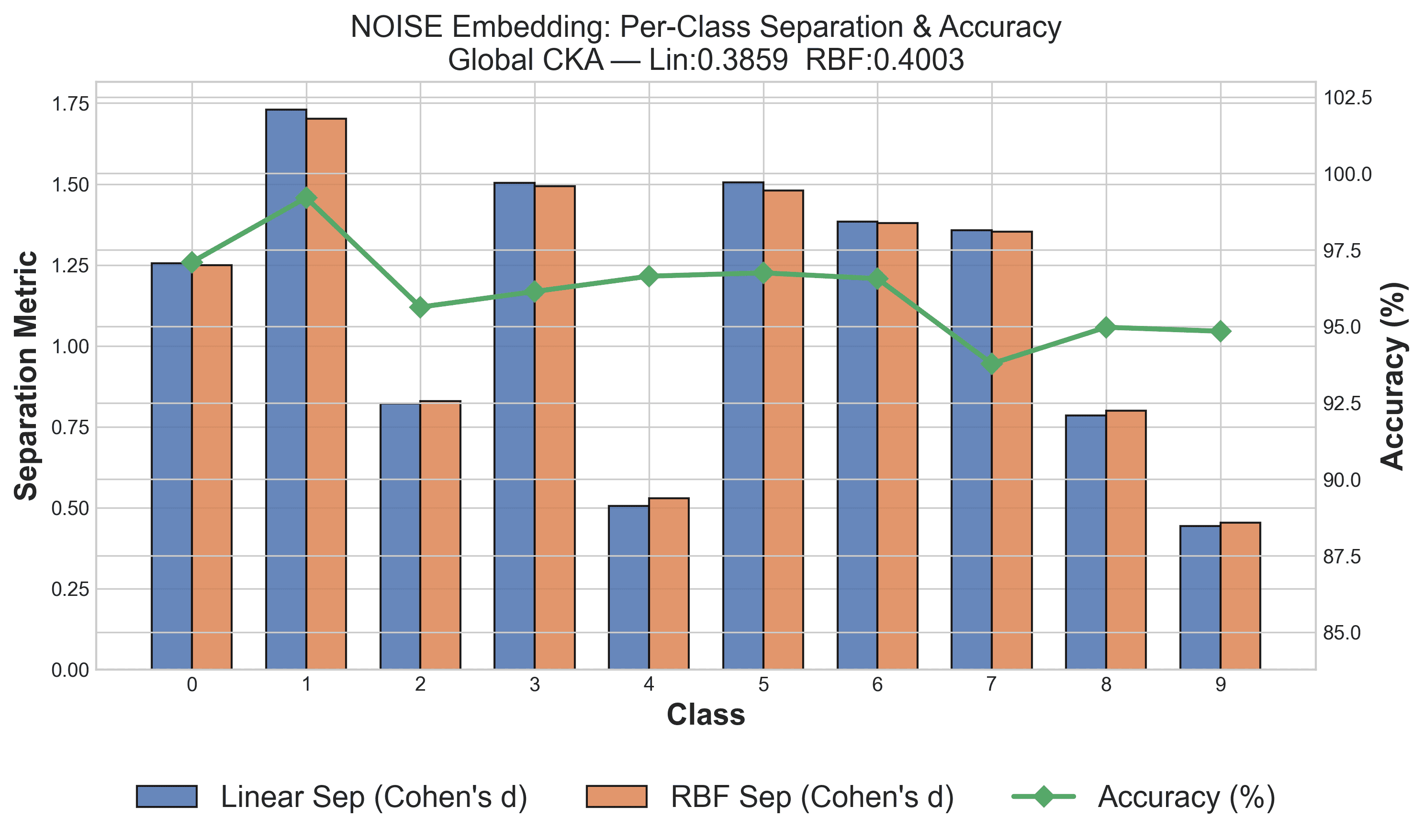}}
\caption{MNIST}
\end{subfigure}
\hfill
\begin{subfigure}{0.45\linewidth}
\centering
\includegraphics[width=\linewidth]{\detokenize{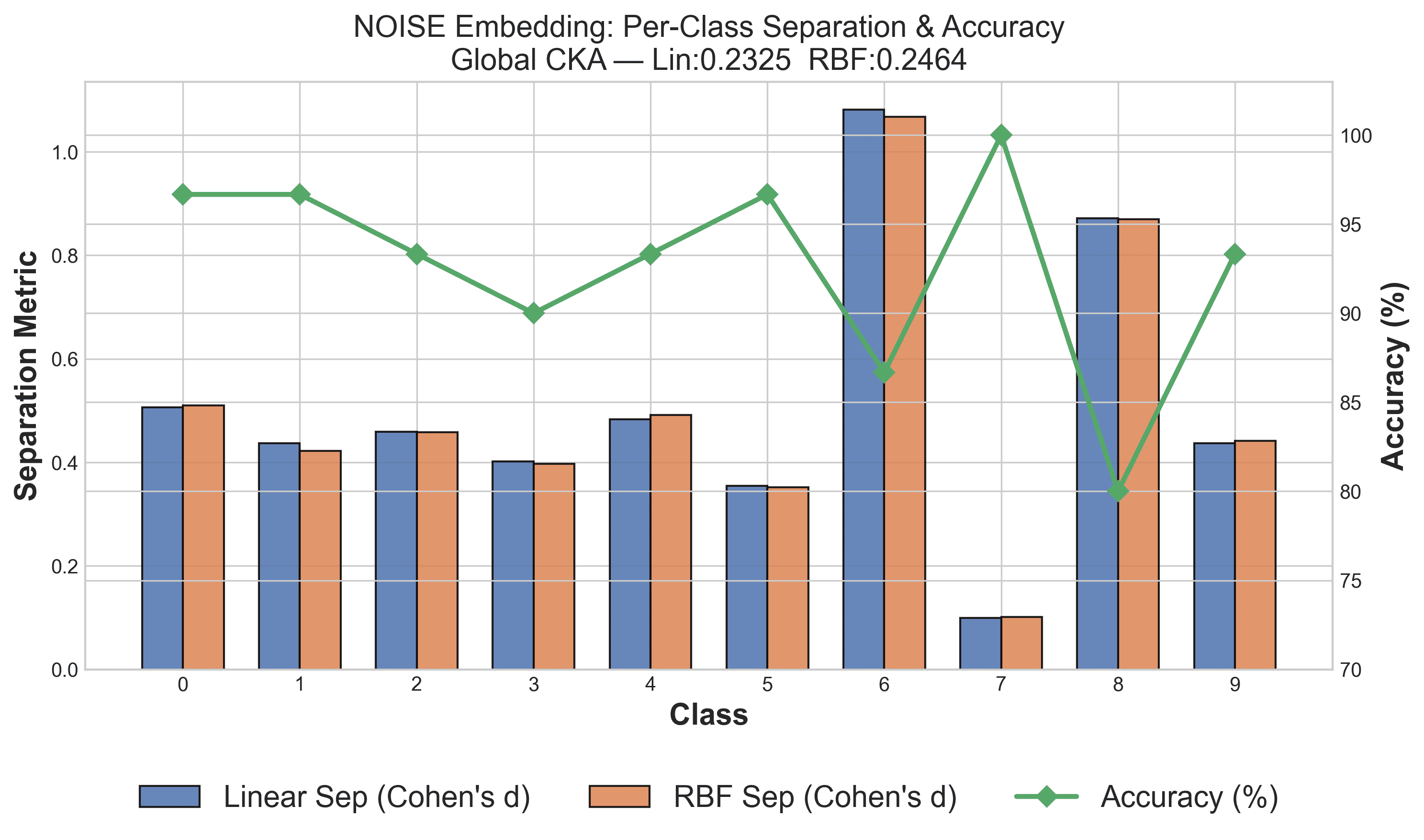}}
\caption{FSDD}
\end{subfigure}

\vspace{0.8em}

\begin{subfigure}{0.45\linewidth}
\centering
\includegraphics[width=\linewidth]{\detokenize{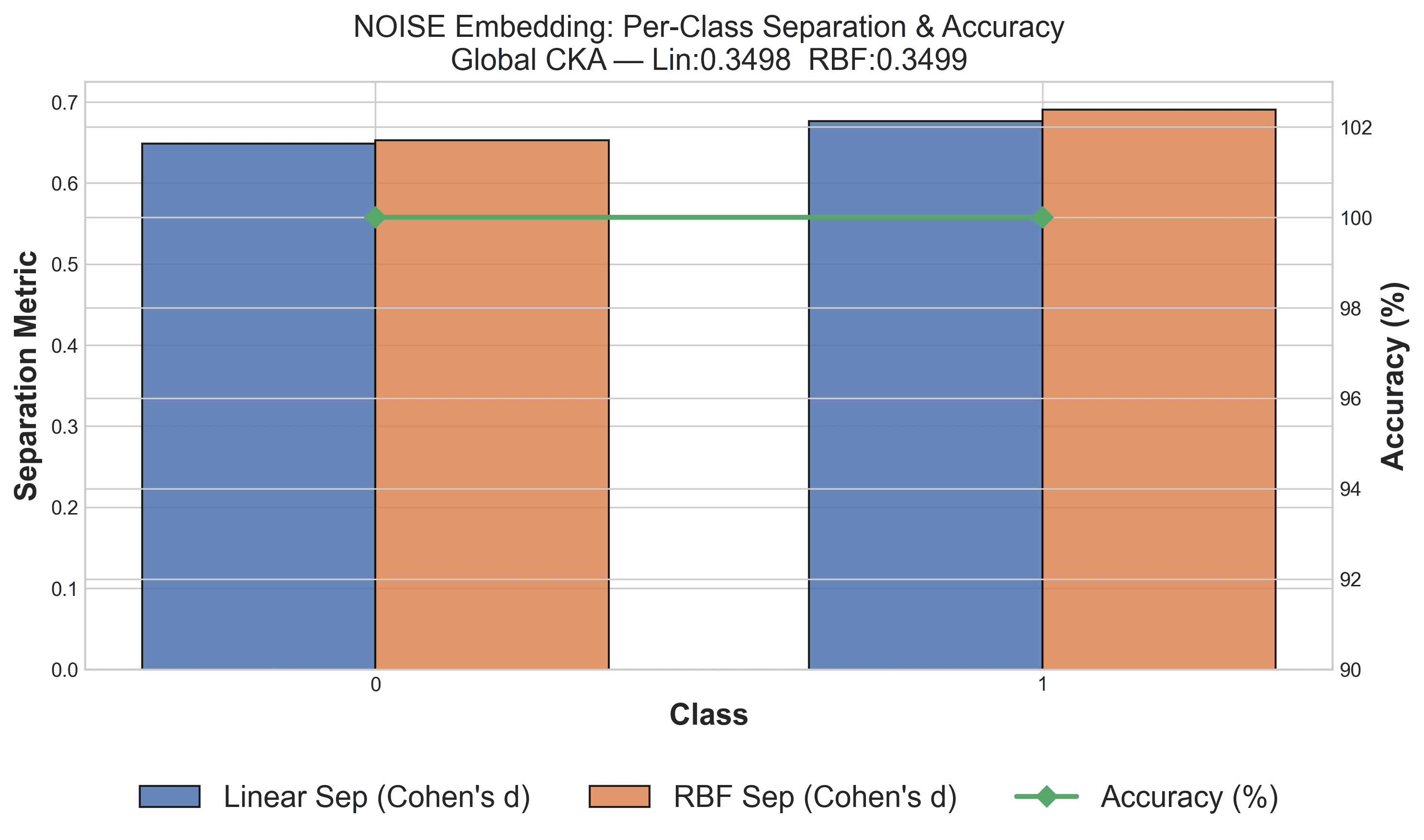}}
\caption{Mushroom}
\end{subfigure}
\hfill

\caption{Per class or per target separation diagnostics for the Noise embedding.}
\label{fig:si separation individual noise}
\end{figure}

\subsection{t-SNE visualizations}

\begin{figure}[H]
\centering
\begin{subfigure}{0.49\linewidth}
\centering
\includegraphics[width=\linewidth]{\detokenize{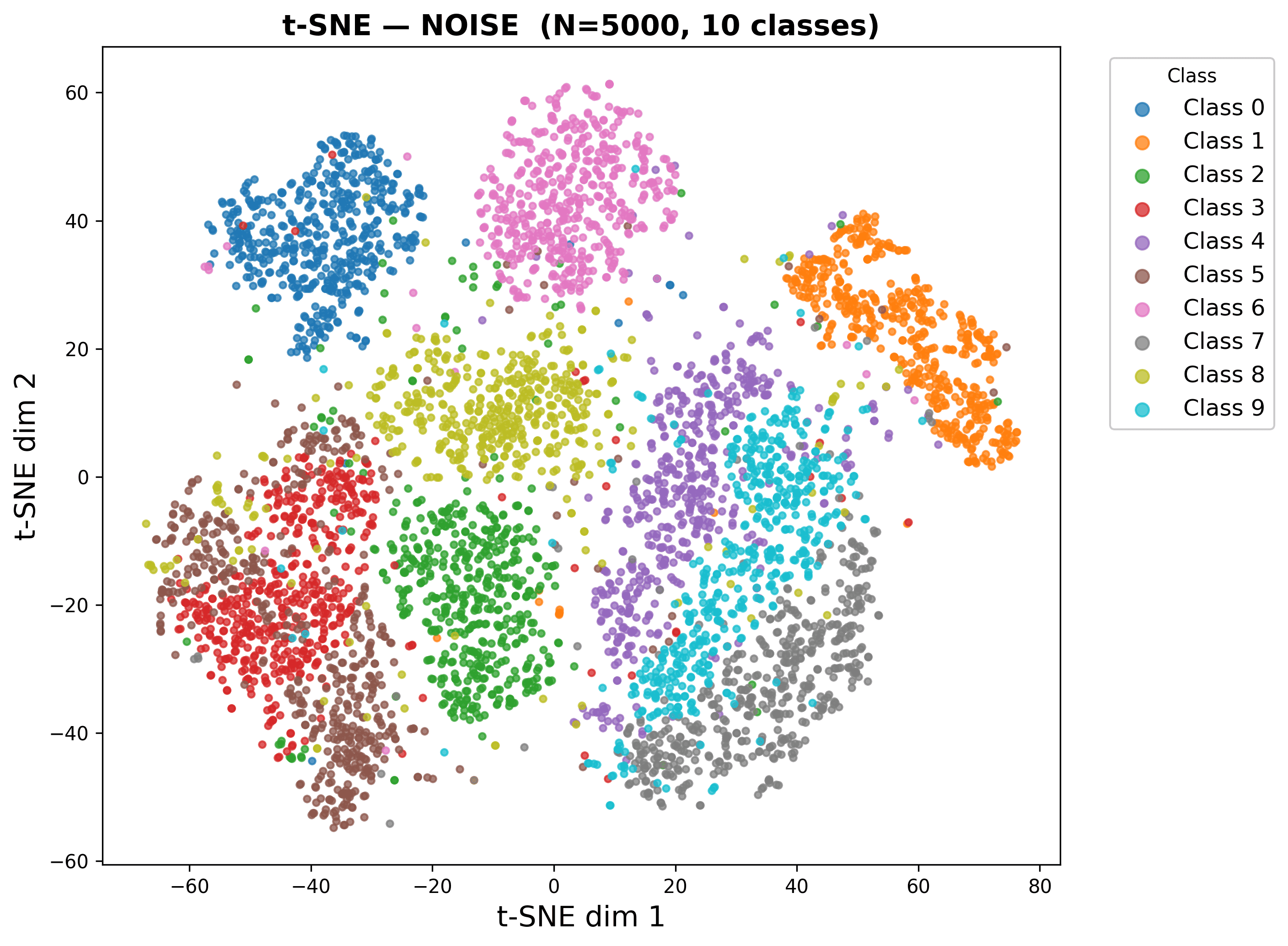}}
\caption{MNIST, Noise Embedding}
\end{subfigure}
\hfill
\begin{subfigure}{0.49\linewidth}
\centering
\includegraphics[width=\linewidth]{\detokenize{f_tsne_mnist.png}}
\caption{MNIST, Fourier Embedding}
\end{subfigure}

\vspace{0.8em}

\begin{subfigure}{0.49\linewidth}
\centering
\includegraphics[width=\linewidth]{\detokenize{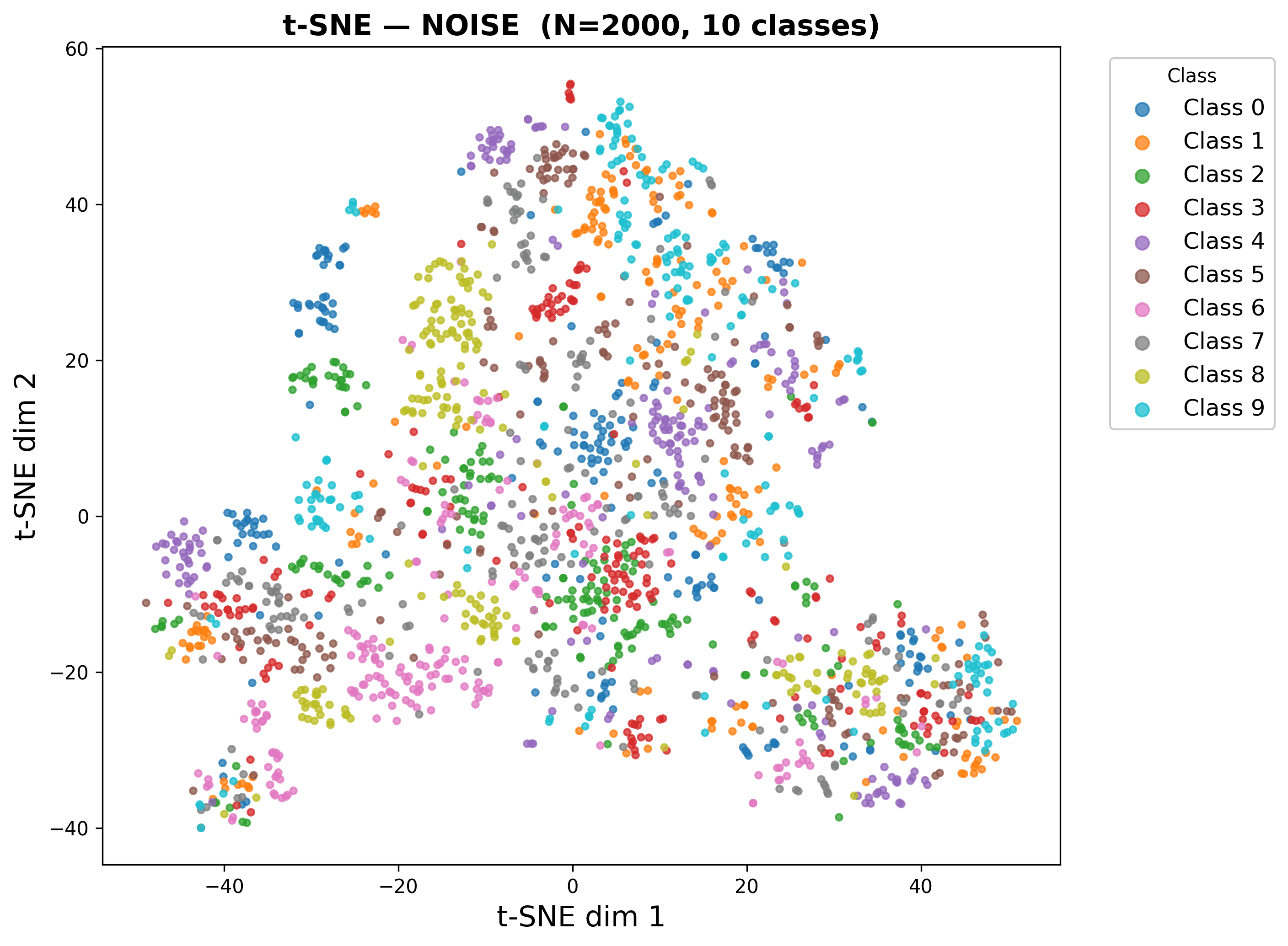}}
\caption{FSDD, Noise Embedding}
\end{subfigure}
\hfill
\begin{subfigure}{0.49\linewidth}
\centering
\includegraphics[width=\linewidth]{\detokenize{f_tsne_fsdd.png}}
\caption{FSDD, Fourier Embedding}
\end{subfigure}

\caption{t-SNE visualizations for MNIST and FSDD optical features under both embeddings. The plots are qualitative projections of the measured high dimensional camera features and should be interpreted alongside the quantitative accuracy, distance, and separation diagnostics.}
\label{fig:si tsne mnist fsdd}
\end{figure}

\begin{figure}[H]
\centering
\begin{subfigure}{0.49\linewidth}
\centering
\includegraphics[width=\linewidth]{\detokenize{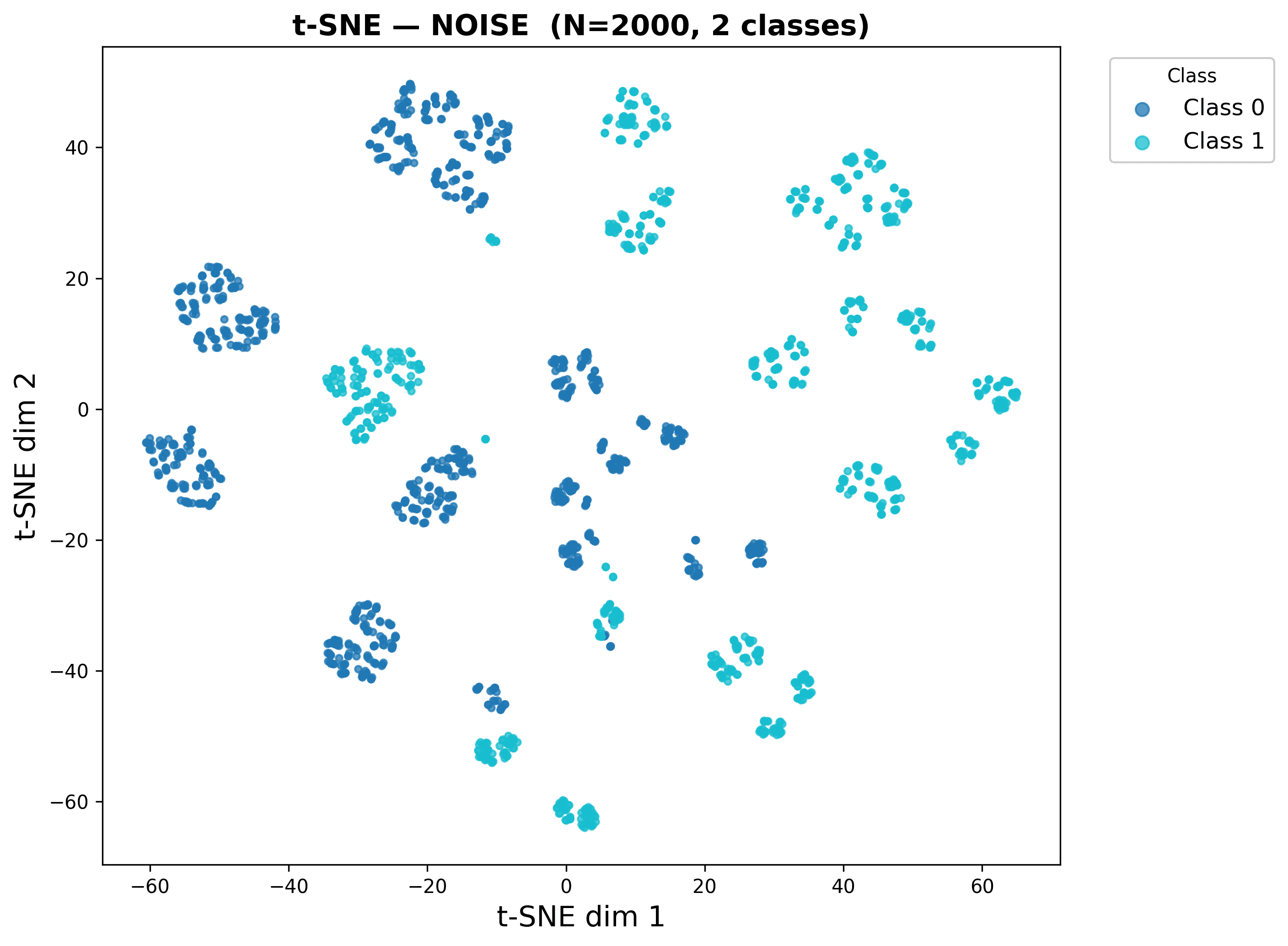}}
\caption{Mushroom, Noise Embedding}
\end{subfigure}
\hfill
\begin{subfigure}{0.49\linewidth}
\centering
\includegraphics[width=\linewidth]{\detokenize{f_tsne_mushroom.png}}
\caption{Mushroom, Fourier Embedding}
\end{subfigure}

\vspace{0.8em}

\begin{subfigure}{0.49\linewidth}
\centering
\includegraphics[width=\linewidth]{\detokenize{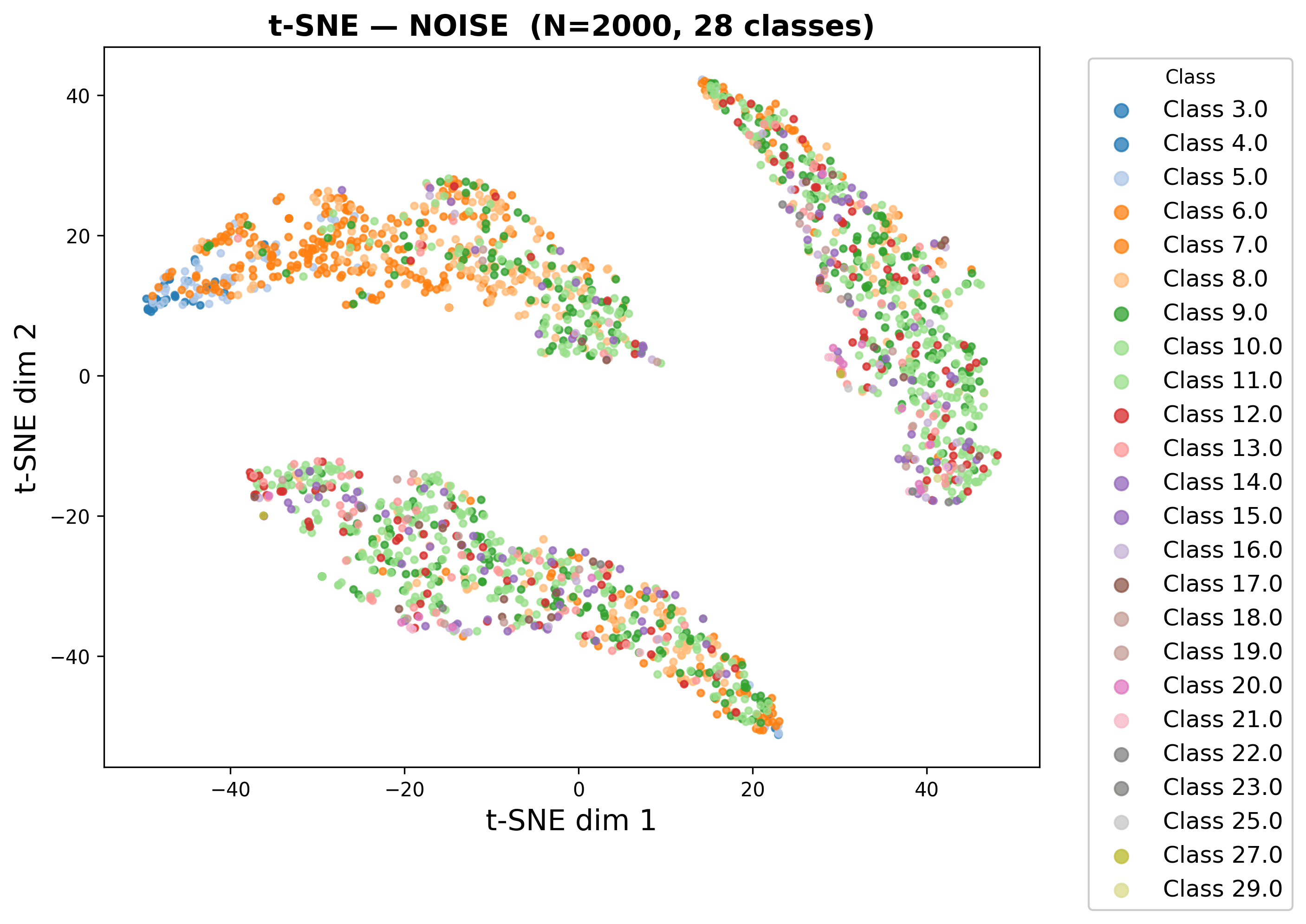}}
\caption{Abalone, Noise Embedding}
\end{subfigure}
\hfill
\begin{subfigure}{0.49\linewidth}
\centering
\includegraphics[width=\linewidth]{\detokenize{f_tsne_abalone.png}}
\caption{Abalone, Fourier Embedding}
\end{subfigure}

\caption{t-SNE visualizations for Mushroom and Abalone optical features under both embeddings. Mushroom forms a more directly separable binary structure, while Abalone forms continuous structures associated with regression targets.}
\label{fig:si tsne mushroom abalone}
\end{figure}

\subsection{Empirical kernel matrices}

\begin{figure}[H]
\centering
\begin{subfigure}{0.85\linewidth}
\centering
\includegraphics[width=\linewidth]{\detokenize{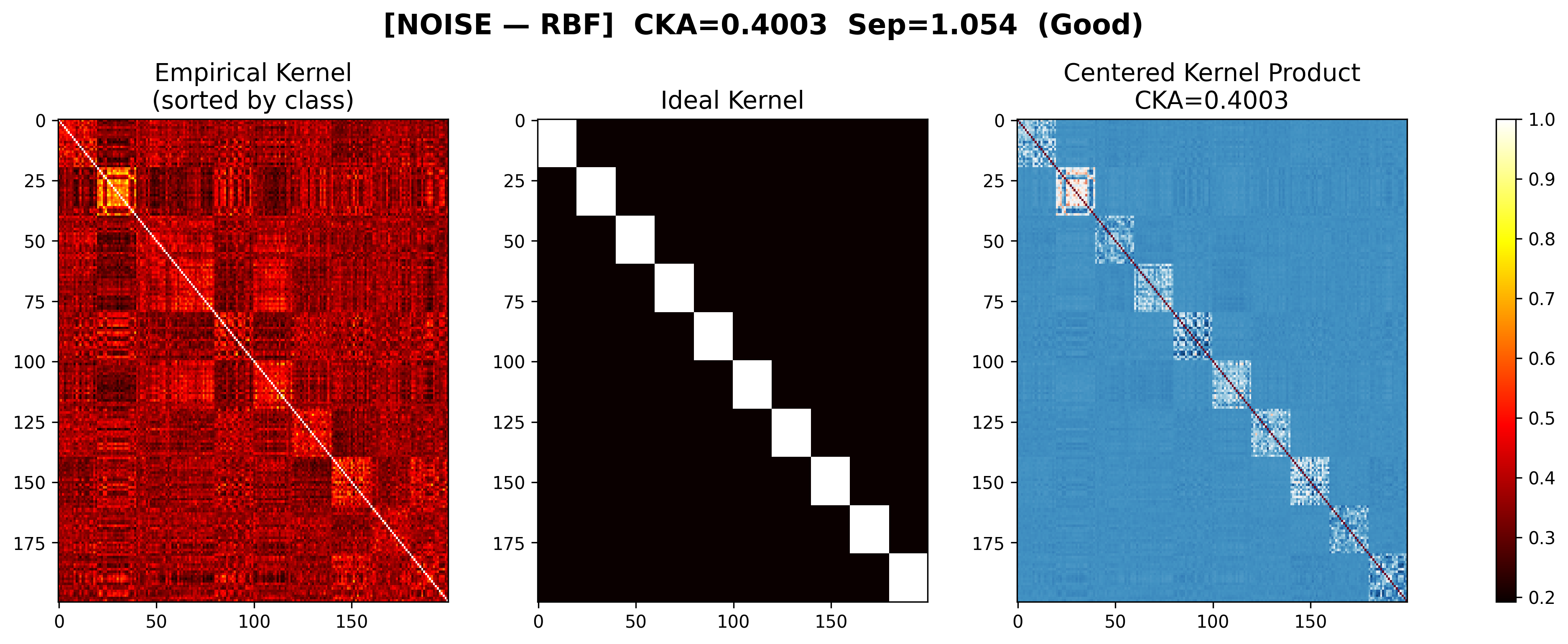}}
\caption{Noise Embedding}
\end{subfigure}
\hfill
\begin{subfigure}{0.85\linewidth}
\centering
\includegraphics[width=\linewidth]{\detokenize{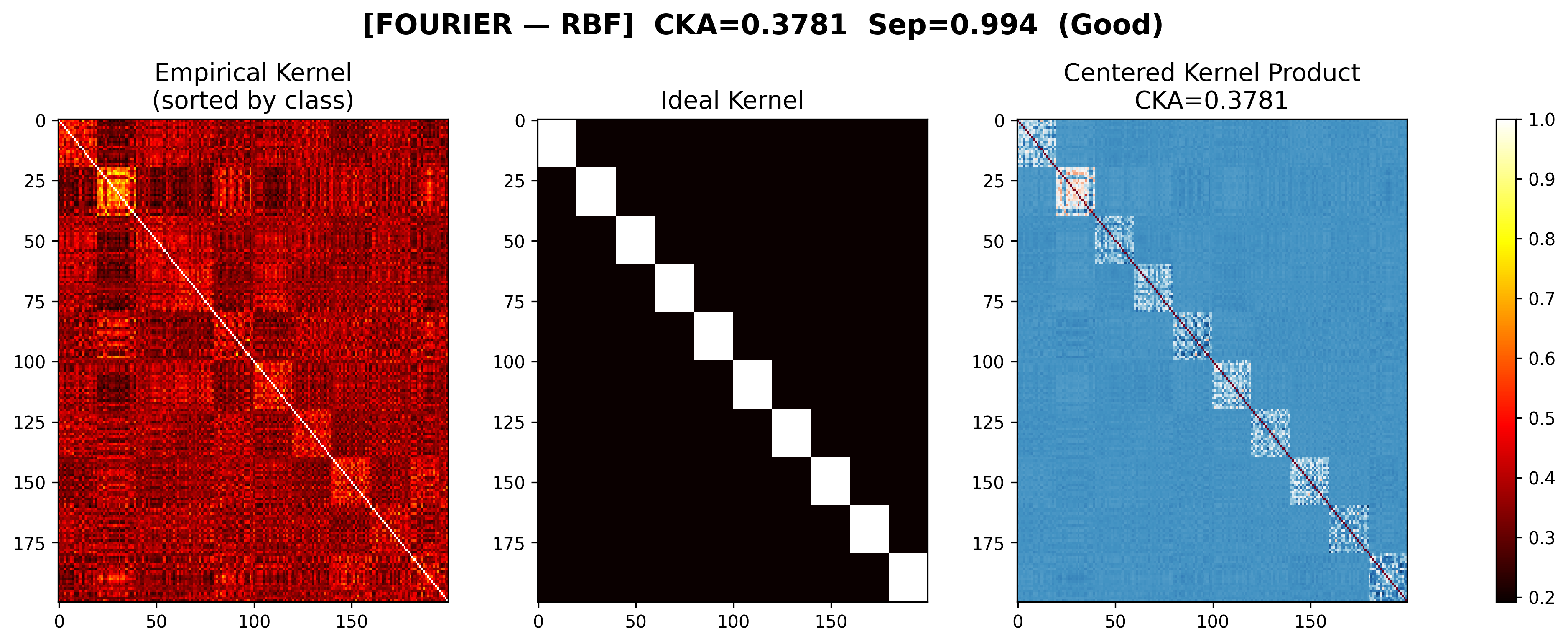}}
\caption{Fourier Embedding}

\end{subfigure}
\caption{Empirical optical kernel matrices for MNIST using (a) Noise Embedding and (b) Fourier embedding.}
\label{fig:si kernel mnist}
\end{figure}

The value labeled ``Sep'' in the heatmap titles of figs.~\ref{fig:si kernel mnist} \ref{fig:si kernel abalone} is the ratio of mean between class to within class kernel similarity derived from the empirical RBF kernel matrix. It is a class separation score quantifying how well the optical kernel organizes samples by class, and is distinct from the kernel fit $r$ reported in supp.~\ref{app:kernel fits} and the distance preservation $r$ reported in Sec.~\ref{subsec:distance preservation empirical} (Fig.~\ref{fig:distance preservation} and 
Supp.~\ref{app:feature diagnostics}, Fig.~\ref{fig:si distance preservation}).

\begin{figure}[H]
\centering
\begin{subfigure}{0.85\linewidth}
\centering
\includegraphics[width=\linewidth]{\detokenize{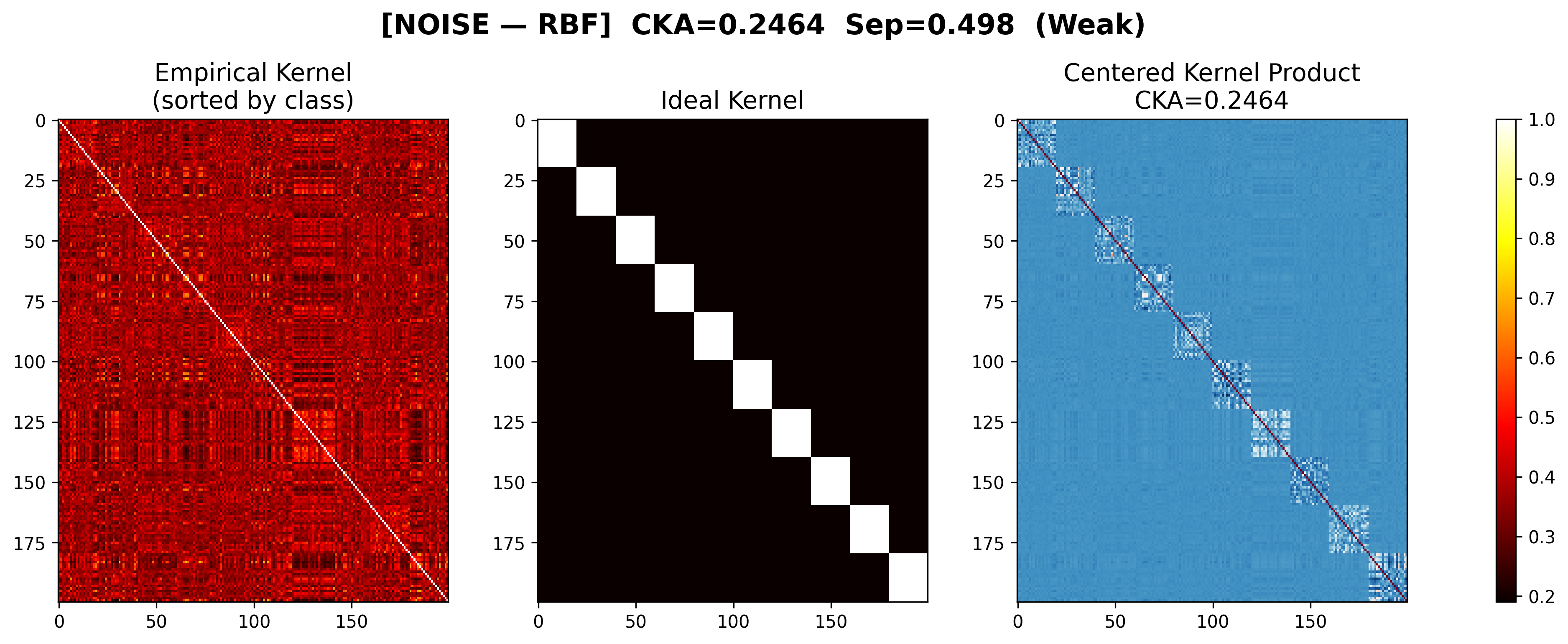}}
\caption{Noise Embedding}
\end{subfigure}
\hfill
\begin{subfigure}{0.85\linewidth}
\centering
\includegraphics[width=\linewidth]{\detokenize{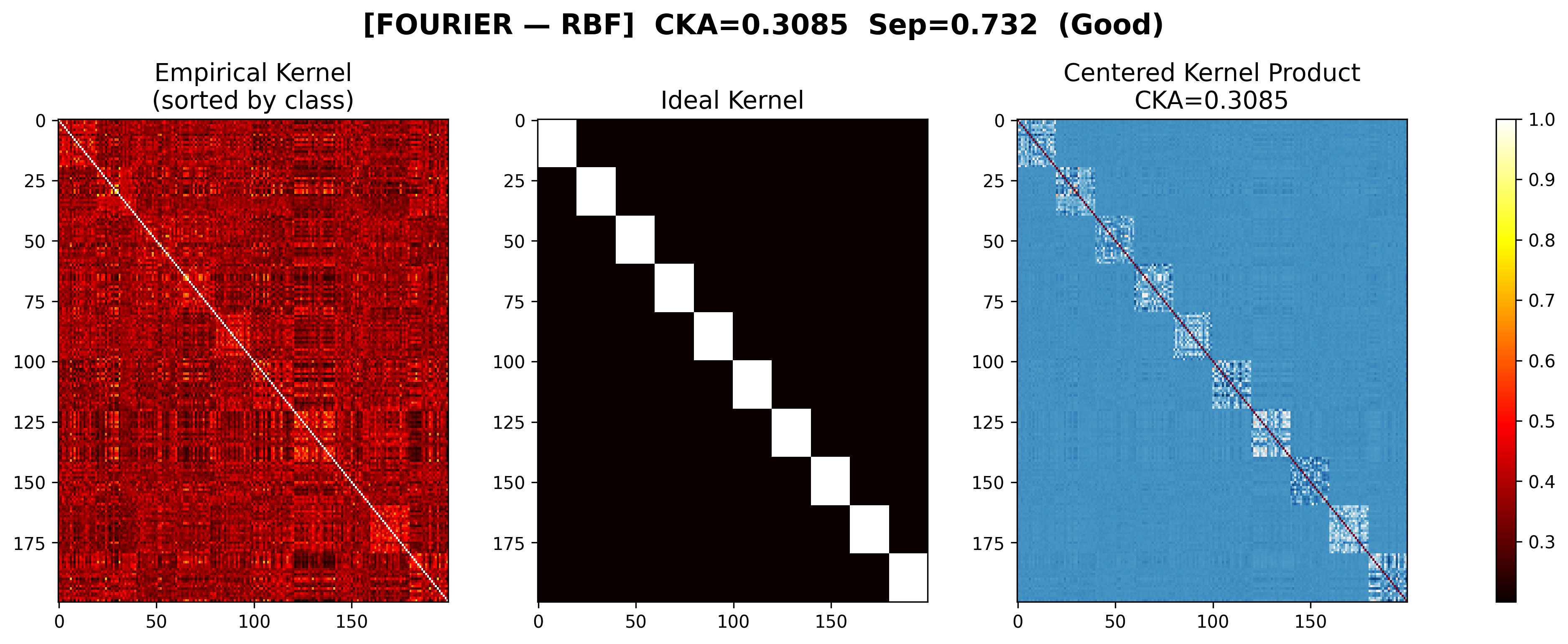}}
\caption{Fourier Embedding}
\end{subfigure}
\hfill

\caption{Empirical optical kernel matrices for FSDD using (a) Noise Embedding and (b) Fourier embedding.}
\label{fig:si kernel fsdd}
\end{figure}

\begin{figure}[H]
\centering
\begin{subfigure}{0.85\linewidth}
\centering
\includegraphics[width=\linewidth]{\detokenize{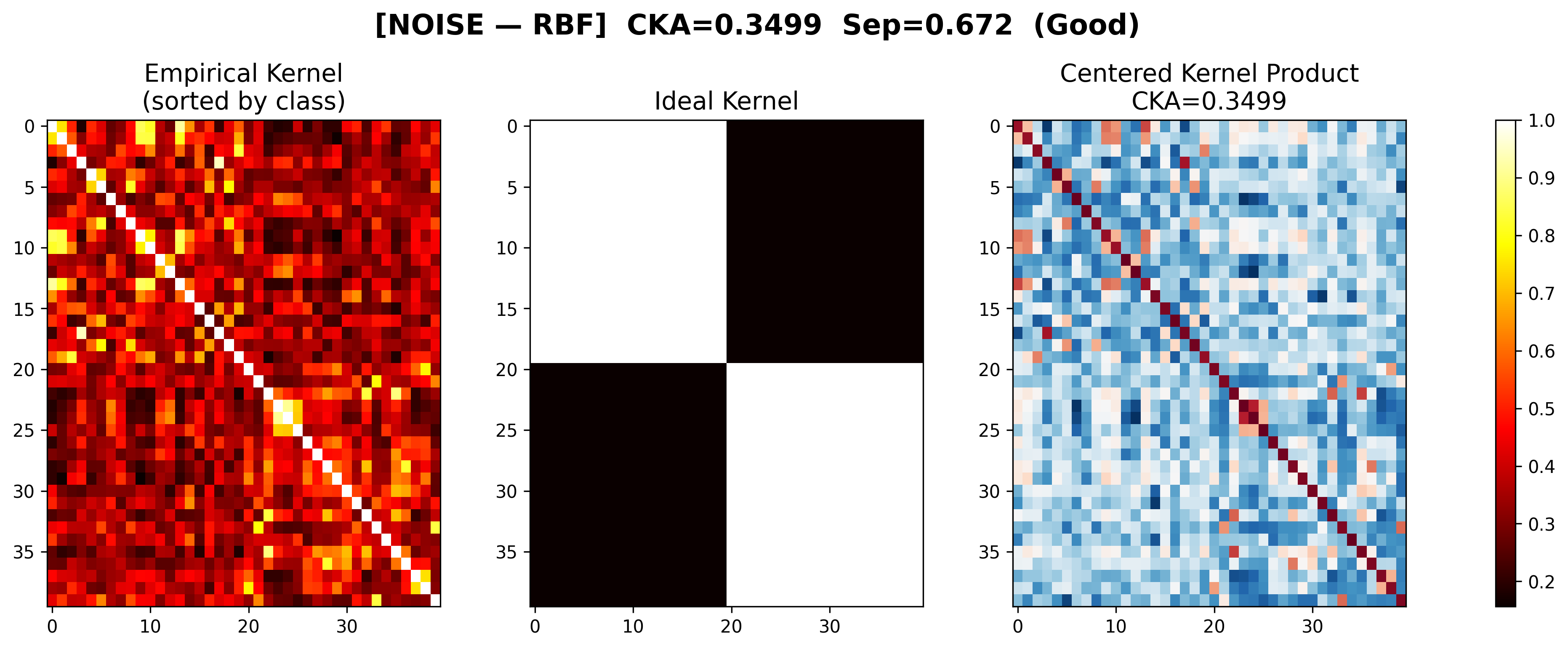}}
\caption{Noise Embedding}
\end{subfigure}
\hfill
\begin{subfigure}{0.85\linewidth}
\centering
\includegraphics[width=\linewidth]{\detokenize{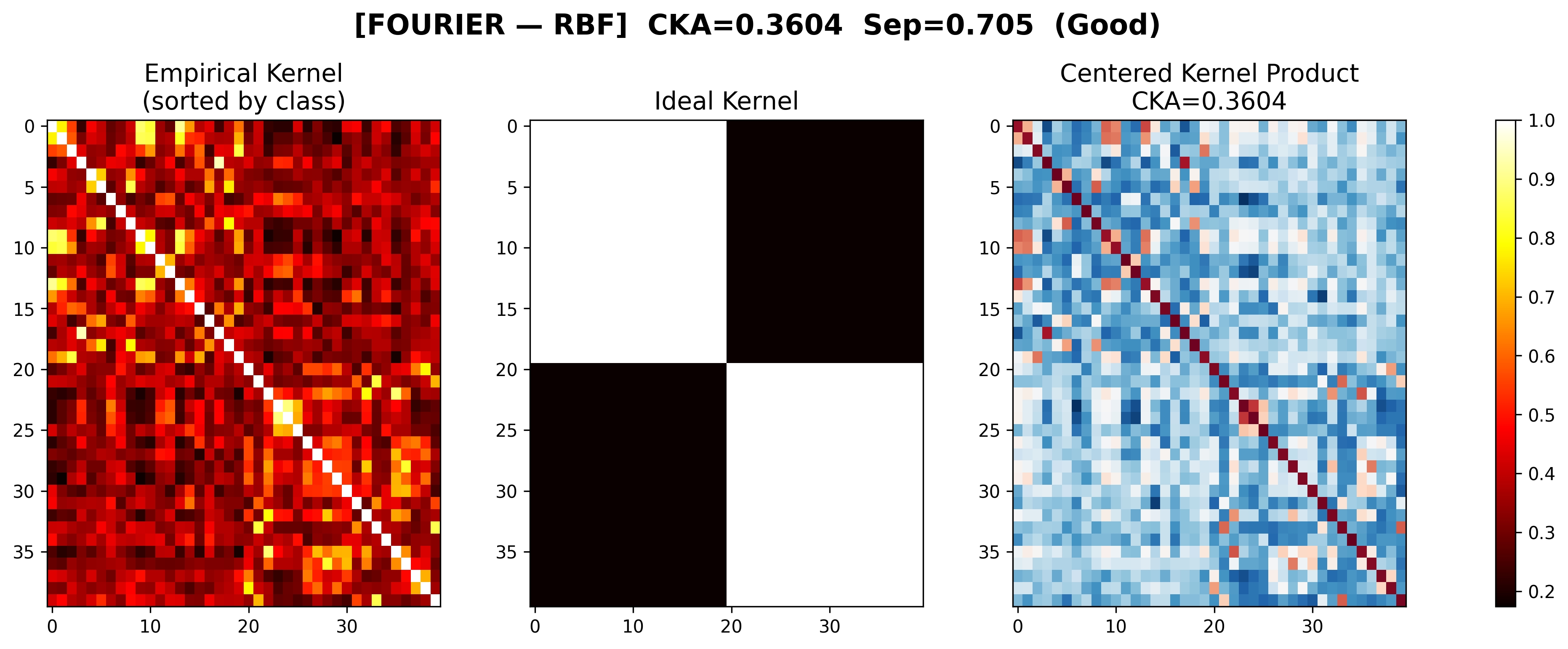}}
\caption{Fourier Embedding}
\end{subfigure}
\hfill

\caption{Empirical optical kernel matrices for Mushroom using (a) Noise Embedding and (b) Fourier Embedding}
\label{fig:si kernel mushroom}
\end{figure}

\begin{figure}[H]
\centering
\begin{subfigure}{0.85\linewidth}
\centering
\includegraphics[width=\linewidth]{\detokenize{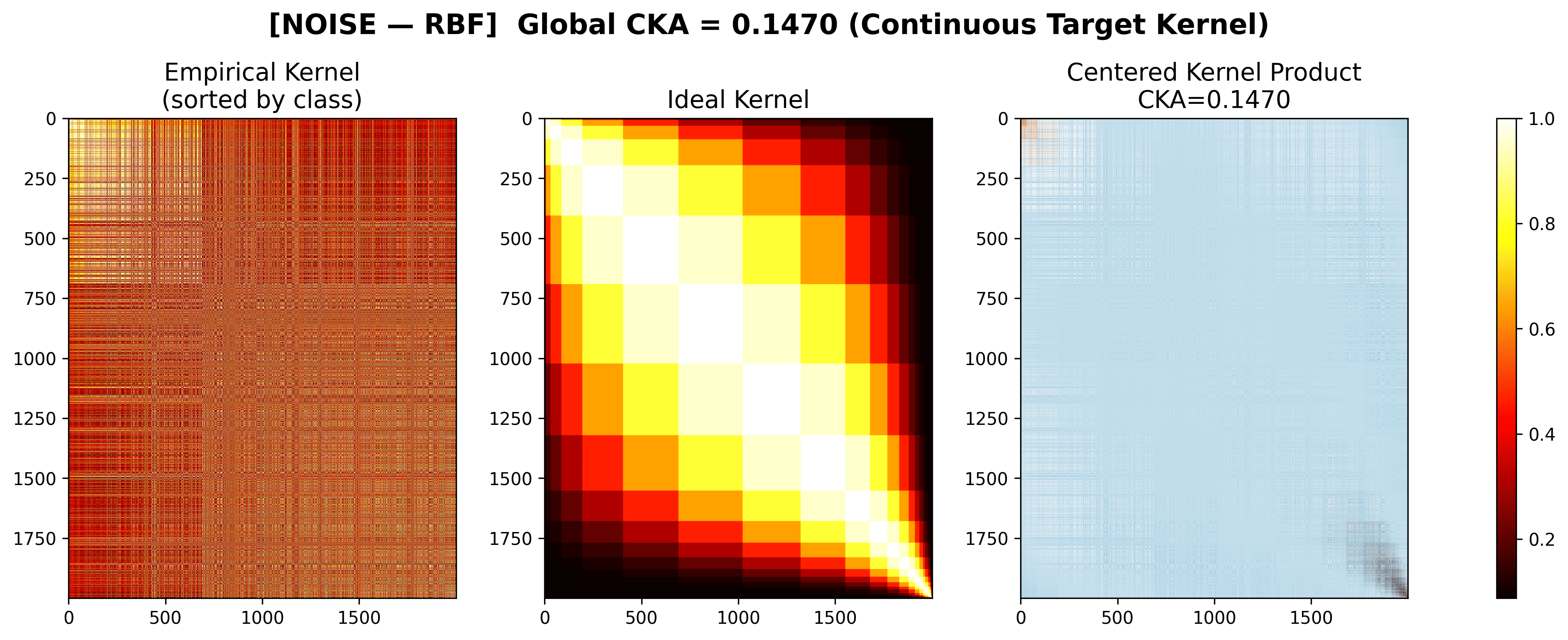}}
\caption{Noise Embedding}
\end{subfigure}
\hfill
\begin{subfigure}{0.85\linewidth}
\centering
\includegraphics[width=\linewidth]{\detokenize{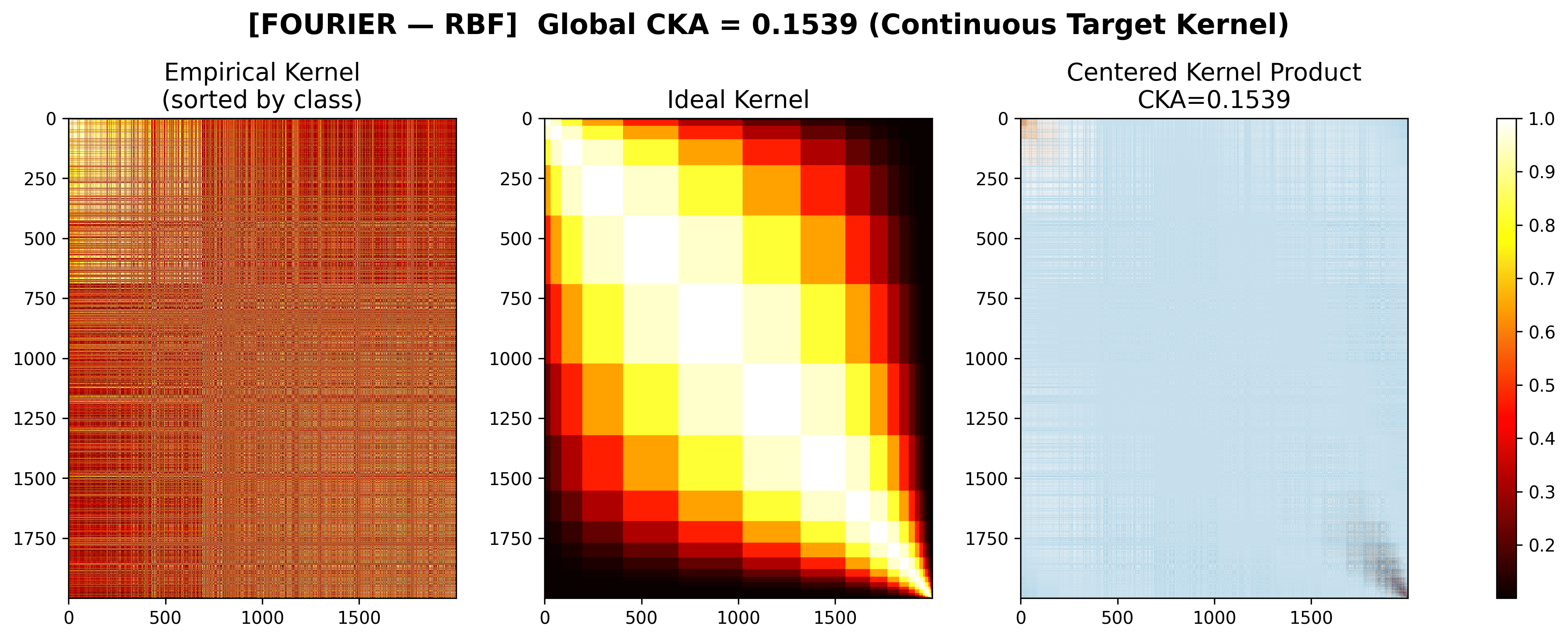}}
\caption{Fourier Embedding}
\end{subfigure}
\hfill

\caption{Empirical optical kernel matrices for Abalone using (a) Noise Embedding and (b) Fourier Embedding}
\label{fig:si kernel abalone}
\end{figure}

\subsection{Summary of CKA diagnostics}

\begin{table}[H]
\centering
\caption{Representative completed CKA diagnostics. CKA is interpreted as an encoding diagnostic rather than as a direct predictor of accuracy.}
\label{tab:cka}
\renewcommand{\arraystretch}{1.15}
\begin{tabular}{@{}lllll@{}}
\toprule
Modality & Embedding & RBF CKA & Performance \\
\midrule
MNIST & Noise & 0.4003 & 96.35\% \\
MNIST & Fourier & 0.3781 & 96.56\% \\
FSDD & Noise & 0.2464 & 93.00\% \\
FSDD & Fourier & 0.3085 & 95.67\% \\
Mushroom & Noise & 0.3336 & 100.00\% \\
Mushroom & Fourier & 0.3454 & 100.00\% \\
Abalone & Noise & 0.1470 & 0.0699 NRMSE \\
Abalone & Fourier & 0.1539 & 0.0704 NRMSE \\
\bottomrule
\end{tabular}
\end{table}

The CKA diagnostics show that global kernel label alignment varies substantially across modalities. MNIST and Mushroom have stronger global alignment, while FSDD and Abalone have weaker CKA values.This reinforces the main empirical conclusion: the optical feature space need not organize all modalities in the same way.Global alignment, distance preservation, t-SNE structure, and final readout performance should therefore be interpreted together rather than reduced to a single scalar diagnostic.

\FloatBarrier

\bibliographystyle{unsrtnat}
\bibliography{references}

\end{document}